\pdfoutput=1
\documentclass[aps,prx,reprint,superscriptaddress,longbibliography,floatfix]{revtex4-2}
\usepackage{makecell}
\usepackage{cellspace}
\usepackage{amsmath}
\usepackage{amsfonts}
\usepackage{amssymb}
\usepackage{graphicx}
\usepackage{svg}
\usepackage{bbding}
\usepackage{mathtools}
\usepackage{centernot} 
\usepackage{amsthm} 
\usepackage{cmap}
\usepackage{xcolor}
\usepackage{bigints}
\usepackage{dsfont} 
\usepackage{esint} 
\usepackage{physics}
\usepackage{natbib}
\usepackage{nicefrac} 
\usepackage{cancel}
\usepackage{bm}
\usepackage{upref}
\usepackage{multirow}
\usepackage{slashed}
\usepackage{paralist}
\usepackage{pifont}

\usepackage{varioref} 
\usepackage[]{hyperref} 
\usepackage{url}
\usepackage[capitalise]{cleveref}

\newcommand{\transp}{\top}
\newcommand{\Km}{\mathcal{K}}
\newcommand{\Um}{\mathcal{U}}

\usepackage{wrapfig}
\usepackage[protrusion=true,tracking=true,expansion,final]{microtype}      
\usepackage{xcolor} 
\usepackage{booktabs} 

\usepackage{chemfig}
\usepackage{adjustbox}

\renewcommand{\cite}[1]{\citep{#1}}

\usepackage[T1]{fontenc}

\vbadness=\maxdimen
\definecolor{citecolor}{HTML}{2980b9}
\definecolor{mydarkblue}{rgb}{0,0.08,0.45}
\definecolor{urlcolor}{rgb}{0,.145,.698}
\definecolor{linkcolor}{rgb}{.71,0.21,0.01}

\hypersetup{ %
	pdftitle={},
	pdfauthor={Evgenii Zheltonozhskii, Barak A. Katzir, Netanel H. Lindner},
	pdfsubject={},
	pdfkeywords={},
	pdfborder=0 0 0,
	pdfpagemode=UseOutlines,
	bookmarksopen=true,
	breaklinks=true,  
	colorlinks=true,
	urlcolor=urlcolor,
	linkcolor=linkcolor,
	citecolor=citecolor,
	filecolor=mydarkblue,
	pdfview=FitH}

\DeclareFontFamily{OMX}{lmex}{}
\DeclareFontShape{OMX}{lmex}{m}{n}{<-> lmex10}{}

\newcommand{\hc}{\mathrm{h.c.}}

\begin{document}

\title{Classification and Signatures of  Dissipative Quantum Hall--Superconductor Junctions }

\author{Evgenii Zheltonozhskii}
\affiliation{\mbox{Physics Department, Technion, 320003 Haifa, Israel}}
 \email{evgeniizh@campus.technion.ac.il}
\author{Barak A. Katzir}
\affiliation{\mbox{Qedma Quantum Computing, Tel Aviv, Israel}}

\author{Yuval Oreg}
\affiliation{\mbox{Department of Condensed Matter Physics, Weizmann Institute of Science, Rehovot 7610001, Israel}}

\author{Erez Berg}
\affiliation{\mbox{Department of Condensed Matter Physics, Weizmann Institute of Science, Rehovot 7610001, Israel}}

\author{Netanel H. Lindner}
\affiliation{\mbox{Physics Department, Technion, 320003 Haifa, Israel}}

\date{\today}

\begin{abstract}
Fractional quantum Hall (FQH) edges with proximity-induced superconductivity can host parafermionic zero modes (PZMs), making them a promising platform for topological quantum computing. 
Recent experiments \cite{gul2021andreev} measured the edge conductance from crossed Andreev reflection (CAR) between counter-propagating FQH edges separated by a narrow superconductor. 
The observed conductance was small and non-quantized, with two distinct behaviors: some states show temperature-independent conductance, while others show conductance that grows as temperature decreases. 
We explain these results by dissipation to sub-gap modes in the superconductor and present a phase diagram for the FQH-superconductor junction. 
We show that for Jain states, temperature-independent non-quantized
conductance occurs only in the presence of an integer mode (such as at
filling $\nu=1\pm\frac{1}{m}$, with odd $m\geq3$) strongly coupled to the dissipative channel. 
The predicted temperature trends qualitatively agree with the experiment, which shows temperature-independent conductance for $\nu=2/3$ and integer states, but not for $\nu=1/3$ and $\nu=2/5$. 
We propose comparing $\nu=4/3$ and $2/3$ to test the predicted influence of the relative chirality of integer and fractional modes.
\end{abstract}


\maketitle 
\section{Introduction}
Topological quantum computing (TQC) uses the topological phases of matter to perform computations that can withstand decoherence and local noise \cite{kitaev2003faulttolerant}. One of the most well-studied TQC platforms is Majorana zero modes (MZMs) \cite{kitaev2001wire}, which host Ising anyons \cite{nayak1996ising,kitaev2006anyons,fendley2007ising}. However, their experimental implementation \cite{mourik2012majoranas,das2012majorana,finck2013zbp,chen2016experimental,ophetveld2020inasnetworks,banerjee2022local,aghaee202620secondparitylifetime} and detection \cite{fukane2008,pikulin2021protocol,valentini2022majoranalike,hess2023trivial} are challenging because of disorder and non-topological states with similar signatures. Furthermore, for Ising anyons, the topologically protected operations are limited to Clifford operators \cite{bravyi2006universal,nayak2007nonabelian,ahlbrecht2009clifford}, which are classically simulable \cite{gottesman1998heisenberg}. 

Parafermionic zero modes (PZMs) \cite{lindner2012fractionalizing,fendley2012parafermion,barkeshli2014parafermions,alicea2016parafermions} generalize MZMs and provide a larger \cite{clarke2013parafermions}, though still non-universal \cite{hutter2016parafermions}, topologically protected gate set.
One proposed way, among others \cite{barkeshli2013twistdefects,klinovaja2014parafermions,zhang2014z4parafermions,karki2022z3kondo}, to realize PZMs is to open the superconducting gap in the fractional quantum Hall (FQH) edge via proximity coupling to a superconductor. 

The simplest example of a gapped FQH edge state consists of a pair of identical counter-propagating edges \cite{levin2013protectededge}. Superconducting coupling between them can produce crossed Andreev reflection (CAR), in which an electron incident from one edge combines with an electron from the opposite edge to form a Cooper pair, leaving a hole that propagates away. As a result, the measured potential of the edge mode propagating away from the grounded superconductor becomes more negative as the injected current increases.
\citet{gul2021andreev} demonstrated superconductor-induced coupling between quantum Hall states, measured the downstream voltage response for several quantum Hall fillings and extracted a normalized CAR signal (Fig.\ 5f in Ref.\ \cite{gul2021andreev}). For $\nu=1/3$ and $2/5$, this signal increases as temperature decreases, whereas for integer states and $\nu=2/3$ it approaches a small, approximately temperature-independent value. An increasing CAR signal corresponds to a more negative downstream conductance $G$. The small conductance values are consistent with substantial charge loss at the edge.

\begin{figure}[tb]
\centering
            \includegraphics[width=\linewidth]{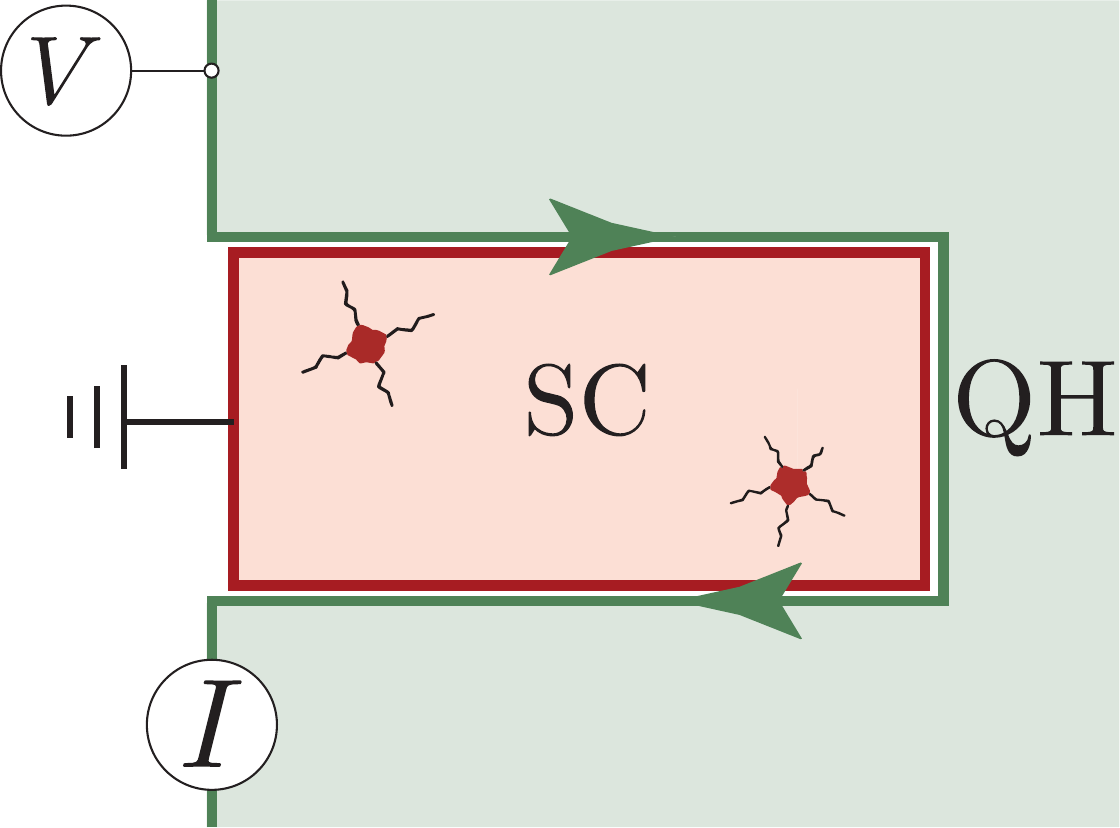}
        \caption{Experimental scheme: a thin grounded superconducting finger with impurities is inserted into a quantum Hall bar to create two counter-propagating edges.}
        \label{fig:rough_model}
\end{figure}

\begin{figure}[tb]
\centering
        \includegraphics[width=\linewidth]{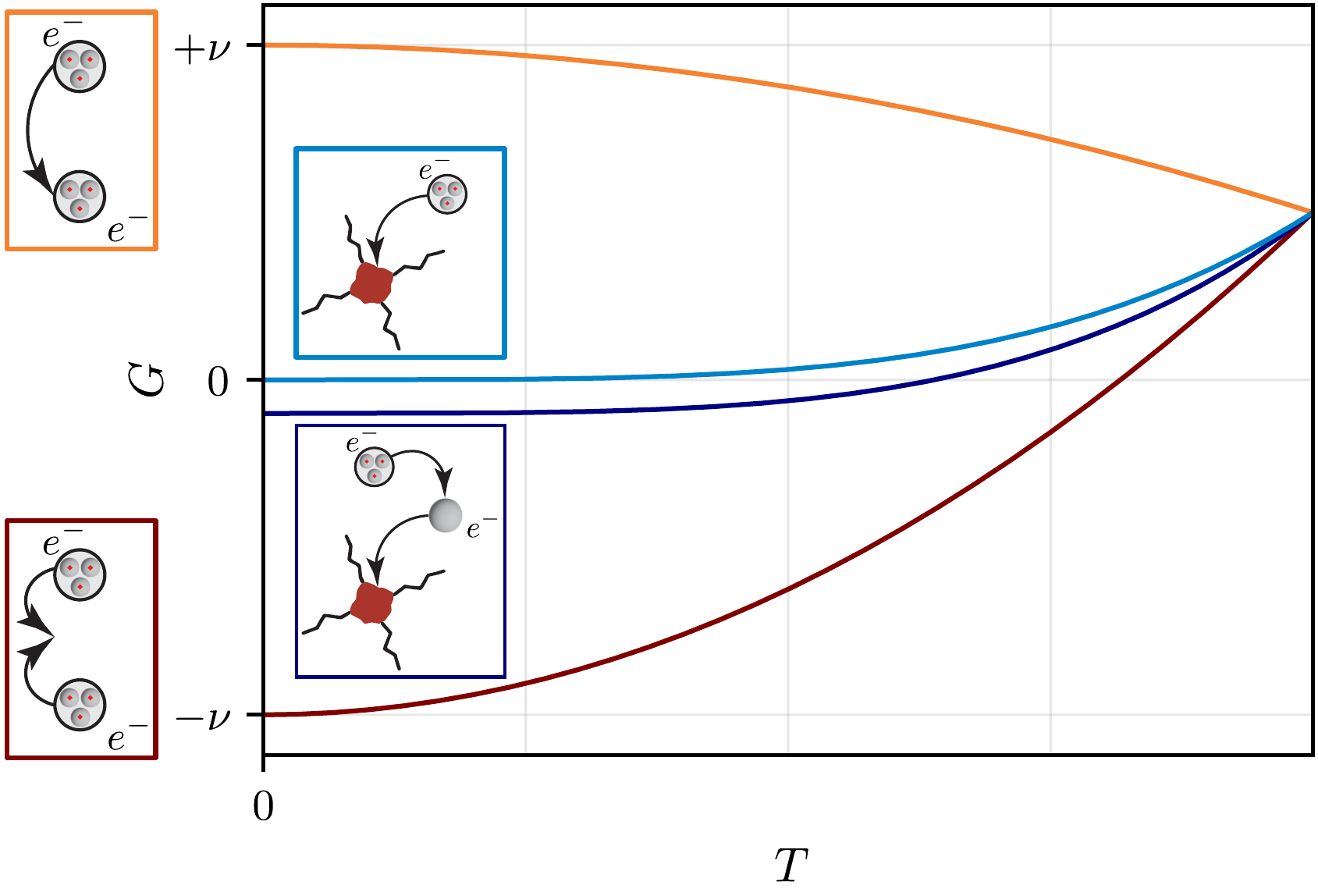}
        \caption{Conductance contribution of the fractional mode in different phases and corresponding leading perturbations at the edge: positive quantized conductance of $+\nu$ for tunneling across the finger (orange), negative conductance of $-\nu$ for CAR (red), zero conductance for dissipation (blue), and non-quantized temperature-independent conductance for coupling to a strongly dissipated integer mode (purple).}
        \label{fig:options}
\end{figure}

These findings have attracted considerable interest  \cite{michelsen2020currentcorr,manesco2021mechanism,kurilovich2022disorder,schiller2022interplay,galambos2022car,tang2022vortex,kurilovich2022criticality,david2022effects,cao2023signatures,zhao2023perfect,hu2024resistance,baba2026topology,bollmann2026phases,jamotte2026interferometer,cao2026interferometric}. In particular, \citet{schiller2022interplay} argued that the difference between integer and $\nu=1/m$ states can be explained by dissipation into subgap states of the superconductor. 
Impurities in the presence of broken time-reversal symmetry or Caroli--De Gennes--Matricon states in the vortex cores \cite{caroli1964subgap} due to high magnetic field can produce such subgap states that accept electrons from the edge in quantum Hall--superconductor devices \cite{rickhaus2012qhesc,amet2016qhesc,lee2017inducing,hashisaka2021andreev,gul2021andreev,hatefipour2021induced,zhao2022loss,vignaud2023evidence,uday2023induced,hatefipour2023andreev,zhao2024thermal}. 
For an integer edge, strong dissipation leads to a regime where the dissipation, normal tunneling, and CAR have the same scaling, yielding a temperature-independent, non-quantized conductance set by microscopic parameters \cite{schiller2022interplay}. For Laughlin $\nu=1/m$ edges, dissipation is irrelevant; the downstream conductance is therefore quantized at $G=-\nu e^2/h$, $+\nu e^2/h$, or $0$. A finite finger can return to ballistic behavior when all accumulated scattering rates vanish upon cooling. These competing processes give the conductance scenarios sketched in \cref{fig:options}.

In this paper, we extend the analysis to multimode edges consisting of a chiral Jain block with or without integer modes. Our model assumes non-zero, momentum-conserving tunneling and pairing amplitudes (i.e., no short-range-correlated disorder); the case in which this does not hold is discussed briefly in \cref{sec:disordered-couplings}. While the edge phase diagram includes multiple phases with different conductance values, the temperature-independent, non-quantized conductance phase appears \emph{if and only if} the edge supports an \emph{integer} mode (defined by the existence of a physical electron operator with absolute conformal spin $1/2$). Examples of systems whose edges include an integer mode are particle-hole conjugates of Jain states such as $2/3$ and $3/5$, and higher-Landau-level fractional states such as $4/3$.

The relative chirality of the modes further refines the picture. For upstream fractional modes (such as $\nu=2/3$), conductance is always small and non-universal, regardless of the phase, due to strong dissipation of the integer mode. For co-propagating edges like $4/3$, in contrast, fractional modes can contribute to conductance, and conductance measurements can distinguish distinct phases experimentally.

Edges that lack an integer mode behave differently. For the lowest-Landau-level Jain sequence $\nu=N/(2pN+1)$, e.g., $1/3$ and $2/5$, which encompasses most observed fractions without an integer mode \cite{jain1989cf,frohlich1997qheclassification,lin2014fqhereview}, the system flows to one of the quantized conductance plateaus. These results are summarized in \cref{tab:intro_summary}.

\begin{table}
    \centering
    \setlength\extrarowheight{2pt}
    
    \caption{Conductance regimes of FQH states with different edge structures: only fractional modes (\textbf{Fractional only}), fractional modes co-propagating with integer modes and charge flow  (\textbf{Integer + Downstream}), and fractional modes counter-propagating relative to integer modes  (\textbf{Integer + Upstream}). At strong dissipation, integer modes contribute $\order{\alpha^{-2}}$ \cite{schiller2022interplay}, where $\alpha$ quantifies the dissipation strength~\eqref{eq:diss_action}, while fractional modes can contribute either quantized or zero conductance.}
     \resizebox{\linewidth}{!}{
\begin{tabular}{lccc} 
    \toprule
   \multirow{3}{*}{\makecell{\textbf{Conductance at}\\  \textbf{Strong Dissipation}}} & \multicolumn{3}{c}{\textbf{Edge Mode Configuration}} \\
    \cmidrule(lr){2-4}
   & \makecell{\textbf{Fractional} \\ \textbf{only}} & \makecell{\textbf{Integer +} \\ \textbf{Downstream}} & \makecell{\textbf{Integer +} \\ \textbf{Upstream}} \\
    & $\nicefrac{1}{3}$, $\nicefrac{2}{5}$ & $\nicefrac{4}{3}$ & $\nicefrac{2}{3}$, $\nicefrac{3}{5}$  \\
    \midrule
     $\order{\alpha^{-2}}$  & \ding{55} & \ding{51} & \ding{51} \\
    Quantized + $\order{\alpha^{-2}}$    & \ding{55} & \ding{51} & \ding{55} \\
    Quantized    & \ding{51} & \ding{55} & \ding{55} \\
    Zero    & \ding{51} & \ding{55} & \ding{55} \\
    \bottomrule
\end{tabular}
    }
    \label{tab:intro_summary}
\end{table}

The rest of the paper is organized as follows: \cref{sec:result_summary} summarizes the results, \cref{sec:setup_model} describes the setup and our model, \cref{seq:cross-scattering} explains the calculations of scaling dimensions in multimode edges, \cref{sec:currents} calculates currents and shows when the temperature-independent non-quantized conductance phase is possible, \cref{sec:example} compares the numerical conductance for $\nu=2/5$, $\nu=2/3$, and $\nu=4/3$, and \cref{sec:conclusion} concludes the paper.

\section{Main Results}
\label{sec:result_summary}

We consider two counter-propagating quantum Hall edges that are locally coupled by a narrow superconducting finger, as sketched in \cref{fig:rough_model}. In the perturbative regime relevant to the experiments, we retain single-electron tunneling, two-electron pairing, and electron loss to the bath. Inter-side tunneling and pairing involve one elementary electron operator on each side; same-side transfer between integer and fractional modes is also retained. We neglect higher multielectron charge-transfer processes, which we expect to be irrelevant in the physical regime considered here, while retaining the lowest-order (four-electron) neutral couplings. 
In addition, intra-edge processes (both backscattering \cite{kane1995impurity} and superconducting coupling) do not conserve momentum but can be relevant in the presence of disorder.

For Laughlin edges, \citet{schiller2022interplay} distinguished superconducting and normal-tunneling gaps from gapless regimes in which all perturbations are irrelevant. In the latter case, the slowest-vanishing rate controls the transport response: CAR, normal tunneling, or dissipation. In the long-finger limit, these give $G=-\nu e^2/h$, $+\nu e^2/h$, and $0$, respectively. ``Gapped'' here refers to the edge; the fermionic bath remains gapless. For a finite finger, cooling below the scale at which its length cuts off the extended interaction produces a point-contact crossover.

\textbf{Chiral Jain States.} For chiral Jain states such as $\nu=2/5$ \cite{jain1989cf}, all modes on one side propagate downstream. Each side of the finger has several electronic modes, so multiple CAR and normal-tunneling terms are possible. We show that in Jain states, a CAR term and a normal-tunneling term cannot be simultaneously relevant. This restriction does not extend to multielectron processes. If either type reaches strong coupling, it can open a partial or complete charge gap. Hence, two types of phases can have a charge gap and quantized conductance: CAR-dominated phases and normal-tunneling-dominated phases. Additional neutral gaps due to neutral inter-edge couplings may change the conductance if the charge sector is partially gapped or entirely gapless. Fractional electron dissipation is always irrelevant, but it can be the leading contribution to transport. Thus, phases with a gapless charge sector can have different quantized conductance values, depending on the scaling dimensions and relation between bare coefficients of the competing processes. Neutral intra-edge disorder \cite{kane1994edge,kane1995impurity} likewise does not transfer charge directly between sides, but can change the scaling dimensions of charge-transfer processes.

\begin{figure*}[t]
\centering
\begin{minipage}[t]{.48\textwidth}
\centering
\includegraphics[width=\linewidth]{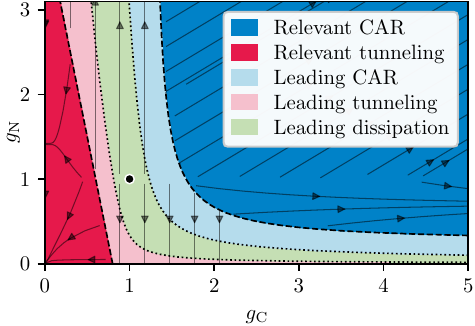}
\caption{ Weak-coupling phase diagram for the $\nu=2/5$
edge ($N=2$, $p=1$) as a function of
inter-side Luttinger parameters of the charge and neutral mode pairs,
$g_{\rm C}$ and $g_{\rm N}$. Dark regions have a relevant elementary CAR or tunneling operator;
dashed curves mark $d_A=2$ and $d_T=2$. In the
gapless regions both perturbations are irrelevant, and the label
identifies the one determining the conductance.
The dot at $(1,1)$ has no inter-side Gaussian interaction. Arrows show the flow from relevant charged and neutral operators, all assigned equal amplitudes; irrelevant terms are omitted.}
\label{fig:pd2d}
\end{minipage}\hfill
\begin{minipage}[t]{.48\textwidth}
\centering
\includegraphics[width=\linewidth]{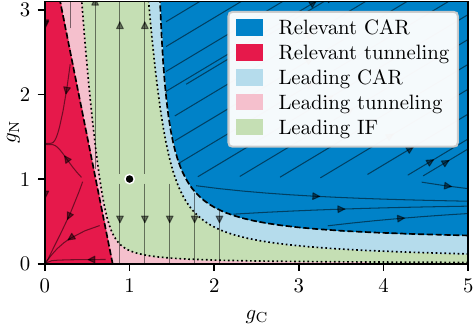}
\caption{Weak-coupling phase diagram for the fractional sector of
$\nu=3/5=1-2/5$, as a function of
inter-side Luttinger parameters of the fractional charge and neutral mode pairs,
$g_{\rm C}$ and $g_{\rm N}$. We set $g_{\mathrm I}=1$ and take the slow-integer limit
$u_{\mathrm I}/u_{\rm f}\to0$.
Dark regions have a relevant elementary CAR or tunneling operator;
dashed curves mark $d_A=2$ and $d_T=2$. Since fractional modes are upstream, the conductance is determined by the integer mode conductance in all cases.
The dot at $(1,1)$ has no inter-side Gaussian interaction. Arrows show the flow from relevant charged and neutral operators, all assigned equal amplitudes; irrelevant terms are omitted.
}
\label{fig:pd35}
\end{minipage}
\end{figure*}

To illustrate the possible phases for a chiral Jain state, we solve analytically a simple model in which RG flow is diagonal in the charge--neutral basis \cite{kane1995impurity}; we parametrize the phase diagram by $N$ Luttinger parameters: $g_{\rm C}$ for the charge mode and $g_i$ ($i=1, \ldots, N-1$) for the neutral modes.
\Cref{fig:pd2d} shows the $\nu=2/5$ case with one neutral pair and parameter $g_{\rm N}$.
The corresponding diagram for its particle--hole conjugate
$\nu=3/5$ includes coupling between integer and fractional modes and is shown
in \cref{fig:pd35}.

\textbf{Quantum Hall states containing integer edge modes.} Strong dissipation suppresses coherent integer--fractional hybridization inside the finger. Near the dissipation-dominated limit $g_{\rm I}=1$, the conductance coefficients for CAR, normal tunneling, and dissipation involving the integer mode are temperature independent, so its contribution is small and non-quantized \cite{schiller2022interplay}.

We consider two experimentally relevant cases. In the first case, all modes co-propagate with the charge-flow direction (e.g., at $\nu=4/3$). In this case, the fractional modes admit all the phases discussed above, with quantized conductance defined by the total filling of the fractional part. In the additional family of phases, the most relevant process involving fractional modes is their coupling to dissipative integer modes. This process suppresses the conductance of the fractional modes. As a result, in such a phase, the integer mode dominates edge transport, yielding a temperature-independent, non-quantized conductance.

The situation changes qualitatively when fractional modes propagate upstream, as in $\nu=2/3$ and $3/5$ \cite{kane1995impurity}. Strong dissipation suppresses integer--fractional hybridization inside the finger (and thus, the edge is not in a Kane--Fisher--Polchinski fixed point).
As a result, upstream fractional modes cannot directly contribute to the downstream current, regardless of the phase. Consequently, the small non-universal response of the integer mode controls the conductance to leading order, even when the fractional sector is gapped. We summarize these results in \cref{tab:sds23_summary}.

The distinction between $\nu=4/3$ and $\nu=2/3$ suggests an experimental verification of our theory. For $\nu=4/3$, the fractional sector can contribute $G_{\rm f}=\pm\nu_{\rm f}e^2/h$, or coupling to the dissipative integer sector can suppress this contribution. For $\nu=2/3$, the total conductance is temperature-independent, non-quantized, and independent of the phase.

\begin{table}[t]
\centering
\caption{\textbf{Contribution of the fractional (Jain) part of the edge to the total conductance}.   The table gives $G_{\rm f}/(e^2/h)$ for three representative edge structures. Both normal-tunneling- and CAR-dominated phases can be gapped or gapless. We assume no strictly neutral pinned operators (\cref{sec:generic-phase-structure}). The total conductance is $G=G_{\rm f}+G_{\rm I}$, where $G_{\rm I}$ is the small, non-universal contribution from the integer mode, if present. The upstream entries retain the small contact-induced fractional contribution; $\alpha$ denotes the integer dissipation strength. In the presence of a neutral gap, the conductance is still quantized, but available values are not limited to $\nu_{\rm f}$.}
\setlength\extrarowheight{2pt}
\setlength{\tabcolsep}{3pt}
\begin{tabular}{lccc}
\toprule
   \multirow{3}{*}{\textbf{Phase}} & \multicolumn{3}{c}{\textbf{Edge Mode Configuration}} \\
    \cmidrule(lr){2-4}
   & \makecell{\textbf{Fractional} \\ \textbf{only}} & \makecell{\textbf{Integer +} \\ \textbf{Downstream}} & \makecell{\textbf{Integer +} \\ \textbf{Upstream}} \\
    & $\nicefrac{1}{3}$, $\nicefrac{2}{5}$ & $\nicefrac{4}{3}$ & $\nicefrac{2}{3}$, $\nicefrac{3}{5}$  \\
    \midrule
    Normal tunneling & $\nu_{\rm f}$ & $\nu_{\rm f}$ & $\order{\alpha^{-2}}$ \\
    CAR & $-\nu_{\rm f}$ & $-\nu_{\rm f}$ & $\order{\alpha^{-2}}$ \\   
    \makecell[l]{Fractional\\dissipation} & $0$ & -- & -- \\
    \makecell[l]{Integer--fractional\\coupling} & -- & $\order{\alpha^{-2}}$ & $\order{\alpha^{-2}}$ \\
\bottomrule
\end{tabular}
\label{tab:sds23_summary}
\end{table}

The observation or absence of CAR alone cannot unambiguously diagnose parafermionic zero modes: a gapless edge can produce a CAR response, and dissipation can mask gapped fractional modes. Therefore, detecting PZMs requires additional experimental protocols.

For weak dissipation, coherent integer--fractional hybridization can survive, and the strong-dissipation analysis above need not apply. This regime can be relevant, for example, in the absence of a magnetic field, specifically in the experiments involving fractional Chern insulators. In this case, we do not expect non-quantized conductance as in the fractional-only case.

Two further settings lie beyond the present model. First are multilayer states, in which the edge modes are weakly coupled, such as $1/5+1/5$ or $1/3+1/3$. These edges can be treated by considering each layer separately when interlayer coupling is small. Similarly, edge reconstruction can produce
additional edge modes \cite{meir1994compositeedge}, for example, two downstream $1/3$ charge channels at $\nu=2/3$ accompanied by upstream neutral modes \cite{wang2013edgereconstruction,sabo2017reconstruction,manna2024shotnoise}.

\section{Quantum field theory of quantum Hall edges}
\label{sec:setup_model}
A quantum Hall edge with $n_{\rm edge}$ modes can be modeled as an $n_{\rm edge}$-component chiral Luttinger liquid governed by the Wen--Zee  action~\cite{wenzee1992classification,wen1995edge,hansson2016quantum}:
\begin{align}
    S_0 &= -\frac{1}{4\pi} \int \dd{x}\dd{t} \qty[\partial_x \vb*{\phi}^\transp  \Km \partial_t \vb*{\phi} +  \partial_x \vb*{\phi}^\transp  \Um \partial_x \vb*{\phi} ] \label{eq:edge_action}
\end{align}
with total filling
\begin{align}
   \nu = \vb{q}^\transp \qty(\Km^{-1})\vb{q}.
\end{align}
Here, $\vb*{\phi}$ is a vector of $n_{\rm edge}$ bosonic operators, $\Km$ is an integer symmetric matrix that encodes the topological order, $\Um$ is a positive-definite coupling matrix, and $\vb{q}$ is the charge vector. We use the real-time convention $\hat{\vb*{\rho}}=-\partial_x\vb*{\phi}/(2\pi)$, so that a positive-$\Km$ mode propagates toward increasing $x$ and $\exp(i\vb m^\transp\vb*\phi)$ creates charge $e\vb m^\transp\Km^{-1}\vb q$.
Since odd-charge local excitations are fermionic and even-charge ones are bosonic, the spin-statistics theorem requires \cite{levin2012abelian}
\begin{align}
    \Km_{ii} \equiv q_i \mod 2,
\end{align}
and the existence of a unit-charge electron demands $ \mathrm{gcd}(q_1, \dots, q_{n_{\rm edge}})=1$. 
The $\Km$ matrix describing the state is not unique. A unimodular basis
change $\mathsf{W}\in\mathrm{SL}(n_{\rm edge},\mathbb{Z})$ gives the equivalent description
$\mathsf W^\transp\Km\mathsf W$ and
$\mathsf W^\transp\vb q$.

In particular, Jain sequence states \cite{halperin1983theory,jain1989cf,kane1995impurity} with $\nu=N/(2pN\pm1)$ are described with the $N\times N$ $\Km$ matrix
\begin{align}
    \Km_{ij} &= 2p \pm \delta_{ij},
\end{align}
and $\vb{q}_j=1$ (symmetric basis).

Quasiparticle excitations at the edge are labeled by dimensionless integer vectors $\vb{m}\in\mathbb{Z}^{n_{\rm edge}}$, and are created by the vertex operators
\begin{align}
    V_{\vb{m}} &= \exp(i \vb{m}\vdot \vb*{\phi}).
\end{align}
Their charge and statistical angle are given by
\begin{align}
 Q_{\vb{m}}/e &= \vb{m}^\transp \Km^{-1} \vb{q}\\
    \theta &= \pi \cdot \vb{m}^\transp \Km^{-1} \vb{m} \label{eq:statistics}.
\end{align}
Since the electronic wavefunction is single-valued, in the symmetric basis ($q_i=1$) electronic operators correspond to $\vb{m} = \Km \vb{n}$ for $\vb{n} \in \mathbb{Z}^{n_{\rm edge}}$ with integer $n_i$ and unit charge, i.e., satisfying $\vb{q} \vdot \vb{n}=1$  \cite{wenzee1992classification}.

For the system described by \cref{eq:edge_action} to be fully gapped, $n_{\rm edge}$ must be even, with $n_{\rm edge}/2$ modes of each chirality. \citet{levin2013protectededge} showed that the necessary and sufficient condition for a gappable edge is the existence of a set of linearly independent integer-valued vectors $\qty{\vb*{\Lambda}_i}_{i=1}^{n_{\rm edge}/2}$ such that
\begin{align}
    \forall i,j \quad \vb*{\Lambda}_i^\transp \Km \vb*{\Lambda}_j = 0.
\end{align}
The gapping operators conserve fermion parity and are given by $\vb{m}_i^\transp = \vb*{\Lambda}^\transp_i \Km$.

The scaling dimension $d_{\vb{m}}$ of an operator
$y_{\vb{m}}V_{\vb{m}}$ defines the RG flow of its
coefficient via
\begin{align}
    \dv{y_{\vb{m}}}{\ell} &= (2-d_{\vb{m}})y_{\vb{m}},
\end{align}
where $\ell=\ln(a/a_0)$ is the RG time, $a$ is the running short-distance cutoff, and $a_0$ is its initial value.
For a general coupling matrix $\Um$, the scaling dimensions can be
calculated using a diagonalizing matrix $W$:
\begin{align}
\tilde{\Km}=W^\transp \Km W,
\\
\tilde{\Um}=W^\transp  \Um W,
\label{eq:diagonalizing-W}
\end{align}
that decouples the components, i.e., both $\tilde{\Km}$ and $\tilde{\Um}$ are diagonal, and $\tilde{\Km}$ has entries $\pm 1$ \cite{kane1995impurity,katzir2020superconducting}. We find this matrix by solving a generalized eigenvalue problem and normalizing the eigenvectors in the $\Km$ metric.
Denoting $\vb*{\lambda} = W^\transp \vb{m}$, the operator
$V_{\vb{m}}$ then has scaling dimension
\begin{align}
d_{\vb{m}}=\frac{\vb*{\lambda}  \vdot \vb*{\lambda} }{2}. \label{eq:sd}
\end{align}
The conformal spin of the operator is
\begin{align}
    s_{\vb{m}}&=\frac{1}{2}\vb{m}^\transp\Km^{-1}\vb{m}
    =\frac{1}{2}\vb{n}^\transp\Km\vb{n}.
\end{align}
For the Jain state $\Km_{ij}=2p+\delta_{ij}$, the smallest
conformal spin of a local electron is $p+\nicefrac{1}{2}$. Following the unitarity bound in 1+1D \cite{ginsparg1988applied}, i.e., the fact that conformal weights are non-negative, the scaling dimension satisfies $d_{\vb{m}}\geq\abs{s_{\vb{m}}}$.
Alternatively, this can be seen from the fact that 
\begin{align}
\qty(W^\transp \vb{m})^\transp \vdot \qty(W^\transp \vb{m})
\geqslant
\abs{
    \qty(W^\transp \vb{m})^\transp
    \tilde{\Km}
    \qty(W^\transp \vb{m})
}
=
\abs{\vb{n}^\transp\Km\vb{n}};
\end{align}
the equality is attained when $\vb{m}$ is fully chiral. 

Similarly, for two vectors $\vb*{\lambda}_1$ and $\vb*{\lambda}_2$ we can write
\begin{align}
    \abs{\vb*{\lambda}_1^\transp \tilde{\Km} \vb*{\lambda}_2 }\leqslant \norm{\vb*{\lambda}_1}\norm{\vb*{\lambda}_2}.
\end{align}
Consider operators $V_{\vb{m}_1}$ and $V_{\vb{m}_2}$. Then, since $W^\transp \Km W = \tilde{\Km}$, we have $\vb*{\lambda}_1^\transp \tilde{\Km} \vb*{\lambda}_2 = \vb{m}_1^\transp  \Km^{-1}  \vb{m}_2$; at the same time $\norm{\vb*{\lambda}_i}^2 = 2d_i$, where $d_i$ is the scaling dimension of $V_{\vb{m}_i}$. This gives a lower bound on a product of their scaling dimensions:
\begin{align}
    2\sqrt{d_1d_2} \geqslant \abs{\vb{m}_1^\transp \Km^{-1} \vb{m}_2}, \label{eq:sd_bound}
\end{align}

\section{Phase diagram of the quantum Hall--superconductor junction}
\label{seq:cross-scattering}

\begin{figure}[t]
\centering
\includegraphics[width=\linewidth]{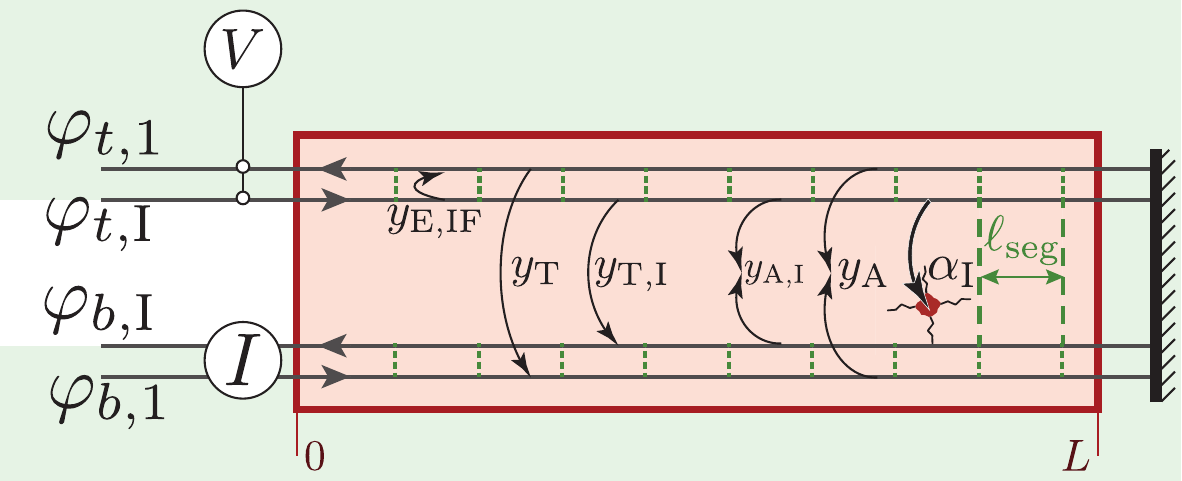}
\caption{Model system for one integer ($\phi_{t/b,\mathrm{I}}$) and one counter-propagating fractional ($\phi_{t/b,1}$) mode: a superconducting finger separates two counter-propagating quantum Hall edges with multiple modes. Arrows visualize possible processes, including tunneling ($y_{\text{T}}$) between the modes, tunneling ($y_{\text{E},\text{IF}}$) between the fractional and integer modes, and crossed Andreev reflection (CAR, denoted by $y_{\text{A}}$), as well as hopping into the dissipative subgap modes ($\alpha_{\text{I}}$) of the superconductor.}
\label{fig:model}
\end{figure}

Our model for a superconducting finger of length $L$ is shown in
\cref{fig:model}. We choose the coordinate $x$ to increase from the
open end of the finger, $x=0$, toward its tip, $x=L$. The two
geometric sides of the finger are denoted by $s\in\{\mathrm{t},\mathrm{b}\}$,
and we introduce the orientation factors
\begin{align}
    \epsilon_{\mathrm{t}}=+1,
    \qquad
    \epsilon_{\mathrm{b}}=-1.
\end{align}
In this section, we denote the single-edge matrix $\Km$ of
\cref{sec:setup_model} by $\Km_{\rm edge}$ to distinguish it from the
matrix of the doubled top--bottom system. The system possesses a symmetry
that is a product of time reversal and mirroring around the plane equidistant
from the two edges, so the matrices on the two sides satisfy
\begin{align}
    \Km_s=\epsilon_s\Km_{\rm edge}, 
    \label{eq:side-K-convention}
\end{align}
where $\Km_{\rm edge}$ refers to the orientation of the top edge.

A source contact is biased by a small voltage $V_{\rm in}$, while the
superconducting finger is grounded. The chemical potential of the incoming charge modes is fixed by the source, and the
downstream differential conductance $G={\mathrm{d}I_{\rm out}}/{\mathrm{d}V_{\rm in}}$ is measured at the drain.
We use $G_{\rm Q}=e^2/h$ for the conductance quantum and
$\sigma_{\mathrm{xy}}=\nu G_{\rm Q}$ for the Hall conductance.
With this convention, full transmission of the charge gives $G=+\nu G_{\rm Q}$, ideal crossed Andreev reflection gives $G=-\nu G_{\rm Q}$, and complete charge loss into the subgap states gives $G=0$.

For ideal equilibrated electrical contacts in linear response,
$I_{\rm out}=\nu G_{\rm Q}V_{\rm out}$. The normalized CAR signal
extracted in Ref.\ \cite{gul2021andreev} is therefore
\begin{align}
    p_{\rm CAR}
    &\equiv-\frac{V_{\rm out}}{V_{\rm in}}
    =-\frac{R_{\rm CAR}}{R_{\rm CAR}+R_{\rm XY}}
    =-\frac{G}{\nu G_{\rm Q}},
    \label{eq:car-observable}
\end{align}
where $R_{\rm CAR}=V_{\rm out}/I_{\rm exc}$ and
$R_{\rm XY}=(V_{\rm in}-V_{\rm out})/I_{\rm exc}$. 

When discussing the integer and fractional sectors separately, we use
$G_c$ for the physical conductance contribution of sector
$c\in\{\mathrm{I},\mathrm{f}\}$, where $\sigma_{\mathrm{xy},c}>0$ is the
magnitude of the sector Hall conductance; chirality signs are kept separately.

We say that an edge contains an \emph{integer mode} if there exists a local unit-charge operator with
$\vb{m}=\Km_{\rm edge}\vb{n}$ for an integer vector $\vb{n}$ satisfying
\begin{align}
\vb{q}_{\rm edge}^\transp\vb{n}&=1,
\qquad
    \abs{\vb{n}^\transp\Km_{\rm edge}\vb{n}}=1.
\end{align}
Here $V_{\vb{m}}=\exp(i\vb{m}\vdot\vb*{\phi})$, and
$\vb{q}_{\rm edge}$ is the single-edge charge vector of
\cref{sec:setup_model}.  Its conformal spin is
$s_{\vb{m}}=\vb{n}^\transp\Km_{\rm edge}\vb{n}/2
=\pm\nicefrac{1}{2}$.
Equivalently, after a unimodular basis change, $\Km_{\rm edge}$ contains
an explicit $\pm1$ block.

We consider edges whose $\Km$-matrix consists of a block corresponding to an $N$-mode chiral Jain state and
$N_{\rm I}$ (possibly zero) integer modes. Thus the number of modes on
one side is $n_{\rm edge}=N_{\rm I}+N$, and the single-edge matrix can be written as follows
\begin{align}
    \Km_{\rm edge}
    &=
    \mathds{1}_{N_{\rm I}}\oplus \chi_{\rm f}\Km_{\rm J},
    \label{eq:K-block-decomposition}
\end{align}
where $\mathds{1}_{N_{\rm I}}$ is $N_{\rm I} \times N_{\rm I}$ identity matrix, $\chi_{\rm f}=\pm1$, and
$\Km_{\rm J}$ is a fully chiral fractional block containing no
integer mode. The
integer blocks have sign $+1$ because in our case they always propagate in the charge propagation direction. For this decomposition, $\nu_{\rm f}=N/(2pN+1)>0$ and the total filling is
$\nu=N_{\rm I}+\chi_{\rm f}\nu_{\rm f}$.
The sign of $\chi_{\rm f}$ determines the
relative propagation directions within one edge. For example,
\begin{align}
    \Km_{\text{edge},4/3}&=(1)\oplus(3),
   \\
    \Km_{\text{edge},2/3}&=(1)\oplus(-3),
\end{align}
so the fractional mode co-propagates with the integer mode at
$\nu=4/3$ and counter-propagates at $\nu=2/3$. Purely fractional
edges are obtained simply by setting $N_{\rm I}=0$. The fields on each side form an $n_{\rm edge}$-component vector $\vb*{\phi}_s$.
The full matrix for the two sides is then
\begin{align}
    \Km_{\rm tot}
    =
    \begin{pmatrix}
       \Km_{\rm edge} & 0\\
        0 & -\Km_{\rm edge}
    \end{pmatrix}.
\end{align}

For the chiral Jain state,
$\Km_{\rm J}=\mathds{1}_N+2p\mathsf{J}_N$, where $\mathsf{J}_N$ is the $N\times N$
matrix of ones and $p\geq1$ is the Jain-sequence integer introduced in
\cref{sec:setup_model}. Using an orthogonal
matrix $O$ we can diagonalize $\Km_{\rm J}$ in the basis of the charge and neutral modes:
\begin{align}
    O^\transp\Km_{\rm J}O
    =
    \mathrm{diag}\qty(
        \underbrace{1,\ldots,1}_{N-1},
        \kappa_{\rm C}
    ), \label{eq:diagonalizing_o}
\end{align}
where $\kappa_{\rm C}=1+2Np$ and the charged mode is chosen as the
last column of $O$.
For the complete edge, we therefore use the block transformation
\begin{align}
    R=\mathds{1}_{N_{\rm I}}\oplus O,
\end{align}
which leaves the already diagonal integer sector unchanged and gives
\begin{align}
    \widehat{\Km}_{\rm edge}
    =
    R^\transp\Km_{\rm edge} R
    =
    \mathrm{diag}\qty(
        \underbrace{1,\ldots,1}_{N_{\rm I}},
        \underbrace{\chi_{\rm f},\ldots,\chi_{\rm f}}_{N-1},
        \chi_{\rm f}\kappa_{\rm C}
    ).
    \label{eq:partially-diagonal-K}
\end{align}
This transformation diagonalizes the single-edge matrix $\Km_{\rm edge}$, but not
necessarily the coupling matrix $\Um$. For the doubled theory we use
$\mathcal{R}=R\oplus R$. 

We define the
local electron $\varphi_{s,j}$ by
\begin{align}
    \vb{m}_{s,j}&=\Km_s\vb{e}_j
    =\epsilon_s\Km_{\rm edge}\vb{e}_j, \label{eq:local_electron}
    \\
    \varphi_{s,j}
    &\equiv
    \vb{e}_j^\transp\Km_{\rm edge}\vb*{\phi}_s,
\end{align}
where $\vb{e}_j\in\mathbb{R}^{n_{\rm edge}}$ is the $j$-th unit vector.
Suppressing Klein factors, the electron creation operator on side $s$
is
\begin{align}
    \Psi^\dagger_{s,j}
    =\frac{1}{\sqrt{2\pi a}}e^{i\epsilon_s\varphi_{s,j}}.
    \label{eq:electron-side-convention}
\end{align}

For the single vector of fields on both sides, $\vb*{\phi}=(\vb*{\phi}_{\rm t}^{\transp},\vb*{\phi}_{\rm b}^{\transp})^{\transp}$, the electron vectors are
\begin{align}
    \vb{m}^{\rm tot}_{{\rm t},j}&=\binom{\vb{m}_{{\rm t},j}}{0},
    &\vb{m}^{\rm tot}_{{\rm b},j}&=\binom{0}{\vb{m}_{{\rm b},j}}.
    \label{eq:doubled-electron-vectors}
\end{align}

With these definitions, we now write the total action of the finger as
\begin{align}
    S &= S_0 + S_{\rm pert}, \label{eq:action}
\end{align}
where the quadratic part of the action $S_0$ is given by \cref{eq:edge_action} with $\Km=\Km_{\rm tot}$. The charged tunneling and pairing operators use the electron representatives in \cref{eq:local_electron}; higher multiparticle charge transfers are expected to be less relevant and are outside the model. We write the perturbations as
\begin{align}
    S_{\rm pert}
    &=
    \sum_{j,k}\qty(S_{T,jk}+S_{A,jk})
    +\sum_{s,j}S_{\alpha,s,j}
    \nonumber\\
    &\quad
    +\sum_{s,j<k} S_{E,s;jk}   
    +\sum_{X\in\{A,T\}}\sum_{\substack{j<k\\l<m}}S_{\mathrm{N},X;jk,lm}.
    \label{eq:full-perturbation-action}
\end{align}
Normal electron tunneling between the two sides,
$\Psi^\dagger_{\rm t,k}\Psi_{\rm b,j}+\hc$, gives
\begin{align}
    S_{T,jk}
    &=
    -\frac{u}{2\pi a^2}\int\dd{x}\dd{t}\,
    y_{T,jk}(x)\nonumber\\
    &\qquad\times\exp(i(
        \varphi_{\rm t,k}
        +
        \varphi_{\rm b,j})) +\hc
    \label{eq:action-tunneling}
\end{align}
where $u$ is the common reference velocity used to normalize
all coherent amplitudes, and $y_{T,jk}(x)$ is a dimensionless position-dependent coefficient. Crossed
Andreev reflection creates or annihilates one electron on each side,
$\Psi^\dagger_{\rm t,k}\Psi^\dagger_{\rm b,j}+\hc$, and is
described by
\begin{align}
    S_{A,jk}
    &=
    -\frac{u}{2\pi a^2}\int\dd{x}\dd{t}\,
    y_{A,jk}(x)
    \nonumber\\
    &\qquad\qquad\times\exp(i(
        \varphi_{\rm t,k}
        -
        \varphi_{\rm b,j}
        -
        \theta_{\rm sc})) +\hc
    \label{eq:action-andreev}
\end{align}
Here $\theta_{\rm sc}$ is the superconducting condensate phase; the
operator transfers charge $2e$ to or from the condensate.

For $N\geq2$, the action also admits local neutral inter-side couplings. Let $\vb r_{jk}=\vb e_j-\vb e_k$ within the fractional
Jain block. Since $\Km_{\rm J}\vb r_{jk}=\vb r_{jk}$ and
$\vb q_{\rm f}^{\transp}\vb r_{jk}=0$, the vertices
\begin{align}
    V_{\mathrm{N},T;jk,lm}
    &=e^{i\chi_{\rm f}(\vb r_{jk}^{\transp}\vb*\phi_{\rm t,f}
       +\vb r_{lm}^{\transp}\vb*\phi_{\rm b,f})},\nonumber\\
    V_{\mathrm{N},A;jk,lm}
    &=e^{i\chi_{\rm f}(\vb r_{jk}^{\transp}\vb*\phi_{\rm t,f}
       -\vb r_{lm}^{\transp}\vb*\phi_{\rm b,f})}
    \label{eq:neutral-inter-side-operators}
\end{align}
with $j<k$ and $l<m$ conserve electric charge separately on each side.
They describe correlated transfers between fractional modes on opposite
sides. Their contributions are
\begin{align}
    S_{\mathrm{N},X;jk,lm}
    &=-\frac{u}{2\pi a^2}\int\dd{x}\dd{t}\,\nonumber\\
    &\qquad\times y_{\mathrm{N},X;jk,lm}(x)V_{\mathrm{N},X;jk,lm}+\hc.
    \label{eq:neutral-inter-side-action}
\end{align} 
Although these interactions involve four electron
operators, they can be relevant.

In general, the coefficients in
\cref{eq:action-tunneling,eq:action-andreev} contain the longitudinal
momentum mismatch between the edges. For $X\in\{A,T\}$, we write
\begin{align}
    y_{X,jk}(x)
    =y^{\mathrm{coh}}_{X,jk}e^{i\delta k_{X,jk}x}
    +\delta y_{X,jk}(x),
\end{align}
where the superscript $\mathrm{coh}$ denotes the coherent component, $\delta k_{X,jk}$ is the momentum mismatch, and
$\delta y_{X,jk}$ has zero spatial average along the finger.
Only the non-oscillating or long-range-correlated component contributes
to the RG equations below. For the CAR terms analyzed below, we assume
that the momentum-mismatch $\delta k_{A,jk}$ vanishes. This assumption can be justified by the product of mirror and time-reversal symmetries alongside translational invariance. For superconducting coupling between the two edges, the superconducting finger width must be on the order of the coherence length. As a result, we do not expect Abrikosov vortices to appear even in a type-II superconductor \cite{langar1967narrow,kupriyanov1975js, likharev1979weaklinks, qui2008numericalphaseslip}.
Therefore, these symmetries can be good approximate symmetries of the system. With these symmetries, we can set the phase
$\delta k_{A,jk}=\theta_{\rm sc}=0$ in \cref{eq:action-andreev}.
In contrast, normal inter-side tunneling generally
requires a potential modulation or disorder that compensates the
momentum mismatch. 

We model the dissipation by coupling electronic modes of each edge to impurity-induced subgap fermionic states in the superconductor, in the presence of broken time-reversal symmetry.  
We approximate their local low-energy density of states as continuous and nearly constant over the
temperature and bias range of interest. We represent this bath by a set of independent 1D chiral metallic wires perpendicular to the finger. Their fields are denoted by $\zeta_{s,j,x}(t)$, where $x$ labels the edge coordinate. The microscopic edge--bath amplitude is $w_{s,j}$:
\begin{align}
    S_{{\rm edge}\text{-}{\rm bath}}
    &=
    -\sum_{s,j}
    \int\dd{x}\dd{t}\,
    \nonumber\\
    &\quad\times\qty[
        \frac{w_{s,j}}{\sqrt{a}}\,
        \zeta^\dagger_{s,j,x}(t)\Psi_{s,j}(x,t)
        +\hc
    ].
\end{align}
Integrating out the bath gives
\begin{align}
    S_{\alpha,s,j}
    &=
    \frac{u\alpha_{s,j}}{\pi a^2}
    \int\dd{x}\dd{t_1}\dd{t_2}\,
    \mathcal{D}_\zeta(t_1-t_2)
    \nonumber\\
    &\qquad\times
    e^{i\epsilon_s[
        \varphi_{s,j}(t_1,x)-\varphi_{s,j}(t_2,x)
    ]},
    \label{eq:diss_action}
\end{align}
where $a$ is the short-distance cutoff, $u$ is a reference velocity,
$\alpha_{s,j}\propto\abs{w_{s,j}}^2$ is the dimensionless effective dissipative
strength, and the bath kernel $\mathcal{D}_\zeta(t)\propto 1/t$ at zero temperature, with the appropriate time-ordering prescription.
Since the bath is non-Markovian,
\cref{eq:diss_action} is local in space but nonlocal in time \cite{breuer2002theory}. In
particular, the bath cannot mediate coherent propagation along the
finger (unlike coupling to an extended integer edge mode).

Normal tunneling between two modes on the same side is described by
\begin{align}
    S_{E,s;jk}
    &=
    -\frac{u}{2\pi a^2}\int\dd{x}\dd{t}\,
    y_{E,s;jk}(x)
    \nonumber\\
    &\qquad\times\exp(i\epsilon_s(
        \varphi_{s,j}-\varphi_{s,k}
)) + \hc
    \label{eq:same-side-equilibration}
\end{align}
It conserves the total charge of that edge but redistributes charge
between its modes. 
Within a Jain block, $j,k$ label fractional modes and their full-edge indices are $N_{\rm I}+j,N_{\rm I}+k$;
$(\Km_{\rm J})_{jm}=\delta_{jm}+2p$, so
\begin{align}
    \varphi_{s,j}-\varphi_{s,k}
    &= 
\chi_{\rm f}(\vb{e}_j-\vb{e}_k)^\transp
\Km_{\rm J}\vb*{\phi}_{s,\rm f}
=\chi_{\rm f}\qty(\phi_{s,j}-\phi_{s,k}),
\end{align}
where $\vb*{\phi}_{s,\rm f}$ is the fractional block of the side-$s$ fields.
As a result, \cref{eq:same-side-equilibration} reduces to the usual neutral
intra-edge operator. Such a term cannot directly open a charge gap,
although it renormalizes the neutral sector and hence the scaling
dimensions of the charge-transferring operators. In the presence of an integer mode, \cref{eq:same-side-equilibration}  also describes the coupling between integer and fractional modes.

Same-side pairing between modes of the same chirality has non-zero conformal spin and cannot open the gap at the edge \cite{levin2013protectededge}. While such pairing can produce electron--hole conversion
\cite{gamayun2017transport,michelsen2020currentcorr,kurilovich2022disorder,tang2022vortex,burset2025conversion,bondarev2026conversion},
and mix the spin, valley, or orbital
content of the edge modes \cite{cuozzo2024su4,maji2026mixing}, it does not influence the RG fixed point of the system.
These terms are also generally
oscillatory because the two co-propagating electrons carry a non-zero total
longitudinal momentum. We therefore omit them from the RG analysis below.
We also omit the pairing of integer and fractional modes of opposite chirality on the same side, since in the presence of strong dissipation of the integer mode it enters the RG flow in the same way as a tunneling term between these two modes, which are included in the model in \cref{eq:same-side-equilibration}.

\subsection{RG flow in the symmetric model}
\label{sec:analytical-rg}

We now analyze the RG flow of the analytically tractable symmetric model, in which all
couplings of a given type have the same coefficient. Here $j,k=1,\ldots,N$
index the fractional block, while $\mathrm{I}$ labels the integer mode; we set
\begin{align}
    y_{A,jj}&=y_A,
    \qquad
    y_{T,jj}=y_T,
    \qquad
    \alpha_{sj}=\alpha_{\rm f}
    \end{align}
    and
    \begin{align}
    y_{\mathrm{N},X;jk}&\equiv y_{\mathrm{N},X;jk,jk}=y_{\mathrm{N},X}
    \end{align}
    for $ j<k$ and $ X\in\{A,T\}$.

The bath amplitudes are equal on both sides $s$. In this model, we set the off-diagonal fractional CAR and tunneling amplitudes $y_{A,jk}$ and $y_{T,jk}$ with $j\neq k$ to zero.
We also set $y_{\mathrm{N},X;jk,lm}=0$ for $(j,k)\ne(l,m)$.
The coupling matrix in the rotated
basis defined in \cref{eq:partially-diagonal-K} is
\begin{align}
    \widehat{\Um}_{\rm tot}
    =
   \mathcal{R}^\transp
    \Um_{\rm tot} \mathcal{R}
\end{align}
We assume that its fractional sector decouples into pairs of top and bottom modes in this basis. We label the charged mode by
$\mathrm{C}$ and the $N-1$ neutral modes by $i=1,\ldots,N-1$. For each pair of one top and one bottom modes, the  $\Km$-matrix is given by $\widehat{\Km}_i=\chi_{\rm f}\kappa_i\sigma_z$, where
\begin{align}
    \kappa_i=
    \begin{cases}
        1, & i=1,\ldots,N-1,\\
        \kappa_{\rm C}, & i=\mathrm{C},
    \end{cases}.
\end{align}
The corresponding $2\times2$ blocks of $\widehat{\Um}_{\rm tot}$ are
parameterized as
\begin{align}
    \widehat{\Um}_i
    =
    \frac{\kappa_i u_i}{2}
    \qty[
        (g_i+g_i^{-1})\mathds{1}_2
        -
        (g_i-g_i^{-1})\sigma_x
    ],
    \label{eq:Ui-parametrization}
\end{align}
where $\sigma_x$ and $\sigma_z$ are Pauli matrices, $u_i>0$ is a velocity, and $g_i>0$ is a Luttinger parameter.
Then, each block is diagonalized by
\begin{align}
    W_i
    =
    \frac{1}{\sqrt{\kappa_i}}
    e^{-\theta_i\sigma_x},
    \label{eq:Wi-boost}
\end{align}
where $g_i=e^{-2\theta_i}$ and $W_{\rm pair}=\bigoplus_iW_i$ acts on the
rotated fields grouped into top--bottom pairs. The full $W$ of
\cref{eq:diagonalizing-W} also includes the charge--neutral rotation
$\mathcal{R}$ and this implicit reordering, giving
\begin{align}
    W_i^\transp\widehat{\Km}_iW_i
    &=\chi_{\rm f}\sigma_z,
    \\
    W_i^\transp\widehat{\Um}_iW_i
    &=u_i\mathds{1}_2.
\end{align}

We denote the scaling dimension $d_{\vb{m}}$ defined in \cref{eq:sd} by
$d_X$ when the vertex belongs to a process of type $X$. For general
$g_i$, the scaling dimensions of the CAR and normal-tunneling operators involving
fractional physical mode $j$ are given in terms of matrix elements of diagonalizing matrix $O$ (\ref{eq:diagonalizing_o})
\begin{align}
    d_{A,j}
    &=
    \sum_{i}
    \kappa_i O_{ji}^2 g_i^{-1},
    \label{eq:general-dim-A}
    \\
    d_{T,j}
    &=
    \sum_{i}
    \kappa_i O_{ji}^2 g_i,
    \label{eq:general-dim-T}
\end{align}
where the sum runs over $\qty{1,\ldots,N-1,\mathrm{C}}$.
We take all neutral Luttinger parameters to be equal, i.e., set
$g_i=g_{\rm N}$ for $i=1,\ldots,N-1$ and denote the charged-mode
parameter by $g_{\rm C}$.
Using
$\sum_{i=1}^{N-1}O_{ji}^2=(N-1)/N$ and $O_{j\mathrm{C}}^2=1/N$, we obtain
\begin{align}
    d_A
    &=
    \frac{N-1}{N}g_{\rm N}^{-1}
    +
    \frac{1+2Np}{N}g_{\rm C}^{-1},
    \label{eq:dim_A}
    \\
    d_T
    &=
    \frac{N-1}{N}g_{\rm N}
    +
    \frac{1+2Np}{N}g_{\rm C}.
    \label{eq:dim_T}
\end{align}
The local electron scaling dimension is
\begin{align}
    d_{e,\mathrm{f}}
    =
    \frac{d_A+d_T}{4}.
    \label{eq:dim-alpha-f}
\end{align}

The neutral vertices in \cref{eq:neutral-inter-side-operators}
have zero conformal spin. Their scaling
dimensions are
\begin{align}
    d_{\mathrm{N},T}&=2g_{\rm N} 
    \label{eq:neutral-dimension-tunneling}\\
    d_{\mathrm{N},A}&=2/g_{\rm N}.
    \label{eq:neutral-dimension-car}
\end{align}
Thus the tunneling neutral coupling is relevant for
$g_{\rm N}<1$, and the superconducting neutral coupling for $g_{\rm N}>1$.
The neutral interaction can therefore open a neutral gap while
the charge-transfer operators are irrelevant.

We use a leading-order RG truncation: tree-level flow for the
amplitudes, quadratic flow of the Gaussian action,
and terms linear in the bath strength. Quadratic corrections to
the amplitude equations are omitted. 
Without integer modes, this gives
\begin{align}
    \dv{y_A}{\ell}
    &=
    (2-d_A)y_A,
    \label{eq:flow-yA}
    \\
    \dv{y_T}{\ell}
    &=
    (2-d_T)y_T,
    \label{eq:flow-yT}
    \\
    \dv{\alpha_{\rm f}}{\ell}
    &=
    \qty(2-2d_{e,\mathrm{f}})\alpha_{\rm f},
    \label{eq:flow-alpha-f}
    \\
    \dv{y_{\mathrm{N},X}}{\ell}
    &=(2-d_{\mathrm{N},X})y_{\mathrm{N},X},
    \label{eq:flow-neutral-couplings}
\end{align}
where $X\in\{A,T\}$.
For the one-loop coefficients of the quadratic action, we evaluate a shell with equal velocities $u_i=u$ for the fractional modes. 
It gives
\begin{align}
    \dv{u_i}{\ell}
    &=
    -\kappa_i\alpha_{\rm f}
    u_i(g_i+g_i^{-1}),
    \label{eq:flow-ui-clean}
    \\
    \dv{g_i}{\ell}
    &=
    \kappa_i
    \qty[
        \abs{y_A}^2-\abs{y_T}^2g_i^2
        +\alpha_{\rm f}(1-g_i^2)
    ]
    \nonumber\\
    &\quad+N(1-\delta_{i\mathrm C})\qty[
        \abs{y_{\mathrm{N},A}}^2
        -\abs{y_{\mathrm{N},T}}^2g_i^2
    ].
    \label{eq:flow-gi-clean}
\end{align}
The RG flow for this model is visualized in \cref{fig:pd2d}.
The superconducting and normal-tunneling terms drive $g_i$ in
opposite directions, whereas dissipation attracts each $g_i$ toward
the decoupled value $g_i=1$. The neutral superconducting and
tunneling couplings likewise increase and decrease $g_{\rm N}$,
respectively, without directly changing $g_{\rm C}$.

\subsubsection{Symmetric model with integer mode}
\label{sec:integer-mode-rg}
When an integer mode is present, its top and bottom components form an
additional $2\times2$ block.  For notational simplicity, we label one such
mode by $\mathrm{I}$ (with electron creation operator $\Psi^\dagger_{s,\mathrm{I}}\sim\exp(i\epsilon_s\varphi_{s,\mathrm{I}})$); multiple integer modes can be treated similarly. Its
top--bottom blocks satisfy $\widehat{\Km}_{\mathrm{I}}=\sigma_z$, and we
parameterize the coupling block by
\begin{align}
    \widehat{\Um}_{\mathrm{I}}
    =
    \frac{u_{\mathrm{I}}}{2}
    \qty[
        (g_{\mathrm{I}}+g_{\mathrm{I}}^{-1})\mathds{1}_2
        -
        (g_{\mathrm{I}}-g_{\mathrm{I}}^{-1})\sigma_x
    ].
    \label{eq:U-integer}
\end{align}
For this block, we abbreviate $y_{X,\mathrm{I}\mathrm{I}}=y_{X,\mathrm{I}}$ for $X\in\{A,T\}$ and,
under the side symmetry, $\alpha_{s,\mathrm{I}}=\alpha_{\mathrm{I}}$.
The corresponding scaling dimensions are
\begin{align}
    d_{A,\mathrm{I}}=g_{\mathrm{I}}^{-1},
    \qquad
    d_{T,\mathrm{I}}=g_{\mathrm{I}},
    \qquad
    d_{e,\mathrm{I}}
    =
    \frac{g_{\mathrm{I}}+g_{\mathrm{I}}^{-1}}{4}.
    \label{eq:integer-dimensions}
\end{align}
At the leading order, the equations for an isolated integer block are
\begin{align}
    \dv{y_{A,\mathrm{I}}}{\ell}
    &=
    (2-g_{\mathrm{I}}^{-1})y_{A,\mathrm{I}},
    \\
    \dv{y_{T,\mathrm{I}}}{\ell}
    &=
    (2-g_{\mathrm{I}})y_{T,\mathrm{I}},
    \\
    \dv{\alpha_{\mathrm{I}}}{\ell}
    &=
    \qty[
        2-\frac{g_{\mathrm{I}}+g_{\mathrm{I}}^{-1}}{2}
    ]\alpha_{\mathrm{I}},
    \label{eq:integer-coupling-flows}
    \\
    \dv{u_{\mathrm{I}}}{\ell}
    &=
    -\alpha_{\mathrm{I}}u_{\mathrm{I}}(g_{\mathrm{I}}+g_{\mathrm{I}}^{-1}),
    \label{eq:integer-u-flow}
    \\
    \dv{g_{\mathrm{I}}}{\ell}
    &=
    \abs{y_{A,\mathrm{I}}}^2-\abs{y_{T,\mathrm{I}}}^2g_{\mathrm{I}}^2
    +\alpha_{\mathrm{I}}(1-g_{\mathrm{I}}^2).
    \label{eq:integer-g-flow}
\end{align}
 Strong dissipation attracts $g_{\mathrm{I}}$
toward one, and the RG
therefore approaches a strong-coupling ray near $g_{\mathrm{I}}=1$. We use $g_{\mathrm I}=1$
as the dissipation-dominated modeling limit \cite{schiller2022interplay}; the perturbative flow does not establish it at arbitrarily strong coupling. At $g_{\mathrm{I}}=1$,
\begin{align}
    d_{A,\mathrm{I}}
    =
    d_{T,\mathrm{I}}
    =
    2d_{e,\mathrm{I}}
    =
    1.
    \label{eq:integer-equal-dimensions}
\end{align}
Consequently, superconductivity, normal tunneling, and dissipation have
the same temperature dependence \cite{schiller2022interplay}. The resulting phase has non-universal properties, including conductance, that depend on the ratios of the three microscopic
rates. Without strong dissipation, the edge conductance resembles the case with only fractional modes.

We now consider same-side and opposite-side normal integer--fractional
transfer:
\begin{align}
    S_{\rm IF}
    &=- \frac{u}{2\pi a^2}\sum_{s,j}\int\dd{x}\dd{t}\,
    \bigl[
        y_{E,{\rm IF}}e^{i\epsilon_s(\varphi_{s,\mathrm{I}}-\varphi_{s,j})}
    \nonumber\\
    &\qquad
        +y_{T,{\rm IF}}e^{i\epsilon_s(\varphi_{s,\mathrm{I}}+\varphi_{\bar{s},j})}
    \bigr]+\hc
    \label{eq:IF-action}
\end{align}
Here $\bar{s}$ is the opposite side and $j$ runs over fractional modes;
each displayed vertex creates an integer electron and annihilates a
fractional electron.
As we show below, both terms have the same scaling dimension but different conformal spins, so they enter the RG equations for the quadratic part of the action differently.

On the manifold of block-diagonal $\Um$, the scaling dimension of this term is the sum of the
two single-electron scaling dimensions,
\begin{align}
    d_{\rm IF}
    =
    \frac{g_{\mathrm{I}}+g_{\mathrm{I}}^{-1}}{4}
    +
    \frac{d_A+d_T}{4}.
    \label{eq:IF-dimension-general}
\end{align}
At the dissipative point $g_{\mathrm{I}}=1$ of the integer mode, this reduces to
\begin{align}
    d_{\rm IF}
    =
    \frac{1}{2}
    +
    \frac{d_{A}+d_{T}}{4},
    \label{eq:IF-dimension}
\end{align}
and
\begin{align}
    \dv{y_{X,{\rm IF}}}{\ell}
    =
    (2-d_{\rm IF})y_{X,{\rm IF}}.
    \label{eq:IF-flow}
\end{align}
The last equation applies to $X\in\{E, T\}$.

These operators have different, generally non-zero conformal spins:
\begin{align}
    s_{E,{\rm IF},s}&=\frac{\epsilon_s}{2}
    \qty[1+\chi_{\rm f}(2p+1)].
    \label{eq:IF-spin}
    \\
    s_{T,{\rm IF},s}
    &=\frac{\epsilon_s}{2}\qty[1-\chi_{\rm f}(2p+1)].
    \label{eq:IF-opposite-spin}
\end{align}

For a general $N$-mode Jain block, an explicit result is available
at $g_{\mathrm{I}}=1$ when the fractional normal modes share a velocity
$u_{\rm f}$. Define $v=u_{\mathrm{I}}/u_{\rm f}$, $\xi=(1-v)/(1+v)$, and the counter-propagating coupling strength
\begin{align}
    Y_{\rm IF}
    &=\begin{cases}
        \abs{y_{E,{\rm IF}}}^2,&\chi_{\rm f}=-1,\\
        \abs{y_{T,{\rm IF}}}^2,&\chi_{\rm f}=+1.
    \end{cases}
    \label{eq:IF-counter-strength}
\end{align}
Using $u=u_{\rm f}$ in the amplitude normalization, direct
second-order contraction gives the additional terms
\begin{align}
    \left.\dv{u_i}{\ell}\right|_{\rm IF}
    &=-\kappa_i u_{\rm f}Y_{\rm IF}\xi^{p-1}
      \frac{(1+v^2)(g_i+g_i^{-1})+4v}{(1+v)^3},
    \label{eq:flow-u-IF}\\
    \left.\dv{g_i}{\ell}\right|_{\rm IF}
    &=\kappa_iY_{\rm IF}\frac{(1-v)^p}{(1+v)^{p+1}}(1-g_i^2).
    \label{eq:flow-g-IF}
\end{align}
The derivation is given in \cref{app:IF-spin-rg}. 
These terms supplement \cref{eq:flow-ui-clean,eq:flow-gi-clean}
at a shell with $u_i=u_{\rm f}$.

At equal velocities ($v=1$), the integer--fractional coupling does not
influence the flow of the Luttinger parameters. For $u_{\mathrm{I}}<u_{\rm f}$, the interaction correction attracts each $g_i$ toward $1$ at order $(1-v)^p$ near $v=1$. 
The $v\to0$ limit has the same scalar form as the dissipation of the fractional mode: without integer-mode propagation, the bath and the integer mode are practically indistinguishable. 
Couplings between co-propagating integer and fractional modes do not contribute to the RG flow of $\Um$ at this order.

Once the RG length scale exceeds the finite finger length $L$, a
spatially extended perturbation becomes a boundary perturbation localized at
an effective 0+1D point contact. 
As it has a smaller spatial dimension, the point-contact threshold is $d_X=1$: the perturbation is
irrelevant for $d_X>1$, relevant for $d_X<1$, and marginal at
$d_X=1$~\cite{kane1992pointcontacts,fendley1995pointcontacts,moore2002edgetunneling}.

The symmetric model exhibits the following charge-transport regimes for the infinite-length downstream fractional Jain block. A finite finger crosses over to a point contact when its length cuts off the interactions.
\begin{enumerate}
\item If superconductivity reaches strong coupling, then
$g_{\rm C},g_{\rm N}\to\infty$ and the charged field is pinned by a superconducting
operator. For a downstream fractional block, this gives
$G_{\rm f}=-\sigma_{\mathrm{xy},{\rm f}}$. 
\item If the normal inter-edge tunneling reaches strong coupling, then
$g_{\rm C}\to0$, and the corresponding charge boundary condition gives
$G_{\rm f}=+\sigma_{\mathrm{xy},{\rm f}}$.
\item When all the charged perturbations remain irrelevant, the charge sector can remain gapless, with conductance determined by the lowest value of $d_T$, $d_A$, and $d_{e,\rm f}+1$: the phase is superconducting-dominated when
$d_A$ is the smallest, tunneling-dominated when
$d_T$ is the smallest, and dissipation-dominated otherwise. For $N>1$, neutral couplings exclude a stable dissipation-dominated phase. 
\item An edge containing an integer mode admits a dissipative strong-coupling regime near $g_{\mathrm{I}}=1$. Since
\cref{eq:integer-equal-dimensions} makes all three processes involving the integer mode
equally relevant, its conductance is non-quantized and non-universal. If coupling between integer and fractional modes is the most relevant process, the fractional modes equilibrate with the integer mode, which sets the total edge conductance. As a result, the total edge conductance is non-quantized and, as discussed in \cref{sec:plateau_in_general}, is temperature-independent.
\end{enumerate}

For upstream fractional modes, strong dissipation of the integer mode suppresses their contribution to the downstream response in all four phases, as derived in \cref{sec:currents}.

\subsection{Generic phase structure}
\label{sec:generic-phase-structure}

We derived the preceding equations for a special block-diagonal coupling matrix $\Um_{\rm tot}$. 
We now identify which features of the resulting phase diagram remain valid for a generic positive-definite $\Um_{\rm tot}$. Conductance in this section refers to the long-finger limit defined in
\cref{sec:plateau_in_general}.

Consider a superconducting operator $\exp(i  (\vb{m}_A)^\transp\vb*{\phi})$ and a normal-tunneling operator $\exp(i  (\vb{m}_T)^\transp\vb*{\phi})$ constructed from (possibly different) elementary single-electron modes of a Jain block. 
Let $\vb{e}_j$ be the $j$-th unit vector in the original $N$-dimensional Jain block. The vectors  $\vb{m}_A$ and $ \vb{m}_T$ can be written as
\begin{align}
    \vb{m}_A
    &=
    \chi_{\rm f}
    \begin{pmatrix}
        \Km_{\rm J}\vb{e}_j\\
        -\Km_{\rm J}\vb{e}_k
    \end{pmatrix},
    \\
    \vb{m}_T
    &=
    \chi_{\rm f}
    \begin{pmatrix}
        \Km_{\rm J}\vb{e}_{j'}\\
        \Km_{\rm J}\vb{e}_{k'}
    \end{pmatrix},
\end{align}
for $j,k,j',k'\in\{1,\ldots,N\}$. Here,
$d_A$ and
$d_T$ denote the scaling dimensions of these selected
operators.
Applying the general bound on the product of the scaling dimensions of two operators, \cref{eq:sd_bound}, gives
\begin{align}
    2\sqrt{d_Ad_T}
    &\geq
    \abs{
        (\vb{m}_A)^\transp
        \qty(\Km_{\rm tot})^{-1}
        \vb{m}_T }
    \nonumber\\
    &=
    \abs{
        (\Km_{\rm J})_{jj'}
        +
        (\Km_{\rm J})_{kk'}}
    \geq 4p,
    \label{eq:Jain-AT-bound}
\end{align}
where we used $\Km_{\rm J}=\mathds{1}_N+2p\mathsf{J}_N$. Hence
\begin{align}
    d_Ad_T\geq 4p^2.
    \label{eq:Jain-product-bound}
\end{align}
Simultaneous relevance would require $d_A<2$ and
$d_T<2$, and therefore
$d_Ad_T<4$, in contradiction with
\cref{eq:Jain-product-bound}, since for fractional Jain states $p\geq 1$. Thus, the single-electron CAR and
normal-tunneling processes cannot be simultaneously relevant. 

For generic $\Um_{\rm tot}$, only some of the charged operators may be relevant and open the gap. For the Jain block, let $M$ be the number of independent pinned charged CAR or tunneling operators,
and let $P$ be the number of additional neutral gapped modes, whose constraints are linearly independent of the pinned charged operators.

If the edge is fully gapped, then $M=N$. Otherwise, as all electrons are charged, we can choose the basis for the remaining $N-M$ modes with one charged mode and $N-M-1$ neutral modes. Since all the remaining operators that transfer charge are irrelevant, this charge mode is necessarily gapless.
Thus $0\leq P\leq N-M-1$ for $M<N$, and $P=0$ for $M=N$; $N-M-P$ pairs of modes remain gapless.

First, if $M=N$, all the modes are gapped, and the edge is either in a CAR phase with $G/G_{\rm Q}=-\nu_{\rm f}$ or in a tunneling phase with $G/G_{\rm Q}=\nu_{\rm f}$.

If $M>0$,  single-electron
dissipation is gapped out. For the CAR gap, charge-tunneling operators are also gapped out, and similarly for the tunneling gap, CAR operators are gapped out. We can omit these gapped-out operators from the low-energy theory. 
Thus, the only possible low-energy charged operators involving the electrons that are not gapped are CAR operators for the CAR gap and tunneling operators for the tunneling gap.
However, if neutral operators that do not commute with these operators are gapped, the corresponding mode remains ballistic, as there are no low-energy perturbations involving this mode. 

We now derive a condition on which modes remain ballistic depending on which neutral operators are pinned. Consider the set of indices of electron operators that are not involved in any gapped charged operator. These indices can be grouped into clusters in the following way: if $V_{\mathrm{N}, X;j k, lm}$ (for $X\in \qty{A, T}$) is gapped, then $(t,j)$ and $(t,k)$ are in the same cluster and $(b,l)$, $(b,m)$ are in the same cluster. As shown in \cref{app:neutral-cluster-response}, top and bottom components have equal size and can be matched by relabeling the bottom indices.
In this labeling, each component has a common $A$ or $T$ type, and all four-electron neutral operators with corresponding top and bottom indices are gapped.

The charged CAR operator $\exp(i(  \varphi_{\rm t,a}  -   \varphi_{\rm b,b} ))$  commutes
with the neutral operator $V_{\mathrm{N},T;jk,lm}$ when
\begin{align}
 \delta_{aj}-\delta_{ak}+\delta_{bl}-\delta_{bm}&=0.
 \label{eq:generic-car-neutral-compatibility}
\end{align}
If the CAR process involves these indices, the only allowed combinations
are $(a,b)=(j,m)$ or $(k,l)$. Thus, for a $T$ cluster of size at least
three, every CAR process involving its indices does not commute with at least one of the pinned neutral operators.

Let $B$ be the number of electrons that are members of clusters of size more than three. For all other $N-B$ electronic modes, the most relevant process is CAR, and in the limit of long finger they contribute the conductance $-\nu_{N-B}$ where $\nu_n=n/(2pn+1)$; if $B=0$, then $\nu_N = \nu_{\rm f}$ is the full CAR conductance, as discussed in \cref{tab:sds23_summary}. The $B$ ballistic modes contribute $\nu_N-\nu_{N-B}$, giving $G/G_{\rm Q}=\nu_N-2\nu_{N-B}$.

A similar situation occurs with a tunneling gap and pinned neutral
operators of CAR type. However, since tunneling and ballistic propagation
give the same charge response, we get a universal value $G/G_{\rm Q}=\nu_{\rm f}$.

For $M=0$, we first project out any pinned neutral modes. For the remaining gapless edge, the ordering of the low-temperature rates determines the response. Apart from the fine-tuned case in which two scaling exponents are equal, the limiting conductance is independent of the bare coefficients $y_r$ and
$G/G_{\rm Q}$ is rational, as shown in
\cref{app:strict-hierarchy-proof}.

As such, the values presented in \cref{tab:sds23_summary} apply to the case $M>0$ and $P=0$; for $P>0$ or a fully gapless edge, while conductance in the absence of an integer mode remains quantized, it can take different rational values.

\subsection{Translation-symmetry breaking}
\label{sec:disordered-couplings}
The discussion above assumes a non-zero momentum-conserving component of each inter-edge coupling. We now briefly describe what happens if this component is absent.

The simplest way to include terms that do not conserve momentum is to introduce random disorder to the coupling coefficients. For a process
$X\in\{A, T\}$ and a fixed mode pair (whose indices are suppressed),
let the overline denote the disorder average and write
\begin{align}
    \overline{\eta_X(x)}&=0,
    \\
    \overline{
        \eta_X(x)\eta_X^*(x')
    }
    &=
    aD_X\delta(x-x').
\end{align}
Here $\eta_X(x)$ is the random local coupling, $D_X$ is its dimensionless variance at cutoff $a$,
and $d_X$ below is the scaling
dimension of the corresponding local vertex operator.
In this case, the Giamarchi--Schulz RG scheme gives 
\begin{align}
    \dv{D_X}{\ell}
    =
    (3-2d_X)D_X,
    \label{eq:random-coupling-flow}
\end{align}
instead of
$\dv{y_X}{\ell}=(2-d_X)y_X$
\cite{giamarchi1988gsrg,kane1994edge}. Thus a random
inter-edge perturbation is relevant only for $d_X<3/2$, rather
than $d_X<2$. Weak, irrelevant disorder remains perturbative and does not destabilize the clean phase away from its boundaries. 
The coherent disordered critical regime for an integer edge discussed by \citet{kurilovich2022criticality} is outside this weak-disorder treatment and our locally equilibrated transport model. 

\section{Transport in multi-mode edges}
\label{sec:currents}
Throughout this section, the bare $\Km$, $\Um$, $\vb*{\phi}$, $\vb q$
denote the symbols in the basis including both edges.

In this section, we derive the linear-response conductance of the
superconducting finger from a recurrence for the generalized mode
chemical potentials $\vb*{\mu}_i$ at segment boundaries $x_i$. The
external electrical voltage is denoted by $V$. A reservoir at voltage
$V$ imposes $\vb*{\mu}=eV\vb{q}$ on its incoming modes; a neutral
mode has no direct electrical bias.
We use the temperature dependence of the recurrence in
\cref{sec:plateau_in_general}  to characterize phases with non-quantized, temperature-independent conductance and solve the recurrence numerically in \cref{sec:example}.

Following \citet{schiller2022interplay}, we define the dephasing length $\ell_{\mathrm{seg}}$, beyond which transport can be described classically, determined by the thermal length and the characteristic interaction length. We divide the finger into
coherent segments $[x_i,x_{i+1}]$ of length
$\ell_{\mathrm{seg}}=x_{i+1}-x_i$, as shown in \cref{fig:model}. We label the boundaries by $i=0,\ldots,N_{\rm seg}$, where $N_{\rm seg}=L/\ell_{\mathrm{seg}}$ and $x_{N_{\rm seg}}=L$. Correlations between different segments are neglected, while spatial correlations within a segment
are retained. We give the detailed derivation of the conductance in \cref{app:full-current} \cite{SupplementaryMaterial}.

Denote by $H_0$ the quadratic part of the Hamiltonian corresponding to action \eqref{eq:action}, and let $r$ label local bosonic terms of form $
    y_r\exp(i\vb{m}_r^\transp\vb*{\phi})$ in the perturbation part of the Hamiltonian. 

The vector of densities of each mode is
$\hat{\vb*{\rho}}=-\partial_x\vb*{\phi}/(2\pi)$; the electrical
charge density is $e\vb{q}^\transp\hat{\vb*{\rho}}$.
With the real-time convention of \cref{eq:edge_action}, the flux
directed along increasing $x$ is
\begin{align}
    \hat{\vb{j}}(x,t)
    &=-\frac{1}{2\pi}\Km^{-1}\Um\partial_x\vb*{\phi}(x,t).
\end{align}
A local transfer rate density $\hat{J}_r$ and its segment integral
$\hat{\mathcal{J}}_r[i]$ are defined as
\begin{align}
    \Km^{-1}\vb{m}_r\hat{J}_r(x,t)
    &=i\comm{\hat{\vb*{\rho}}(x,t)}{H_r},\\
    \hat{\mathcal{J}}_r[i](t)
    &=\int_{x_i}^{x_{i+1}}\dd{x}\,\hat{J}_r(x,t).
\end{align}
Positive $\hat{J}_r$ removes density in the direction
$\Km^{-1}\vb m_r$. It is a rate per unit length, whereas
$\hat{\mathcal{J}}_r[i]$ is a rate for the entire segment.
The steady-state continuity equation is
\begin{align}
    -\qty(
        \expval{\hat{\vb{j}}(x_{i+1})}
        -\expval{\hat{\vb{j}}(x_i)}
    )
    &=\sum_r\Km^{-1}\vb{m}_r
      \expval{\hat{\mathcal{J}}_r[i]},
    \label{eq:kirchhoff}
\end{align}

Linear response relates the flux and transfer rates to the mode
chemical potentials:
\begin{align}
    \expval{\hat{\vb{j}}(x_i)}
    &=\frac{1}{h}\Km^{-1}\vb*{\mu}_i,
    \label{eq:mode-flux-potential}
    \\
    \expval{\hat{\mathcal{J}}_r[i]}
    &=\frac{\ell_{\mathrm{seg}}}{e^2}\gamma_r(T)
    \vb{m}_r^\transp\Km^{-1}\vb*{\mu}_i \label{eq:coherent-current-ev},
\end{align}
where $\vb*{\mu}_i$ is the chemical potential at $x_i$, $\gamma_r(T)$ encodes the temperature dependence of the current and has units of conductance per unit length. The exact expression for $\gamma_r(T)$ is given in \cref{appeq:current_final_2}. \cref{eq:coherent-current-ev} gives a generic expression for the current due to an interaction term.
For a spatially uniform segment, as shown in Supplementary Material \cite{SupplementaryMaterial}, the linear response gives two useful limits for tunneling and CAR terms with nonzero weight in both propagation directions:
\begin{align}
    \gamma_r(T)
    &\propto \begin{cases}
    \frac{\ell_{\mathrm{seg}}}{u}\abs{y_r}^2T^{2d_r-2} &\ell_{\mathrm{seg}}\ll\ell_T\\
    \abs{y_r}^2T^{2d_r-3}&\ell_{\mathrm{seg}}\gg\ell_T
    \end{cases},
    \label{eq:coherent-rate}
\end{align}
where $\ell_T=u/(\pi T)$ is the reference thermal length for $\hbar=k_{\rm B}=1$, and $u$ is the reference velocity. For a fully chiral vertex at equal mode velocities, the short-segment
expression remains valid at every segment length.
For CAR or normal
tunneling, $d_r$ is the corresponding dimension $d_A$ or
$d_T$ defined in \cref{eq:general-dim-A,eq:general-dim-T}. In an
edge containing an integer mode with strong dissipation, the dissipation fixes the coherence length, and we take
$\ell_{\mathrm{seg}}=\ell_{\alpha,{\rm I}}\ll\ell_T$. Here $\ell_{\alpha,i}=a/(2\pi\alpha_i)$ denotes the bare characteristic dissipation length of mode or sector $i$; $i={\rm I}$ labels the integer sector. For a non-interacting integer mode, $a$ and $\alpha_{\mathrm{I}}$ have the same RG scaling; $\ell_{\alpha,{\rm I}}$ is therefore temperature independent at fixed microscopic parameters. For a fully fractional edge,
where dissipation is irrelevant, thermal dephasing instead sets the
coherence scale. In the numerical section we use,
$\ell_{\rm seg}\sim \ell_T$.

For $\ell_{\mathrm{seg}}\gg\ell_T$, the spatial correlation integral \eqref{appeq:kubo-segment} samples an area $\ell_{\mathrm{seg}}\times\ell_T$ because coherence
extends only over a thermal-length window. A short segment instead
gives $\ell_{\mathrm{seg}}^2$. Since $\ell_T\sim 1/T$, the long-segment
rate contains one additional inverse power of $T$.
In case the coupling is not long-range correlated, for example, if they are disordered with short-range variance $\overline{y_r(x)y_r^*(x')} \sim \delta(x-x')$, the temperature dependence is $T^{2d_r-2}$ independent of $\ell_{\mathrm{seg}}$. 

 For currents associated with the dissipation  process of electron $j$ at side $s$, the temperature dependence is
\begin{align}
    \gamma_{\alpha,s,j}(T)
    &\propto 
    \alpha_{s,j}T^{2d_{e,s,j}-1},
    \label{eq:dissipative-rate-scaling}
\end{align}
where $d_{e,s,j}$ is the scaling dimension of the dissipated electron and the embedded vector $\vb{m}^{\rm tot}_{s,j}$ from \cref{eq:doubled-electron-vectors} is used in \cref{eq:coherent-current-ev}.

Define a matrix
\begin{align}
    \mathcal{M}_r&=\vb{m}_r\qty(\Km^{-1}\vb{m}_r)^\transp.
\end{align}
With $G_{\mathrm{Q}}=e^2/h$ the conductance quantum, \cref{eq:kirchhoff} becomes the linear recurrence for $\vb*{\mu}$:
\begin{align}
    \frac{\vb*{\mu}_{i+1}-\vb*{\mu}_i}{\ell_{\mathrm{seg}}}
    &=\mathcal{A}(T)\vb*{\mu}_i  \label{eq:diff}\\
    \mathcal{A}(T)
    &=-\frac{1}{G_{\mathrm{Q}}}\sum_r\gamma_r(T)\mathcal{M}_r. \label{eq:recurrence}
\end{align}

The boundary conditions are imposed on incoming modes. Every mode is
continuous around the tip, so
\begin{align}
    \mu_{{\rm t},j,N_{\rm seg}}&=\mu_{{\rm b},j,N_{\rm seg}}.
\end{align}
The boundary conditions at $x=0$ split into two cases. For a fully downstream edge (e.g., $\nu=2/5$ or $\nu=4/3$), the source whose voltage is $V_{\rm in}$ fixes
$\vb*{\mu}_{{\rm t},0}=eV_{\rm in}\vb{q}_{\rm edge}$. In the symmetric basis
used here, $q_j=1$, this reduces to
$\mu_{{\rm t},j,0}=eV_{\rm in}$ for every incoming top mode.
For an edge with upstream fractional modes, such as $\nu=2/3$, we assume
equilibrated contacts at $x=0$ with the exterior edge at the Kane--Fisher fixed point \cite{kane1995impurity}.
For $\nu=2/3$, the incoming components fixed by the boundary conditions are
the top charge mode $\mu_{{\rm t},\rho}$ and the bottom neutral mode $\mu_{{\rm b},\sigma})$.

The recurrence~\eqref{eq:diff}, along with the boundary conditions, can be solved numerically or, in some cases, analytically. To extract the conductance,
denote the solution by $\vb*{\mu}_i(V_{\rm in})$, and let
$\vb*{\mu}_{{\rm b},i}$ be its bottom-side block.
Electrical current is obtained by projecting onto the charge vector $\vb{q}$:
\begin{align}
    I(x_i)&=e\vb{q}^\transp\expval{\hat{\vb{j}}(x_i)}
    =\frac{e}{h}\vb{q}^\transp\Km^{-1}\vb*{\mu}_i.
    \label{eq:charge-projected-current}
\end{align}
With the outgoing
current oriented away from the tip, the conductance is
\begin{align}
    G
    &\equiv
    \eval{\dv{I_{\rm out}}{V_{\rm in}}}_{V_{\rm in}=0} =\frac{e}{h}\vb{q}_{\rm edge}^{\transp}\Km_{\rm edge}^{-1}
    \frac{\vb*{\mu}_{{\rm b},0}(V_{\rm in}) }{V_{\rm in}}.
\label{eq:differential-conductance}
\end{align}

\subsection{Quantization and the role of integer modes in a general FQH system}
\label{sec:plateau_in_general}

We now allow a general electronic $\Km_{\rm edge}$ and charge vector
$\vb q_{\rm edge}$. Within the perturbative RG and transport description,
generic stable phases without an integer mode have quantized conductance
under the conditions below, i.e., a discrete rational multiple of $G_{\rm Q}$.
An integer mode instead permits a continuously tunable, temperature-independent conductance.

\begin{figure*}[t]
\centering
\begin{minipage}[t]{.48\textwidth}
\vspace{0pt}
\centering
    \includegraphics[width=\linewidth]{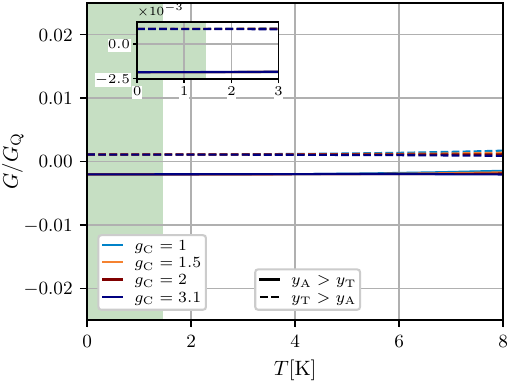}
    \caption{Downstream differential conductance $G/G_{\rm Q}$,
    $G_{\rm Q}=e^2/h$, versus temperature for a $\nu=2/3$ edge coupled
    along a $1\,\mu\mathrm{m}$ grounded superconducting finger.
    Inside the finger, a downstream integer mode with $g_{\rm I}=1$ and an upstream
    fractional mode exchange electrons and can lose charge into
    superconducting subgap states. The values of the parameters are given in
    \cref{tab:numerical-lengths}. Negative $G$ means
    net CAR response. Strong dissipation of the integer mode results in small,
    non-universal low-temperature conductance. Because the fractional mode is upstream, its phase only weakly influences the conductance.
    Colors label $g_{\rm C}=1,1.5,2,3.1$; solid and dashed curves use the
    two columns of \cref{tab:numerical-lengths}. Pale-green shading is a visual guide for the
    finite-size effects. }
    \label{fig:numerics23}
\end{minipage}\hfill
\begin{minipage}[t]{.48\textwidth}
\vspace{0pt}
\centering
    \includegraphics[width=\linewidth]{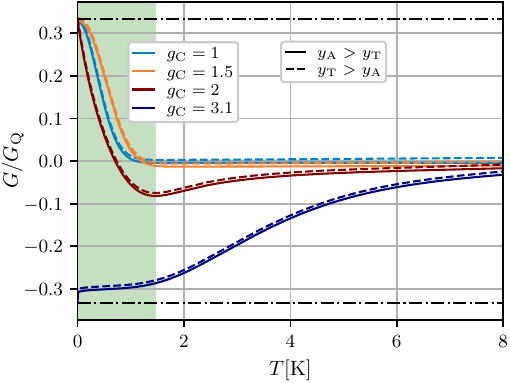}
    \caption{Downstream differential conductance $G/G_{\rm Q}$,
    $G_{\rm Q}=e^2/h$, versus temperature for a $\nu=4/3$ edge coupled
    along a $1\,\mu\mathrm{m}$ grounded superconducting finger.
    The integer mode with $g_{\rm I}=1$ and the fractional mode co-propagate on each side.
    The values of the parameters are given in
    \cref{tab:numerical-lengths}.
    Negative $G$ signals a net CAR response. Dash-dotted horizontal lines
    at $G/G_{\rm Q}=\pm1/3$ mark ideal fractional transmission and CAR,
    excluding the small integer contribution. Strong dissipation of the integer mode results in small,
    non-universal low-temperature conductance; in other phases the contribution of the fractional mode is quantized and total conductance is close to quantized values $\pm 1/3$.
    Colors label $g_{\rm C}=1,1.5,2,3.1$; solid and dashed curves use the
    two columns of \cref{tab:numerical-lengths}. Pale-green shading is a visual guide to the
    finite-size effects.}
    \label{fig:numerics43}
\end{minipage}
\end{figure*}

The recurrence \eqref{eq:diff} has a solution of the form
\begin{align}
    \vb*{\mu}_{N_{\rm seg}} &= \mathcal{P}(L,T) \vb*{\mu}_0\\
    \mathcal{P}(L,T) &= \qty[\mathds{1} + \ell_{\mathrm{seg}}\mathcal{A}(T)]^{N_{\rm seg}}.
\end{align}
Choose a basis in which the contacts fix $\vb*\mu_{\rm in}(0)$ and
the tip condition is $\vb*\mu_{\rm in}(L)=\vb*\mu_{\rm out}(L)$. Then
\begin{align}
    \vb*{\mu}_{\text{out},0} = (\mathcal{P}_{\text{out},\text{out}}-\mathcal{P}_{\text{in},\text{out}})^{-1}(\mathcal{P}_{\text{in},\text{in}}-\mathcal{P}_{\text{out},\text{in}}) \vb*{\mu}_{\text{in},0}, \label{eq:formal_solution}
\end{align}
where $\mathcal{P}_{\text{out},\text{out}}-\mathcal{P}_{\text{in},\text{out}}$ is necessarily invertible for the recurrence to have a unique solution.

We take the long-finger limit: $L\gamma_r/G_{\rm Q}\gg1$ for the
processes needed to determine the response, while scattering within
each segment remains weak. This limit precedes $T\to0$.
We use the fixed contact and tip conditions described above, and retain a
finite set of local processes.

For a set of operators with different temperature scaling, the proof in
\cref{app:strict-hierarchy-proof} establishes a rational conductance $G/G_{\rm Q}$ for any nonsingular $\Km_{\rm edge}$.

Without relevant dissipation, the fixed point at which competing CAR and tunneling operators are equally relevant 
 is not attractive. First, non-zero overlap
$\vb{m}_A^\transp\Km^{-1}\vb{m}_T$ prevents simultaneous pinning.
Second, a generic perturbation of $\Um$ will make the scaling
dimensions unequal. If the operators are relevant, the system will flow toward the corresponding fixed point; if they are irrelevant, it will remain at the perturbed point. In both cases, RG flow does not attract the system back to equality, rendering the fixed point unstable.

Relevant dissipation can change this conclusion. In the absence of an
integer mode, every local
unit-charge electron on one edge has half-integer conformal spin with $\abs{s_e}\geqslant3/2$. The bound $d_e\geqslant\abs{s_e}$ then makes its dissipation irrelevant. Generic stable phases therefore have quantized conductance.

For an IQH edge, strong dissipation attracts $g_{\rm I}$ toward one.
In the dissipation-dominated modeling limit
described in \cref{sec:integer-mode-rg} at $g_{\rm I}=1$, scaling dimensions of all three operators are equal
$d_{A,\rm I}=d_{T,\rm I}=2d_{e,\rm I}=1$,  and the conductance coefficients of all three processes involving the integer mode
scale as $\gamma_r\sim T^0$. Their finite ratios give a non-quantized,
temperature-independent conductance.

For a generic edge hosting an integer mode, this mechanism extends to
the fractional sector when dominant integer--fractional transfers couple every fractional electron to the dissipated
integer mode. In this case, the response inherits its temperature-independent, non-quantized character. Otherwise, the fractional modes contribute a quantized amount. In both cases, the integer mode contributes $\order{\alpha^{-2}}$ to conductance.

\section{Conductance of \texorpdfstring{$\nu=2/3$, $\nu=4/3$, and $\nu=2/5$}{ν=2/3, ν=4/3, and ν=2/5} edges}
\label{sec:example}

\begin{figure}[tp]
    \centering
    \includegraphics[width=\linewidth]{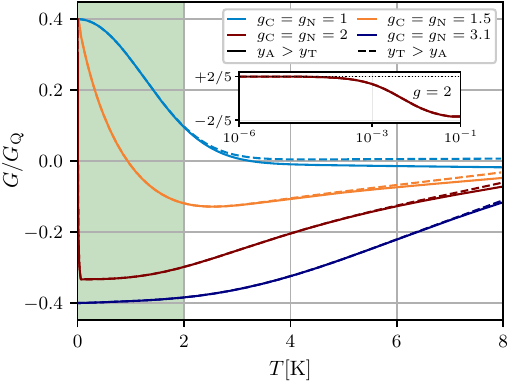}
    \caption{Downstream differential conductance $G/G_{\rm Q}$,
    $G_{\rm Q}=e^2/h$, versus temperature for a purely fractional
    $\nu=2/5$ Jain edge coupled along a $1\,\mu\mathrm m$ grounded
    superconducting finger. 
    Positive $G$ indicates electron transmission and negative $G$ a
    net crossed-Andreev-reflection (CAR) response; ideal limits are
    $G/G_{\rm Q}=\pm2/5$. Solid and dashed curves have the same bare
    CAR strength, with stronger normal tunneling in the dashed set
    (\cref{tab:numerical-lengths}).
    For $g<3$, CAR is negligible at low temperatures (logarithmic inset: $g=2$, $T$ in kelvin); for $g\geqslant 3$, the point-contact CAR rate remains finite.
    Pale-green shading marks the visual guide to the finite-length regime. }
    \label{fig:numerics45}
\end{figure}

We numerically solve the finite-segment recurrence in \cref{eq:diff} for three representative
edge structures: an integer mode with an upstream fractional mode at
$\nu=2/3$, an integer mode with a downstream fractional mode at $\nu=4/3$,
and the purely fractional Jain edge at $\nu=2/5$ with two fractional downstream modes.
In \cref{fig:numerics23,fig:numerics43,fig:numerics45} we plot the downstream differential conductance
$G=\mathrm{d}I_{\rm out}/\mathrm{d}V_{\rm in}$ in units of
$G_{\rm Q}=e^2/h$. Positive and negative $G$ indicate transmitted
electron charge and a net CAR response,
respectively; the superconducting finger is grounded.

\paragraph{Parameters and numerical procedure.}
In our simulations, we use finger length $L=1\,\mu\mathrm{m}$, similar to experimental values, and ultraviolet cutoff
$a=7\,\mathrm{nm}$, which is a magnetic length at the relevant magnetic field. All normal-mode velocities and the bath velocity
are fixed at $u=v_b=2.5\times10^4\,\mathrm{m/s}$.
The Luttinger parameters are held fixed during each temperature sweep:
$g_{\rm I}=1$ and $g_{\rm C}\in \qty{1,1.5,2,3.1}$ for the edges with an integer mode,
and $g_{\rm C}=g_{\rm N}\in \qty{1,1.5,2,3.1}$ for $\nu=2/5$.
We consider the case where all tunneling amplitudes are spatially uniform and have zero longitudinal momentum
mismatch. 
We include same-sector normal tunneling and CAR, dissipation in every
sector, and electron transfer between integer and fractional sectors, both on the same side and across the finger
when an integer mode is present.  

For $\nu=2/3$ and $4/3$, integer-mode dissipation is the most relevant process. We choose the segment length to be the temperature-independent bare dissipation length of the integer mode, $\ell_{\rm seg}=\ell_{\alpha,{\rm I}}=10\,\mathrm{nm}$. For $\nu=2/5$, the relevant length scale is the thermal length, and we use
\begin{align}
    \ell_{\rm seg}&=\min[L,\max(a,\ell_T)].
\end{align}
The incoming boundary conditions are those of
\cref{eq:differential-conductance}:
for downstream edges the source fixes all top incoming potentials to
$eV_{\rm in}$. At $\nu=2/3$, the
edge outside the finger is assumed to reach the
Kane--Fisher--Polchinski (KFP) fixed point; thus the incoming charge mode is biased, and
the incoming neutral mode is at zero potential. The corresponding modes are
continuous around the tip. 

With $r$ labeling a coherent electron-transfer process and
$c\in\{\mathrm I,\mathrm f\}$ a sector, the dimensionless bare couplings
are specified by their characteristic lengths 
\begin{align}
    \ell_r^{-1}&=\frac{\abs{y_r}^2}{4\pi a},
    \\\ell_{\alpha,c}^{-1}&=\frac{2\pi\alpha_c}{a}.
    \label{eq:numerical-bare-lengths}
\end{align}
All coupling lengths are listed in \cref{tab:numerical-lengths}; shorter
lengths mean stronger bare couplings. Each entry applies to every
retained symmetry-related representative. The solid and dashed curves differ
in same-sector normal tunneling and its corresponding neutral amplitude;
all other parameters are held fixed.

\begin{table}[tbp]
    \centering
    \caption{Bare characteristic coupling lengths used in the conductance plots in \cref{fig:numerics23,fig:numerics43,fig:numerics45}.
    $T$ denotes normal electron
    tunneling, $A$ crossed Andreev reflection, $E$ same-side electron
    transfer, and $\alpha$ dissipation into the superconducting bath.
    I and f label the integer and fractional sectors; IF denotes transfer
    between them. Solid and dashed columns specify the two curve sets.
    Shorter lengths correspond to stronger bare coupling.}
    \setlength{\tabcolsep}{3pt}
    \begin{tabular}{lcrr}
        \toprule
        Process & Symbol & \makecell{Solid\\($\mu\mathrm m$)} & \makecell{Dashed\\($\mu\mathrm m$)}\\
        \midrule
        Integer tunneling & $\ell_{T,{\rm I}}$ & $13$ & $6.3$\\
        Integer CAR & $\ell_{A,{\rm I}}$ & $7.7$ & $7.7$\\
        Fractional tunneling & $\ell_{T,{\rm f}}$ & $15$ & $6.3$\\
        Fractional CAR & $\ell_{A,{\rm f}}$ & $7.7$ & $7.7$\\
        Neutral $T$ transfer ($2/5$) & $\ell_{\mathrm N,T}$ & $1500$ & $630$\\
        Neutral $A$ transfer ($2/5$) & $\ell_{\mathrm N,A}$ & $770$ & $770$\\
        Integer dissipation & $\ell_{\alpha,{\rm I}}$ & $0.01$ & $0.01$\\
        Fractional dissipation & $\ell_{\alpha,{\rm f}}$ & $0.28$ & $0.28$\\
        Same-side IF transfer & $\ell_{E,{\rm IF}}$ & $0.28$ & $0.28$\\
        Inter-side IF tunneling & $\ell_{T,{\rm IF}}$ & $31$ & $31$\\
        Inter-side IF CAR & $\ell_{A,{\rm IF}}$ & $1400$ & $1400$\\
        \bottomrule
    \end{tabular}
    \label{tab:numerical-lengths}
\end{table}

\begin{figure}[tbp]
    \centering
    \includegraphics[width=\linewidth]{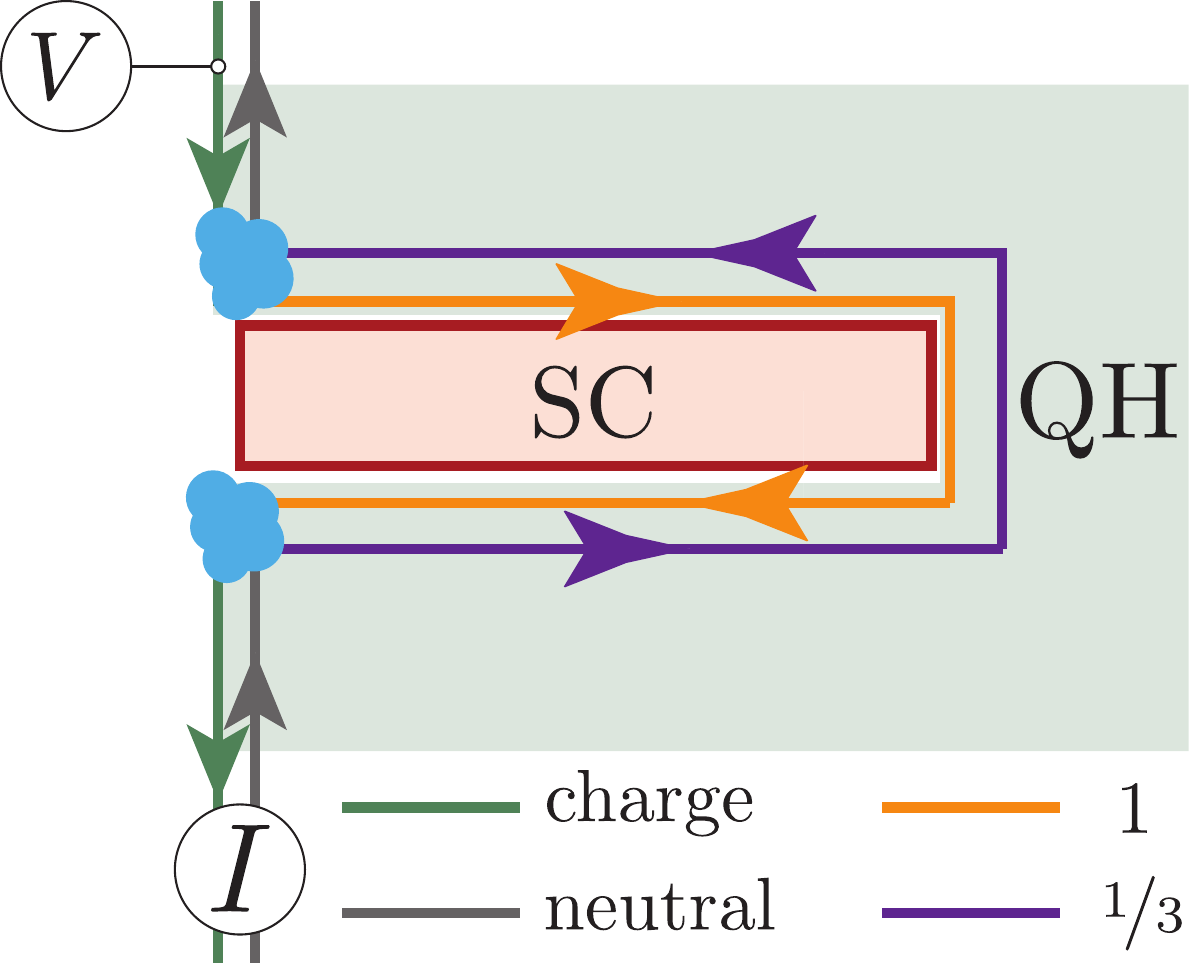}
    \caption{Quantum Hall (QH) edge at a $\nu=2/3$ in a strong-dissipation regime in the superconducting finger.
    Outside the finger, equilibration produces the
    Kane--Fisher--Polchinski fixed point: a downstream charge mode
    (green) and an upstream neutral mode (gray). Inside, the integer (orange)
    and fractional (purple) modes remain distinct and propagate in opposite directions. The blue regions at the
    open end, $x=0$, indicate matching between these two descriptions.}
    \label{fig:equi}
\end{figure}

\subsection{\texorpdfstring{$\nu=2/3$ and $\nu=4/3$}{ν=2/3 and ν=4/3}}

The two fillings use the same microscopic parameters and differ in the
chirality of the fractional mode:
\begin{align}
    \Km_{{\rm edge},2/3}&=\mathrm{diag}(1,-3),
    &\Km_{{\rm edge},4/3}&=\mathrm{diag}(1,3).
\end{align}
Strong dissipation of the integer mode favors $g_{\rm I}=1$. 
The small low-temperature conductance in \cref{fig:numerics23} depends on the bare coefficient values of the
normal tunneling, CAR, and dissipation of integer modes, with subleading fractional-sector corrections.
The upstream fractional mode contributes through transfer to the integer sector and through the boundary matching illustrated in \cref{fig:equi};
strong dissipation of the integer mode suppresses its contribution.

At $\nu=4/3$, the fractional mode can couple to or decouple from the
integer mode. In the decoupled long-finger limit, its conductance approaches
$+G_{\rm Q}/3$ for normal transmission or $-G_{\rm Q}/3$ for CAR.
These ideal fractional values are the horizontal lines in
\cref{fig:numerics43}; the integer mode adds a small non-quantized
contribution. At $g_{\rm C}=1$, integer--fractional coupling
has scaling dimension $2$, below $d_A=d_T=3$. Thus, tunneling into the dissipated integer
mode suppresses transport through the fractional mode at intermediate temperatures.
On further cooling, scattering through this process becomes negligible for a finite finger. For $g_{\rm C}=3.1$,
$d_A=30/31<1$, so at fixed coherence length the fractional CAR coefficient
grows as $T^{-2/31}$ on cooling.

The distinction between these two edges suggests an experimental test: strong dissipation of the integer mode leaves a small, non-quantized downstream response at $\nu=2/3$, whereas at $\nu=4/3$ the contribution from the fractional mode can approach $\pm G_{\rm Q}/3$. The total $\nu=4/3$ conductance retains the non-universal contribution from the integer mode and is therefore only approximately quantized when that contribution is small.
We also provide a perturbative analytic solution in \cref{appsec:23and43sol}.

\subsection{\texorpdfstring{$\nu=2/5$}{ν=2/5}}

For the Jain $\nu=2/5$ edge, we set $g\equiv g_{\rm C}=g_{\rm N}$
and use $g \in \qty{ 1,1.5,2,3.1} $. There is no integer mode, and hence the conductance is quantized at low temperatures. 

For $g=1$, the coherent rates scale as $T^3$, while the dissipation rate scales as $T^2$. Dissipation is therefore
the slowest-vanishing process and suppresses the current. On cooling, that rate decreases, and transmission is restored. 
At $g=1.5$, the CAR bulk scaling
dimension is $2$, i.e., it is marginal. In both cases, the calculated conductance becomes
positive at low temperature. At $g=2$, $d_A=3/2$ and $d_T=6$.
Once $\ell_T>L$, the finger effectively behaves as a point contact; the CAR rate
vanishes with cooling, and $G$ returns to $+(2/5)G_{\rm Q}$.
For $g=3.1$, $d_A=30/31<1$, so the point-contact CAR coefficient grows as
$T^{-2/31}$ and the conductance approaches $-(2/5)G_{\rm Q}$.
\section{Conclusion}
\label{sec:conclusion}
In summary, this paper studies proximitized counter-propagating fractional quantum Hall edges with multiple modes. We classify all the possible phases of a system in experimentally relevant settings involving Jain states with or without additional integer modes. We show that at low temperatures, the conductance is quantized in the absence of an integer mode; however, this does not signal a superconducting gap opening, since gapless phases can show the same behavior. With an integer mode, an additional phase appears with non-quantized, temperature-independent conductance. These conclusions match the experimental evidence \cite{gul2021andreev} and explain the observed difference between $\nu=2/3$ and $\nu=2/5$ edges.

An integer mode strongly coupled to the subgap bath can produce a small, non-quantized, approximately temperature-independent conductance. Upstream fractional modes, as at $\nu=2/3$, affect this response only weakly; downstream fractional modes, as at $\nu=4/3$, can contribute a substantial CAR or tunneling signal with a small offset attributed to the integer mode. The resulting temperature dependencies qualitatively agree with the experiments of \citet{gul2021andreev}. Further experiments comparing $\nu=2/3$ and $\nu=4/3$ can test the proposed theory.

\begin{acknowledgments}
 EZ is supported by the Adams Fellowships Program of the Israel Academy of Sciences and Humanities. This work was supported by ISF-MAFAT Quantum Science and Technology Grant no. 2478/24. YO and EB acknowledge support from CRC 183 of the Deutsche Forschungsgemeinschaft (Project C02) and a research grant from Magnus Konow in honor of his mother Olga Konow Rappaport.

\end{acknowledgments}

\bibliography{CAR}

@article{caroli1964subgap,
  title = {Bound Fermion states on a vortex line in a type {II} superconductor},
  author = {Caroli, Christiane  and De Gennes, Pierre-Gilles  and Matricon, Jean},
  year = 1964,
  month = may,
  journal = {Physics Letters},
  volume = 9,
  number = 4,
  pages = {307--309},
  doi = {https://doi.org/10.1016/0031-9163(64)90375-0},
  issn = {0031-9163},
  url = {https://www.sciencedirect.com/science/article/pii/0031916364903750},
}

@article{langar1967narrow,
  title = {Intrinsic Resistive Transition in Narrow Superconducting Channels},
  author = {Langer, James S. and Ambegaokar, Vinay},
  year = 1967,
  month = dec,
  journal = {Physical Review},
  publisher = {American Physical Society},
  volume = 164,
  pages = {498--510},
  issue = 2,
  numpages = {0},
  doi = {10.1103/PhysRev.164.498},
  url = {https://link.aps.org/doi/10.1103/PhysRev.164.498},
}

@article{likharev1979weaklinks,
  title = {Superconducting weak links},
  author = {Likharev, Konstantin K.},
  year = 1979,
  month = jan,
  journal = {Reviews of Modern Physics},
  publisher = {American Physical Society},
  volume = 51,
  pages = {101--159},
  issue = 1,
  numpages = {0},
  doi = {10.1103/RevModPhys.51.101},
  url = {https://link.aps.org/doi/10.1103/RevModPhys.51.101},
}

@article{halperin1983theory,
  title = {Theory of the quantized {Hall} conductance},
  author = {Halperin, Bertrand I.},
  year = 1983,
  month = mar,
  journal = {Helvetica Physica Acta},
  publisher = {Birkhauser Verlag Ag Viadukstrasse 40-44, Po Box 133, Ch-4010 Basel, Switzerland},
  volume = 56,
  number = {1-3},
  pages = {75--102},
  doi = {10.5169/seals-115362},
}

@article{giamarchi1988gsrg,
  title = {Anderson localization and interactions in one-dimensional metals},
  author = {Giamarchi, Thierry and Schulz, Heinz J.},
  year = 1988,
  month = jan,
  journal = {Physical Review B},
  publisher = {American Physical Society},
  volume = 37,
  pages = {325--340},
  issue = 1,
  numpages = {0},
  doi = {10.1103/PhysRevB.37.325},
  url = {https://link.aps.org/doi/10.1103/PhysRevB.37.325},
}

@article{ginsparg1988applied,
  title = {Applied Conformal Field Theory},
  author = {Paul Ginsparg},
  year = 1988,
  month = nov,
  journal = {arXiv preprint},
  url = {https://arxiv.org/abs/hep-th/9108028},
  eprint = {hep-th/9108028},
  archiveprefix = {arXiv},
  primaryclass = {hep-th},
}

@article{jain1989cf,
  title = {Composite-fermion approach for the fractional quantum {Hall} effect},
  author = {Jain, Jainendra K.},
  year = 1989,
  month = jul,
  journal = {Physical Review Letters},
  publisher = {American Physical Society},
  volume = 63,
  pages = {199--202},
  issue = 2,
  doi = {10.1103/PhysRevLett.63.199},
  url = {https://link.aps.org/doi/10.1103/PhysRevLett.63.199},
}

@article{wenzee1992classification,
  title = {Classification of {Abelian} quantum {Hall} states and matrix formulation of topological fluids},
  author = {Wen, Xiao-Gang and Zee, Anthony},
  year = 1992,
  month = jul,
  journal = {Physical Review B},
  publisher = {American Physical Society},
  volume = 46,
  pages = {2290--2301},
  issue = 4,
  numpages = {0},
  doi = {10.1103/PhysRevB.46.2290},
  url = {https://link.aps.org/doi/10.1103/PhysRevB.46.2290},
}

@article{kane1992pointcontacts,
  title = {Transmission through barriers and resonant tunneling in an interacting one-dimensional electron gas},
  author = {Kane, Charles L. and Fisher, Matthew P. A.},
  year = 1992,
  month = dec,
  journal = {Physical Review B},
  publisher = {American Physical Society},
  volume = 46,
  pages = {15233--15262},
  issue = 23,
  numpages = {0},
  doi = {10.1103/PhysRevB.46.15233},
  url = {https://link.aps.org/doi/10.1103/PhysRevB.46.15233},
}

@article{meir1994compositeedge,
  title = {Composite edge states in the \ensuremath{\nu}=2/3 fractional quantum {Hall} regime},
  author = {Meir, Yigal},
  year = 1994,
  month = apr,
  journal = {Physical Revew Letters},
  publisher = {American Physical Society},
  volume = 72,
  pages = {2624--2627},
  issue = 16,
  numpages = {0},
  doi = {10.1103/PhysRevLett.72.2624},
  url = {https://arxiv.org/abs/cond-mat/9311049},
  eprint = {cond-mat/9311049},
  archiveprefix = {arXiv},
  primaryclass = {cond-mat},
}

@article{kane1994edge,
  title = {Randomness at the edge: Theory of quantum {Hall} transport at filling $\nu$=2/3},
  author = {Kane, Charles L. and Fisher, Matthew P. A. and Polchinski, Joseph},
  year = 1994,
  month = jun,
  journal = {Physical Review Letters},
  publisher = {American Physical Society},
  volume = 72,
  pages = {4129--4132},
  issue = 26,
  doi = {10.1103/PhysRevLett.72.4129},
  url = {https://arxiv.org/abs/cond-mat/9402108},
  eprint = {cond-mat/9402108},
  archiveprefix = {arXiv},
  primaryclass = {cond-mat},
}

@article{kane1995impurity,
  title = {Impurity scattering and transport of fractional quantum {Hall} edge states},
  author = {Kane, Charles L. and Fisher, Matthew P. A.},
  year = 1995,
  month = may,
  journal = {Physical Review B},
  publisher = {American Physical Society},
  volume = 51,
  pages = {13449--13466},
  issue = 19,
  numpages = {0},
  doi = {10.1103/PhysRevB.51.13449},
  url = {https://arxiv.org/abs/cond-mat/9409028},
  eprint = {cond-mat/9409028},
  archiveprefix = {arXiv},
  primaryclass = {cond-mat},
}

@article{fendley1995pointcontacts,
  title = {Exact nonequilibrium transport through point contacts in quantum wires and fractional quantum {Hall} devices},
  author = {Fendley, Paul and Ludwig, Andreas W. W. and Saleur, Hubert},
  year = 1995,
  month = sep,
  journal = {Physical Review B},
  publisher = {American Physical Society},
  volume = 52,
  pages = {8934--8950},
  issue = 12,
  doi = {10.1103/PhysRevB.52.8934},
  url = {https://arxiv.org/abs/cond-mat/9503172},
  eprint = {cond-mat/9503172},
  archiveprefix = {arXiv},
  primaryclass = {cond-mat},
}

@article{wen1995edge,
  title = {Topological orders and edge excitations in fractional quantum {Hall} states},
  author = {Xiao-Gang Wen},
  year = 1995,
  month = sep,
  journal = {Advances in Physics},
  publisher = {Taylor & Francis},
  volume = 44,
  number = 5,
  pages = {405--473},
  doi = {10.1080/00018739500101566},
  url = {https://arxiv.org/abs/cond-mat/9506066},
  eprint = {cond-mat/9506066},
  archiveprefix = {arXiv},
  primaryclass = {cond-mat},
}

@article{nayak1996ising,
  title = {$2n$-quasihole states realize $2n-1$-dimensional spinor braiding statistics in paired quantum {Hall} states},
  author = {Chetan Nayak and Frank Wilczek},
  year = 1996,
  month = nov,
  journal = {Nuclear Physics B},
  volume = 479,
  number = 3,
  pages = {529--553},
  doi = {https://doi.org/10.1016/0550-3213(96)00430-0},
  issn = {0550-3213},
  url = {https://arxiv.org/abs/cond-mat/9605145},
  eprint = {cond-mat/9605145},
  archiveprefix = {arXiv},
  primaryclass = {cond-mat},
}

@article{frohlich1997qheclassification,
  title = {A classification of quantum {Hall} fluids},
  author = {Fröhlich, Jürg and Studer, Urban M. and Thiran, Emmanuel},
  year = 1997,
  month = feb,
  journal = {Journal of Statistical Physics},
  volume = 86,
  number = 3,
  pages = {821--897},
  doi = {10.1007/BF02199122},
  issn = {1572-9613},
  url = {https://arxiv.org/abs/cond-mat/9503113},
  eprint = {cond-mat/9503113},
  archiveprefix = {arXiv},
  primaryclass = {cond-mat},
}

@article{gottesman1998heisenberg,
  title = {The {Heisenberg} Representation of Quantum Computers},
  author = {Daniel Gottesman},
  year = 1998,
  month = jul,
  journal = {arXiv preprint},
  url = {https://arxiv.org/abs/quant-ph/9807006},
  eprint = {quant-ph/9807006},
  archiveprefix = {arXiv},
  primaryclass = {quant-ph},
}

@article{delft1998bosonization,
  title = {Bosonization for beginners --- refermionization for experts},
  author = {von Delft, Jan and Schoeller, Herbert},
  year = 1998,
  month = dec,
  journal = {Annalen der Physik},
  volume = 510,
  number = 4,
  pages = {225--305},
  doi = {https://doi.org/10.1002/andp.19985100401},
  url = {https://arxiv.org/abs/cond-mat/9805275},
  eprint = {cond-mat/9805275},
  archiveprefix = {arXiv},
  primaryclass = {cond-mat.str-el},
}

@article{kitaev2001wire,
  title = {Unpaired {Majorana} fermions in quantum wires},
  author = {Alexei {\relax Yu}. Kitaev},
  year = 2001,
  month = oct,
  journal = {Physics-Uspekhi},
  publisher = {Uspekhi Fizicheskikh Nauk ({UFN}) Journal},
  volume = 44,
  number = {10s},
  pages = {131--136},
  doi = {10.1070/1063-7869/44/10s/s29},
  url = {https://doi.org/10.1070/1063-7869/44/10s/s29},
  eprint = {cond-mat/0010440},
  archiveprefix = {arXiv},
  primaryclass = {cond-mat.mes-hall},
}

@article{moore2002edgetunneling,
  title = {Critical points in edge tunneling between generic fractional quantum {Hall} states},
  author = {Moore, Joel E. and Wen, Xiao-Gang},
  year = 2002,
  month = sep,
  journal = {Physical Review B},
  publisher = {American Physical Society},
  volume = 66,
  pages = 115305,
  issue = 11,
  numpages = 24,
  doi = {10.1103/PhysRevB.66.115305},
  url = {https://arxiv.org/abs/cond-mat/0106387},
}

@article{kitaev2003faulttolerant,
  title = {Fault-tolerant quantum computation by anyons},
  author = {Alexei {\relax Yu}. Kitaev},
  year = 2003,
  month = jan,
  journal = {Annals of Physics},
  volume = 303,
  number = 1,
  pages = {2--30},
  doi = {https://doi.org/10.1016/S0003-4916(02)00018-0},
  issn = {0003-4916},
  url = {https://arxiv.org/abs/quant-ph/9707021},
  eprint = {quant-ph/9707021},
  archiveprefix = {arXiv},
  primaryclass = {quant-ph},
}

@article{kitaev2006anyons,
  title = {Anyons in an exactly solved model and beyond},
  author = {Alexei {\relax Yu}. Kitaev},
  year = 2006,
  month = jan,
  journal = {Annals of Physics},
  volume = 321,
  number = 1,
  pages = {2--111},
  doi = {https://doi.org/10.1016/j.aop.2005.10.005},
  issn = {0003-4916},
  url = {https://arxiv.org/abs/cond-mat/0506438},
  eprint = {cond-mat/0506438},
  archiveprefix = {arXiv},
  primaryclass = {cond-mat.mes-hall},
  note = {{January} Special Issue},
}

@article{bravyi2006universal,
  title = {Universal quantum computation with the $\nu=5/2$ fractional quantum {Hall} state},
  author = {Bravyi, Sergey},
  year = 2006,
  month = apr,
  journal = {Physical Review A},
  publisher = {American Physical Society},
  volume = 73,
  pages = {042313},
  issue = 4,
  numpages = 16,
  doi = {10.1103/PhysRevA.73.042313},
  url = {https://arxiv.org/abs/quant-ph/0511178},
  eprint = {quant-ph/0511178},
  archiveprefix = {arXiv},
  primaryclass = {quant-ph},
}

@article{fendley2007ising,
  title = {Edge states and tunneling of non-{Abelian} quasiparticles in the $\nu=5/2$ quantum {Hall} state and $p+ip$ superconductors},
  author = {Fendley, Paul and Fisher, Matthew P. A. and Nayak, Chetan},
  year = 2007,
  month = jan,
  journal = {Physical Review B},
  publisher = {American Physical Society},
  volume = 75,
  pages = {045317},
  issue = 4,
  numpages = 24,
  doi = {10.1103/PhysRevB.75.045317},
  url = {https://arxiv.org/abs/cond-mat/0607431},
  eprint = {cond-mat/0607431},
  archiveprefix = {arXiv},
  primaryclass = {cond-mat.mes-hall},
}

@article{qui2008numericalphaseslip,
  title = {Numerical study of the phase slip in two-dimensional superconducting strips},
  author = {Qiu, Chunyin and Qian, Tiezheng},
  year = 2008,
  month = {May},
  journal = {Physical Review B},
  publisher = {American Physical Society},
  volume = 77,
  pages = 174517,
  issue = 17,
  numpages = 8,
  doi = {10.1103/PhysRevB.77.174517},
  url = {https://link.aps.org/doi/10.1103/PhysRevB.77.174517},
}

@article{fukane2008,
  title = {Superconducting Proximity Effect and {Majorana} Fermions at the Surface of a Topological Insulator},
  author = {Fu, Liang and Kane, Charles L.},
  year = 2008,
  month = mar,
  journal = {Physical Review Letters},
  publisher = {American Physical Society},
  volume = 100,
  pages = {096407},
  issue = 9,
  numpages = 4,
  doi = {10.1103/PhysRevLett.100.096407},
  url = {https://arxiv.org/abs/0707.1692},
  eprint = {0707.1692},
  archiveprefix = {arXiv},
  primaryclass = {cond-mat.str-el},
}

@article{nayak2007nonabelian,
  title = {Non-{Abelian} anyons and topological quantum computation},
  author = {Nayak, Chetan and Simon, Steven H. and Stern, Ady and Freedman, Michael and Das Sarma, Sankar},
  year = 2008,
  month = sep,
  journal = {Reviews of Modern Physics},
  publisher = {American Physical Society},
  volume = 80,
  pages = {1083--1159},
  issue = 3,
  numpages = {0},
  doi = {10.1103/RevModPhys.80.1083},
  url = {https://arxiv.org/abs/0707.1889},
  eprint = {0707.1889},
  archiveprefix = {arXiv},
  primaryclass = {cond-mat.str-el},
}

@article{ahlbrecht2009clifford,
  title = {Implementation of {Clifford} gates in the {Ising}-anyon topological quantum computer},
  author = {Ahlbrecht, Andr\'e and Georgiev, Lachezar S. and Werner, Reinhard F.},
  year = 2009,
  month = mar,
  journal = {Physical Review  A},
  publisher = {American Physical Society},
  volume = 79,
  pages = {032311},
  issue = 3,
  numpages = 16,
  doi = {10.1103/PhysRevA.79.032311},
  url = {https://arxiv.org/abs/0812.2338},
  eprint = {0812.2338},
  archiveprefix = {arXiv},
  primaryclass = {quant-ph},
}

@article{rickhaus2012qhesc,
  title = {Quantum {Hall} Effect in Graphene with Superconducting Electrodes},
  author = {Rickhaus, Peter and Weiss, Markus and Marot, Laurent and Sch\"{o}nenberger, Christian},
  year = 2012,
  month = mar,
  journal = {Nano Letters},
  volume = 12,
  number = 4,
  pages = {1942--1945},
  doi = {10.1021/nl204415s},
  url = {https://arxiv.org/abs/1303.3394},
  eprint = {1303.3394},
  archiveprefix = {arXiv},
  primaryclass = {cond-mat.mes-hall},
}

@article{mourik2012majoranas,
  title = {Signatures of {Majorana} Fermions in Hybrid Superconductor-Semiconductor Nanowire Devices},
  author = {Mourik, Vincent and Zuo, Kun and Frolov, Sergey M. and Plissard, S\'{e}bastien R. and Bakkers, Erik P. A. M. and Kouwenhoven, Leo P.},
  year = 2012,
  month = apr,
  journal = {Science},
  volume = 336,
  number = 6084,
  pages = {1003--1007},
  doi = {10.1126/science.1222360},
  url = {https://arxiv.org/abs/1204.2792},
  eprint = {1204.2792},
  archiveprefix = {arXiv},
  primaryclass = {cond-mat.mes-hall},
}

@article{levin2012abelian,
  title = {Classification and analysis of two-dimensional {Abelian} fractional topological insulators},
  author = {Levin, Michael and Stern, Ady},
  year = 2012,
  month = sep,
  journal = {Physical Review B},
  publisher = {American Physical Society},
  volume = 86,
  pages = 115131,
  issue = 11,
  numpages = 17,
  doi = {10.1103/PhysRevB.86.115131},
  url = {https://arxiv.org/abs/1205.1244},
  eprint = {1205.1244},
  archiveprefix = {arXiv},
  primaryclass = {cond-mat.str-el},
}

@article{lindner2012fractionalizing,
  title = {Fractionalizing {Majorana} Fermions: Non-{Abelian} Statistics on the Edges of {Abelian} Quantum {Hall} States},
  author = {Lindner, Netanel H. and Berg, Erez and Refael, Gil and Stern, Ady},
  year = 2012,
  month = oct,
  journal = {Physical Review X},
  publisher = {American Physical Society},
  volume = 2,
  pages = {041002},
  issue = 4,
  numpages = 22,
  doi = {10.1103/PhysRevX.2.041002},
  url = {https://link.aps.org/doi/10.1103/PhysRevX.2.041002},
  eprint = {1204.5733},
  archiveprefix = {arXiv},
  primaryclass = {cond-mat.mes-hall},
}

@article{fendley2012parafermion,
  title = {Parafermionic edge zero modes in {$\mathbb{Z}_n$}-invariant spin chains},
  author = {Paul Fendley},
  year = 2012,
  month = nov,
  journal = {Journal of Statistical Mechanics: Theory and Experiment},
  publisher = {{IOP} Publishing},
  volume = 2012,
  number = 11,
  pages = {P11020},
  doi = {10.1088/1742-5468/2012/11/p11020},
  url = {https://arxiv.org/abs/1209.0472},
  eprint = {1209.0472},
  archiveprefix = {arXiv},
  primaryclass = {cond-mat.str-el},
}

@article{das2012majorana,
  title = {Zero-bias peaks and splitting in an {Al}--{InAs} nanowire topological superconductor as a signature of {Majorana} fermions},
  author = {Das, Anindya and Ronen, Yuval and Most, Yonatan and Oreg, Yuval and Heiblum, Moty and Shtrikman, Hadas},
  year = 2012,
  month = dec,
  journal = {Nature Physics},
  volume = 8,
  number = 12,
  pages = {887--895},
  doi = {10.1038/nphys2479},
  isbn = {1745-2481},
  url = {https://doi.org/10.1038/nphys2479},
}

@article{barkeshli2013twistdefects,
  title = {Twist defects and projective {non-Abelian} braiding statistics},
  author = {Barkeshli, Maissam and Jian, Chao-Ming and Qi, Xiao-Liang},
  year = 2013,
  month = jan,
  journal = {Physical Review B},
  publisher = {American Physical Society},
  volume = 87,
  pages = {045130},
  issue = 4,
  numpages = 23,
  doi = {10.1103/PhysRevB.87.045130},
  url = {https://arxiv.org/abs/1208.4834},
  eprint = {1208.4834},
  archiveprefix = {arXiv},
  primaryclass = {cond-mat.str-el},
}

@article{clarke2013parafermions,
  title = {Exotic {non-Abelian} anyons from conventional fractional quantum {Hall} states},
  author = {Clarke, David J. and Alicea, Jason and Shtengel, Kirill},
  year = 2013,
  month = jan,
  journal = {Nature Communications},
  volume = 4,
  number = 1,
  pages = 1348,
  doi = {10.1038/ncomms2340},
  isbn = {2041-1723},
  url = {https://arxiv.org/abs/1204.5479},
  eprint = {1204.5479},
  archiveprefix = {arXiv},
  primaryclass = {cond-mat.str-el},
}

@article{finck2013zbp,
  title = {Anomalous Modulation of a Zero-Bias Peak in a Hybrid Nanowire-Superconductor Device},
  author = {Finck, Aaron D. K. and Van Harlingen, Dale J. and Mohseni, Parsian K. and Jung, Kyooho and Li, Xiuling},
  year = 2013,
  month = mar,
  journal = {Physical Review Letters},
  publisher = {American Physical Society},
  volume = 110,
  pages = 126406,
  issue = 12,
  numpages = 5,
  doi = {10.1103/PhysRevLett.110.126406},
  url = {https://arxiv.org/abs/1212.1101},
  eprint = {1212.1101},
  archiveprefix = {arXiv},
  primaryclass = {cond-mat.mes-hall},
}

@article{levin2013protectededge,
  title = {Protected Edge Modes without Symmetry},
  author = {Levin, Michael},
  year = 2013,
  month = may,
  journal = {Physical Review X},
  publisher = {American Physical Society},
  volume = 3,
  pages = {021009},
  issue = 2,
  numpages = 18,
  doi = {10.1103/PhysRevX.3.021009},
  url = {https://link.aps.org/doi/10.1103/PhysRevX.3.021009},
  eprint = {1301.7355},
  archiveprefix = {arXiv},
  primaryclass = {cond-mat.str-el},
}

@article{wang2013edgereconstruction,
  title = {Edge Reconstruction in the $\ensuremath{\nu}\mathbf{=}2/3$ Fractional Quantum {Hall} State},
  author = {Wang, Jianhui and Meir, Yigal and Gefen, Yuval},
  year = 2013,
  month = dec,
  journal = {Phys. Rev. Lett.},
  publisher = {American Physical Society},
  volume = 111,
  pages = 246803,
  issue = 24,
  numpages = 5,
  doi = {10.1103/PhysRevLett.111.246803},
  url = {https://arxiv.org/abs/1305.6029},
  eprint = {1305.6029},
  archiveprefix = {arXiv},
  primaryclass = {cond-mat.str-el},
}

@article{klinovaja2014parafermions,
  title = {Parafermions in an Interacting Nanowire Bundle},
  author = {Klinovaja, Jelena and Loss, Daniel},
  year = 2014,
  month = jun,
  journal = {Physical Review Letters},
  publisher = {American Physical Society},
  volume = 112,
  pages = 246403,
  issue = 24,
  numpages = 5,
  doi = {10.1103/PhysRevLett.112.246403},
  url = {https://arxiv.org/abs/1311.3259},
  eprint = {1311.3259},
  archiveprefix = {arXiv},
  primaryclass = {cond-mat.mes-hall},
}

@article{zhang2014z4parafermions,
  title = {Time-Reversal-Invariant ${Z}_{4}$ Fractional {Josephson} Effect},
  author = {Zhang, Fan and Kane, Charles L.},
  year = 2014,
  month = jul,
  journal = {Physical Review Letters},
  publisher = {American Physical Society},
  volume = 113,
  pages = {036401},
  issue = 3,
  numpages = 5,
  doi = {10.1103/PhysRevLett.113.036401},
  url = {https://arxiv.org/abs/1404.1072},
  eprint = {1404.1072},
  archiveprefix = {arXiv},
  primaryclass = {cond-mat.supr-con},
}

@article{barkeshli2014parafermions,
  title = {Synthetic Topological Qubits in Conventional Bilayer Quantum {Hall} Systems},
  author = {Barkeshli, Maissam and Qi, Xiao-Liang},
  year = 2014,
  month = nov,
  journal = {Physical Review X},
  publisher = {American Physical Society},
  volume = 4,
  pages = {041035},
  issue = 4,
  numpages = 16,
  doi = {10.1103/PhysRevX.4.041035},
  url = {https://link.aps.org/doi/10.1103/PhysRevX.4.041035},
  eprint = {1302.2673},
  archiveprefix = {arXiv},
  primaryclass = {cond-mat.mes-hall},
}

@article{lin2014fqhereview,
  title = {Recent experimental progress of fractional quantum {Hall} effect: 5/2 filling state and graphene},
  author = {Lin, Xi and Du, Rui-Rui and Xie, Xincheng},
  year = 2014,
  month = nov,
  journal = {National Science Review},
  volume = 1,
  number = 4,
  pages = {564--579},
  doi = {10.1093/nsr/nwu071},
  issn = {2095-5138},
  url = {https://arxiv.org/abs/1501.00073},
  eprint = {1501.00073},
  archiveprefix = {arXiv},
  primaryclass = {cond-mat.mes-hall},
}

@article{alicea2016parafermions,
  title = {Topological Phases with Parafermions: Theory and Blueprints},
  author = {Alicea, Jason and Fendley, Paul},
  year = 2016,
  month = mar,
  journal = {Annual Review of Condensed Matter Physics},
  volume = 7,
  number = 1,
  pages = {119--139},
  doi = {10.1146/annurev-conmatphys-031115-011336},
  url = {https://arxiv.org/abs/1504.02476},
  eprint = {1504.02476},
  archiveprefix = {arXiv},
  primaryclass = {cond-mat.str-el},
}

@article{hutter2016parafermions,
  title = {Quantum computing with parafermions},
  author = {Hutter, Adrian and Loss, Daniel},
  year = 2016,
  month = mar,
  journal = {Physical Review B},
  publisher = {American Physical Society},
  volume = 93,
  pages = 125105,
  issue = 12,
  numpages = 7,
  doi = {10.1103/PhysRevB.93.125105},
  url = {https://arxiv.org/abs/1511.02704},
  eprint = {1511.02704},
  archiveprefix = {arXiv},
  primaryclass = {quant-ph},
}

@article{amet2016qhesc,
  title = {Supercurrent in the quantum {Hall} regime},
  author = {Amet, Francois and Ke, Chung T. and Borzenets, Ivan V. and Wang, Jiyingmei and Watanabe, Kenji and Taniguchi, Takashi and Deacon, Russell S. and Yamamoto, Michihisa and Bomze, Yuriy and Tarucha, Seigo  and Finkelstein, Gleb},
  year = 2016,
  month = may,
  journal = {Science},
  volume = 352,
  number = 6288,
  pages = {966--969},
  doi = {10.1126/science.aad6203},
  url = {https://arxiv.org/abs/1512.09083},
  eprint = {1512.09083},
  archiveprefix = {arXiv},
  primaryclass = {cond-mat.mes-hall},
}

@article{chen2016experimental,
  title = {Experimental phase diagram of zero-bias conductance peaks in superconductor/semiconductor nanowire devices},
  author = {Jun Chen  and Peng Yu  and John Stenger  and Moïra Hocevar  and Diana Car  and S\'{e}bastien R. Plissard  and Erik P. A. M. Bakkers  and Tudor D. Stanescu  and Sergey M. Frolov},
  year = 2017,
  journal = {Science Advances},
  volume = 3,
  number = 9,
  pages = {e1701476},
  doi = {10.1126/sciadv.1701476},
  url = {https://www.science.org/doi/abs/10.1126/sciadv.1701476},
  eprint = {1610.04555},
  archiveprefix = {arXiv},
  primaryclass = {cond-mat.mes-hall},
}

@article{hansson2016quantum,
  title = {Quantum {Hall} physics: Hierarchies and conformal field theory techniques},
  author = {Hansson, Thors H. and Hermanns, Maria and Simon, Steven H. and Viefers, Susanne F.},
  year = 2017,
  month = may,
  journal = {Reviews of Modern Physics},
  publisher = {American Physical Society},
  volume = 89,
  pages = {025005},
  issue = 2,
  numpages = 61,
  doi = {10.1103/RevModPhys.89.025005},
  url = {https://arxiv.org/abs/1601.01697},
  eprint = {1601.01697},
  archiveprefix = {arXiv},
  primaryclass = {cond-mat.str-el},
}

@article{sabo2017reconstruction,
  title = {Edge reconstruction in fractional quantum {Hall} states},
  author = {Sabo, Ron and Gurman, Itamar and Rosenblatt, Amir and Lafont, Fabien and Banitt, Daniel and Park, Jinhong and Heiblum, Moty and Gefen, Yuval and Umansky, Vladimir and Mahalu, Diana},
  year = 2017,
  month = may,
  journal = {Nature Physics},
  volume = 13,
  number = 5,
  pages = {491--496},
  doi = {10.1038/nphys4010},
  isbn = {1745-2481},
  url = {https://arxiv.org/abs/1603.06908},
  eprint = {1603.06908},
  archiveprefix = {arXiv},
  primaryclass = {cond-mat.mes-hall},
}

@article{lee2017inducing,
  title = {Inducing superconducting correlation in quantum {Hall} edge states},
  author = {Lee, Gil-Ho and Huang, Ko-Fan and Efetov, Dmitri K. and Wei, Di S. and Hart, Sean and Taniguchi, Takashi and Watanabe, Kenji and Yacoby, Amir and Kim, Philip},
  year = 2017,
  month = jul,
  journal = {Nature Physics},
  volume = 13,
  number = 7,
  pages = {693--698},
  doi = {10.1038/nphys4084},
  isbn = {1745-2481},
  url = {https://arxiv.org/abs/1609.08104},
  eprint = {1609.08104},
  archiveprefix = {arXiv},
  primaryclass = {cond-mat.mes-hall},
}

@article{gamayun2017transport,
  title = {Two-terminal transport along a proximity-induced superconducting quantum {Hall} edge},
  author = {Gamayun, Oleksandr and Hutasoit, Jimmy A. and Cheianov, Vadim V.},
  year = 2017,
  month = dec,
  journal = {Physical Review B},
  publisher = {American Physical Society},
  volume = 96,
  pages = 241104,
  issue = 24,
  numpages = 5,
  doi = {10.1103/PhysRevB.96.241104},
  url = {https://arxiv.org/abs/1705.01546},
  eprint = {1705.01546},
  archiveprefix = {arXiv},
  primaryclass = {cond-mat.mes-hall},
}

@article{ophetveld2020inasnetworks,
  title = {In-plane selective area {InSb--Al} nanowire quantum networks},
  author = {Op het Veld, Roy L. M. and Xu, Di and Schaller, Vanessa and Verheijen, Marcel A. and Peters, Stan M. E. and Jung, Jason and Tong, Chuyao and Wang, Qingzhen and de Moor, Michiel W. A. and Hesselmann, Bart and Vermeulen, Kiefer and Bommer, Jouri D. S. and Sue Lee, Joon and Sarikov, Andrey and Pendharkar, Mihir and Marzegalli, Anna and Koelling, Sebastian and Kouwenhoven, Leo P. and Miglio, Leo and Palmstr{\o}m, Chris J. and Zhang, Hao and Bakkers, Erik P. A. M.},
  year = 2020,
  month = mar,
  journal = {Communications Physics},
  volume = 3,
  number = 1,
  pages = 59,
  doi = {10.1038/s42005-020-0324-4},
  isbn = {2399-3650},
  url = {https://doi.org/10.1038/s42005-020-0324-4},
  eprint = {2103.06793},
  archiveprefix = {arXiv},
  primaryclass = {cond-mat.mes-hall},
}

@article{katzir2020superconducting,
  title = {Superconducting fractional quantum {Hall} edges via repulsive interactions},
  author = {Barak A. Katzir and Ady Stern and Erez Berg and Netanel H. Lindner},
  year = 2020,
  month = nov,
  journal = {arXiv preprint},
  url = {https://arxiv.org/abs/2011.13950},
  eprint = {2011.13950},
  archiveprefix = {arXiv},
  primaryclass = {cond-mat.mes-hall},
}

@article{pikulin2021protocol,
  title = {Protocol to identify a topological superconducting phase in a three-terminal device},
  author = {Dmitry I. Pikulin and Bernard van Heck and Torsten Karzig and Esteban A. Martinez and Bas Nijholt and Tom Laeven and Georg W. Winkler and John D. Watson and Sebastian Heedt and Mine Temurhan and Vicky Svidenko and Roman M. Lutchyn and Mason Thomas and de Lange, Gijs  and Lucas Casparis and Chetan Nayak},
  year = 2021,
  month = mar,
  journal = {arXiv preprint},
  url = {https://arxiv.org/abs/2103.12217},
  eprint = {2103.12217},
  archiveprefix = {arXiv},
  primaryclass = {cond-mat.mes-hall},
}

@article{hashisaka2021andreev,
  title = {Andreev reflection of fractional quantum {Hall} quasiparticles},
  author = {Hashisaka, Masayuki and Jonckheere, Thibaut and Akiho, Takafumi and Sasaki, Satoshi and Rech, Jerome and Martin, Thierry and Muraki, Koji},
  year = 2021,
  month = may,
  journal = {Nature Communications},
  volume = 12,
  number = 1,
  pages = 2794,
  doi = {10.1038/s41467-021-23160-6},
  isbn = {2041-1723},
  url = {https://doi.org/10.1038/s41467-021-23160-6},
  eprint = {2105.08916},
  archiveprefix = {arXiv},
  primaryclass = {cond-mat.mes-hall},
}

@article{gul2021andreev,
  title = {Andreev reflection in the fractional quantum {Hall} state},
  author = {{\"{O}}nder G\"ul and Yuval Ronen and Si Young Lee and Hassan Shapourian and Jonathan Zauberman and Young Hee Lee and Kenji Watanabe and Takashi Taniguchi and Ashvin Vishwanath and Amir Yacoby and Philip Kim},
  year = 2022,
  month = jun,
  journal = {Physical Review X},
  publisher = {American Physical Society},
  volume = 12,
  pages = {021057},
  issue = 2,
  numpages = 23,
  doi = {10.1103/PhysRevX.12.021057},
  url = {https://link.aps.org/doi/10.1103/PhysRevX.12.021057},
  eprint = {2009.07836},
  archiveprefix = {arXiv},
  primaryclass = {cond-mat.mes-hall},
}

@article{hatefipour2021induced,
  title = {Induced Superconducting Pairing in Integer Quantum {Hall} Edge States},
  author = {Hatefipour, Mehdi and Cuozzo, Joseph J. and Kanter, Jesse and Strickland, William M. and Allemang, Christopher R. and Lu, Tzu-Ming and Rossi, Enrico and Shabani, Javad},
  year = 2022,
  month = jul,
  journal = {Nano Letters},
  volume = 22,
  number = 15,
  pages = {6173--6178},
  doi = {10.1021/acs.nanolett.2c01413},
  url = {https://arxiv.org/abs/2108.08899},
  eprint = {2108.08899},
  archiveprefix = {arXiv},
  primaryclass = {cond-mat.mes-hall},
}

@article{galambos2022car,
  title = {Crossed {Andreev} reflection in spin-polarized chiral edge states due to {Meissner} effect},
  author = {Galambos, Tam{\'a}s Haidekker and Ronetti, Flavio and Het{\'e}nyi, Bence and Loss, Daniel and Klinovaja, Jelena},
  year = 2022,
  month = aug,
  journal = {Physical Review B},
  publisher = {American Physical Society},
  volume = 106,
  pages = {075410},
  issue = 7,
  numpages = 19,
  doi = {10.1103/PhysRevB.106.075410},
  url = {https://arxiv.org/abs/2203.05894},
  eprint = {2203.05894},
  archiveprefix = {arXiv},
  primaryclass = {cond-mat.mes-hall},
}

@article{manesco2021mechanism,
  title = {Mechanisms of {Andreev} reflection in quantum {Hall} graphene},
  author = {Antonio L. R. Manesco and Ian Matthias Flór and Chun-Xiao Liu and Anton R. Akhmerov},
  year = 2022,
  month = sep,
  journal = {SciPost Physics Core},
  publisher = {SciPost},
  volume = 5,
  pages = {045},
  doi = {10.21468/SciPostPhysCore.5.3.045},
  url = {https://scipost.org/10.21468/SciPostPhysCore.5.3.045},
  eprint = {2103.06722},
  archiveprefix = {arXiv},
  primaryclass = {cond-mat.mes-hall},
}

@article{tang2022vortex,
  title = {Vortex-enabled {Andreev} processes in quantum {Hall}--superconductor hybrids},
  author = {Tang, Yuchen and Knapp, Christina and Alicea, Jason},
  year = 2022,
  month = dec,
  journal = {Physical Review B},
  publisher = {American Physical Society},
  volume = 106,
  pages = 245411,
  issue = 24,
  numpages = 15,
  doi = {10.1103/PhysRevB.106.245411},
  url = {https://arxiv.org/abs/2207.10687},
  eprint = {2207.10687},
  archiveprefix = {arXiv},
  primaryclass = {cond-mat.mes-hall},
}

@article{valentini2022majoranalike,
  title = {{Majorana-like} {Coulomb} spectroscopy in the absence of zero-bias peaks},
  author = {Valentini, Marco and Borovkov, Maksim and Prada, Elsa and Mart{\'\i}-S{\'a}nchez, Sara and Botifoll, Marc and Hofmann, Andrea and Arbiol, Jordi and Aguado, Ram{\'o}n and San-Jose, Pablo and Katsaros, Georgios},
  year = 2022,
  month = dec,
  journal = {Nature},
  volume = 612,
  number = 7940,
  pages = {442--447},
  doi = {10.1038/s41586-022-05382-w},
  isbn = {1476-4687},
  url = {https://doi.org/10.1038/s41586-022-05382-w},
  eprint = {2203.07829},
  archiveprefix = {arXiv},
  primaryclass = {cond-mat.mes-hall},
}

@article{michelsen2020currentcorr,
  title = {Supercurrent-enabled {Andreev} reflection in a chiral quantum {Hall} edge state},
  author = {Michelsen, Andreas Bock and Recher, Patrik and Braunecker, Bernd and Schmidt, Thomas L.},
  year = 2023,
  month = jan,
  journal = {Physical Review Research},
  publisher = {American Physical Society},
  volume = 5,
  pages = {013066},
  issue = 1,
  numpages = 14,
  doi = {10.1103/PhysRevResearch.5.013066},
  url = {https://link.aps.org/doi/10.1103/PhysRevResearch.5.013066},
  eprint = {2004.10279},
  archiveprefix = {arXiv},
  primaryclass = {cond-mat.mes-hall},
}

@article{vignaud2023evidence,
  title = {Evidence for chiral supercurrent in quantum {Hall} {Josephson} junctions},
  author = {Vignaud, Hadrien and Perconte, David and Yang, Wenmin and Kousar, Bilal and Wagner, Edouard and Gay, Fr{\'e}d{\'e}ric and Watanabe, Kenji and Taniguchi, Takashi and Courtois, Herv{\'e} and Han, Zheng and Sellier, Hermann and Sac{\'e}p{\'e}, Benjamin},
  year = 2023,
  month = jan,
  journal = {Nature},
  volume = 624,
  number = 7992,
  pages = {545--550},
  doi = {10.1038/s41586-023-06764-4},
  isbn = {1476-4687},
  url = {https://arxiv.org/abs/2305.01766},
  eprint = {2305.01766},
  archiveprefix = {arXiv},
  primaryclass = {cond-mat.mes-hall},
}

@article{banerjee2022local,
  title = {Local and Nonlocal Transport Spectroscopy in Planar {Josephson} Junctions},
  author = {Abhishek Banerjee and Omri Lesser and Md Ahnaf Rahman and Candice Thomas and Tian Wang and Michael J. Manfra and Erez Berg and Yuval Oreg and Ady Stern and Charles M. Marcus},
  year = 2023,
  month = feb,
  journal = {Physical Review Letters},
  publisher = {American Physical Society},
  volume = 130,
  pages = {096202},
  issue = 9,
  numpages = 6,
  doi = {10.1103/PhysRevLett.130.096202},
  url = {https://arxiv.org/abs/2205.09419},
  eprint = {2205.09419},
  archiveprefix = {arXiv},
  primaryclass = {cond-mat.mes-hall},
}

@article{david2022effects,
  title = {Geometrical effects on the downstream conductance in quantum-{Hall}--superconductor hybrid systems},
  author = {David, Anthony and Meyer, Julia S. and Houzet, Manuel},
  year = 2023,
  month = mar,
  journal = {Physical Review B},
  publisher = {American Physical Society},
  volume = 107,
  pages = 125416,
  issue = 12,
  numpages = 9,
  doi = {10.1103/PhysRevB.107.125416},
  url = {https://arxiv.org/abs/2210.16867},
  eprint = {2210.16867},
  archiveprefix = {arXiv},
  primaryclass = {cond-mat.mes-hall},
}

@article{karki2022z3kondo,
  title = {{${\mathbb{Z}}_{3}$} Parafermion in the Double Charge {Kondo} Model},
  author = {Karki, Deepak B. and Boulat, Edouard and Pouse, Winston and Goldhaber-Gordon, David and Mitchell, Andrew K. and Mora, Christophe},
  year = 2023,
  month = apr,
  journal = {Physical Review Letters},
  publisher = {American Physical Society},
  volume = 130,
  pages = 146201,
  issue = 14,
  numpages = 7,
  doi = {10.1103/PhysRevLett.130.146201},
  url = {https://arxiv.org/abs/2210.04937},
  eprint = {2210.04937},
  archiveprefix = {arXiv},
  primaryclass = {cond-mat.mes-hall},
}

@article{kurilovich2022disorder,
  title = {Disorder-enabled {Andreev} reflection of a quantum {Hall} edge},
  author = {Kurilovich, Vladislav D. and Raines, Zachary M. and Glazman, Leonid I.},
  year = 2023,
  month = apr,
  journal = {Nature Communications},
  volume = 14,
  number = 1,
  pages = 2237,
  doi = {10.1038/s41467-023-37794-1},
  isbn = {2041-1723},
  url = {https://doi.org/10.1038/s41467-023-37794-1},
  eprint = {2201.00273},
  archiveprefix = {arXiv},
  primaryclass = {cond-mat.mes-hall},
}

@article{schiller2022interplay,
  title = {Superconductivity and fermionic dissipation in quantum {Hall} edges},
  author = {Schiller, Noam and Katzir, Barak A. and Stern, Ady and Berg, Erez and Lindner, Netanel H. and Oreg, Yuval},
  year = 2023,
  month = apr,
  journal = {Physical Review B},
  publisher = {American Physical Society},
  volume = 107,
  pages = {L161105},
  issue = 16,
  numpages = 6,
  doi = {10.1103/PhysRevB.107.L161105},
  url = {https://arxiv.org/abs/2202.10475},
  eprint = {2202.10475},
  archiveprefix = {arXiv},
  primaryclass = {cond-mat.mes-hall},
}

@article{hess2023trivial,
  title = {Trivial {Andreev} Band Mimicking Topological Bulk Gap Reopening in the Nonlocal Conductance of Long {Rashba} Nanowires},
  author = {Hess, Richard and Legg, Henry F. and Loss, Daniel and Klinovaja, Jelena},
  year = 2023,
  month = may,
  journal = {Physical Review Letters},
  publisher = {American Physical Society},
  volume = 130,
  pages = 207001,
  issue = 20,
  numpages = 6,
  doi = {10.1103/PhysRevLett.130.207001},
  url = {https://arxiv.org/abs/2210.03507},
  eprint = {2210.03507},
  archiveprefix = {arXiv},
  primaryclass = {cond-mat.mes-hall},
}

@article{kurilovich2022criticality,
  title = {Criticality in the Crossed {Andreev} Reflection of a Quantum {Hall} Edge},
  author = {Kurilovich, Vladislav D. and Glazman, Leonid I.},
  year = 2023,
  month = sep,
  journal = {Physical Review X},
  publisher = {American Physical Society},
  volume = 13,
  pages = {031027},
  issue = 3,
  numpages = 11,
  doi = {10.1103/PhysRevX.13.031027},
  url = {https://link.aps.org/doi/10.1103/PhysRevX.13.031027},
  eprint = {2209.12932},
  archiveprefix = {arXiv},
  primaryclass = {cond-mat.mes-hall},
}

@article{zhao2022loss,
  title = {Loss and Decoherence at the Quantum {Hall}-Superconductor Interface},
  author = {Zhao, Lingfei and Iftikhar, Zubair and Larson, Trevyn F. Q. and Arnault, Ethan G. and Watanabe, Kenji and Taniguchi, Takashi and Amet, Fran\c{c}ois and Finkelstein, Gleb},
  year = 2023,
  month = oct,
  journal = {Physical Review Letters},
  publisher = {American Physical Society},
  volume = 131,
  pages = 176604,
  issue = 17,
  numpages = 5,
  doi = {10.1103/PhysRevLett.131.176604},
  url = {https://arxiv.org/abs/2210.04842},
  eprint = {2210.04842},
  archiveprefix = {arXiv},
  primaryclass = {cond-mat.mes-hall},
}

@article{zhao2023perfect,
  title = {Perfect crossed {Andreev} reflection in the proximitized graphene/superconductor/proximitized graphene junctions},
  author = {Zhao, Shu-Chang and Gao, Lu and Cheng, Qiang and Sun, Qing-Feng},
  year = 2023,
  month = oct,
  journal = {Physical Review B},
  publisher = {American Physical Society},
  volume = 108,
  pages = 134511,
  issue = 13,
  numpages = 9,
  doi = {10.1103/PhysRevB.108.134511},
  url = {https://arxiv.org/abs/2310.17301},
  eprint = {2310.17301},
  archiveprefix = {arXiv},
  primaryclass = {cond-mat.supr-con},
}

@article{hatefipour2023andreev,
  title = {{Andreev} reflection of quantum {Hall} states through a quantum point contact},
  author = {Hatefipour, Mehdi and Cuozzo, Joseph J. and Levy, Ido and Strickland, William M. and Langone, Dylan and Rossi, Enrico and Shabani, Javad},
  year = 2024,
  month = jan,
  journal = {Physical Review B},
  publisher = {American Physical Society},
  volume = 109,
  pages = {035430},
  issue = 3,
  numpages = 6,
  doi = {10.1103/PhysRevB.109.035430},
  url = {https://arxiv.org/abs/2309.01856},
  eprint = {2309.01856},
  archiveprefix = {arXiv},
  primaryclass = {cond-mat.mes-hall},
}

@article{cao2023signatures,
  title = {Signatures of parafermion zero modes in fractional quantum {Hall}--superconductor heterostructures},
  author = {Cao, Junyi and Kou, Angela and Fradkin, Eduardo},
  year = 2024,
  month = apr,
  journal = {Physical Review B},
  publisher = {American Physical Society},
  volume = 109,
  pages = {L161106},
  issue = 16,
  numpages = 6,
  doi = {10.1103/PhysRevB.109.L161106},
  url = {https://arxiv.org/abs/2309.14411},
  eprint = {2309.14411},
  archiveprefix = {arXiv},
  primaryclass = {cond-mat.str-el},
}

@article{hu2024resistance,
  title = {Resistance Distribution of Decoherent Quantum {Hall}-Superconductor Edges},
  author = {Yichen Hu and Jing Wang and Biao Lian},
  year = 2024,
  month = may,
  journal = {arXiv preprint},
  url = {https://arxiv.org/abs/2405.17550},
  eprint = {2405.17550},
  archiveprefix = {arXiv},
  primaryclass = {cond-mat.mes-hall},
}

@article{cuozzo2024su4,
  title = {{SU(4)} symmetry breaking and induced superconductivity in graphene quantum {Hall} edges},
  author = {Cuozzo, Joseph J. and Rossi, Enrico},
  year = 2024,
  month = jul,
  journal = {Physical Review B},
  volume = 110,
  pages = {024518},
  doi = {10.1103/PhysRevB.110.024518},
  url = {https://arxiv.org/abs/2306.12483},
  eprint = {2306.12483},
  archiveprefix = {arXiv},
  primaryclass = {cond-mat.mes-hall},
}

@article{uday2023induced,
  title = {Induced superconducting correlations in a quantum anomalous {Hall} insulator},
  author = {Uday, Anjana and Lippertz, Gertjan and Moors, Kristof and Legg, Henry F. and Joris, Rikkie and Bliesener, Andrea and Pereira, Lino M. C. and Taskin, Alexey A. and Ando, Yoichi},
  year = 2024,
  month = jul,
  journal = {Nature Physics},
  doi = {10.1038/s41567-024-02574-1},
  isbn = {1745-2481},
  url = {https://doi.org/10.1038/s41567-024-02574-1},
  eprint = {2307.08578},
  archiveprefix = {arXiv},
  primaryclass = {cond-mat.mes-hall},
}

@article{manna2024shotnoise,
  title = {Shot noise as a diagnostic in the   $\nu=2/3$ fractional quantum {Hall} edge zoo},
  author = {Manna, Sourav and Das, Ankur and Gefen, Yuval and Goldstein, Moshe},
  year = 2024,
  month = dec,
  journal = {Low Temperature Physics},
  volume = 50,
  number = 12,
  pages = {1113--1122},
  doi = {10.1063/10.0034344},
  issn = {1063-777X},
  url = {https://arxiv.org/abs/2307.05175},
  eprint = {2307.05175},
  archiveprefix = {arXiv},
  primaryclass = {cond-mat.mes-hall},
}

@article{zhao2024thermal,
  title = {Thermal Properties of the Superconductor--Quantum Hall Interface},
  author = {Zhao, Lingfei and Larson, Trevyn F. Q. and Iftikhar, Zubair and Chiles, John and Watanabe, Kenji and Taniguchi, Takashi and Amet, Fran{\ifmmode \mbox{\c{c}}\else \c{c}\fi{}}ois and Finkelstein, Gleb},
  year = 2025,
  month = feb,
  journal = {Physical Review Letters},
  publisher = {American Physical Society},
  volume = 134,
  pages = {066001},
  issue = 6,
  numpages = 6,
  doi = {10.1103/PhysRevLett.134.066001},
  url = {https://arxiv.org/abs/2406.00850},
  eprint = {2406.00850},
  archiveprefix = {arXiv},
  primaryclass = {cond-mat.supr-con},
}

@article{burset2025conversion,
  title = {On-Demand and Tunable {Andreev} Conversion of Single-Electron Charge Pulses},
  author = {Burset, Pablo and Roussel, Benjamin and Moskalets, Michael and Flindt, Christian},
  year = 2025,
  month = dec,
  journal = {Physical Review Letters},
  volume = 135,
  pages = 246303,
  doi = {10.1103/lfsm-z1sl},
  url = {https://arxiv.org/abs/2312.13145},
  eprint = {2312.13145},
  archiveprefix = {arXiv},
  primaryclass = {cond-mat.mes-hall},
}

@article{bondarev2026conversion,
  title = {Quantum {Hall} {Andreev} conversion in graphene nanostructures},
  author = {Bondarev, Alexey and Klein, William H. and Baranger, Harold U.},
  year = 2026,
  month = feb,
  journal = {Physical Review B},
  volume = 113,
  pages = {085417},
  doi = {10.1103/kp8y-y288},
  url = {https://arxiv.org/abs/2507.02114},
  eprint = {2507.02114},
  archiveprefix = {arXiv},
  primaryclass = {cond-mat.mes-hall},
}

@article{baba2026topology,
  title = {Emergent Topology from {Landau} Level Mixing in Quantum {Hall}-Superconductor Nanostructures},
  author = {Baba, Yuriko and Yeyati, Alfredo Levy and Burset, Pablo},
  year = 2026,
  month = apr,
  journal = {Physical Review Letters},
  publisher = {American Physical Society},
  volume = 136,
  pages = 176601,
  doi = {10.1103/v68s-tgxs},
  url = {https://arxiv.org/abs/2507.14074},
  eprint = {2507.14074},
  archiveprefix = {arXiv},
  primaryclass = {cond-mat.mes-hall},
}

@article{jamotte2026interferometer,
  title = {Detecting crossed {Andreev} reflection in a quantum {Hall} interferometer with a superconducting beam splitter},
  author = {Jamotte, Maxime and Menei, Tom and Kumar, Manohar and Zyuzin, Alexander and Schmidt, Thomas L.},
  year = 2026,
  month = apr,
  journal = {arXiv preprint},
  url = {https://arxiv.org/abs/2604.09463},
  eprint = {2604.09463},
  archiveprefix = {arXiv},
  primaryclass = {cond-mat.mes-hall},
}

@article{bollmann2026phases,
  title = {Phases of quasi-one-dimensional fractional quantum anomalous {Hall} -- superconductor heterostructures},
  author = {Bollmann, Steffen and Haller, Andreas and Väyrynen, Jukka I. and Schmidt, Thomas L. and König, Elio J.},
  year = 2026,
  month = may,
  journal = {npj Quantum Materials},
  volume = 11,
  pages = 43,
  doi = {10.1038/s41535-026-00897-1},
  url = {https://arxiv.org/abs/2510.26686},
  eprint = {2510.26686},
  archiveprefix = {arXiv},
  primaryclass = {cond-mat.str-el},
}

@article{maji2026mixing,
  title = {Spin and orbital mixing of edge states in a quantum {Hall} system proximitized by a superconductor},
  author = {Maji, Shalini and Nowak, Michal P.},
  year = 2026,
  month = may,
  journal = {arXiv preprint},
  url = {https://arxiv.org/abs/2605.18411},
  eprint = {2605.18411},
  archiveprefix = {arXiv},
  primaryclass = {cond-mat.mes-hall},
}

@article{aghaee202620secondparitylifetime,
  title = {20 Second Parity Lifetime in an {InAs}--{Pb} Tetron Device},
  author = {Morteza Aghaee and Zulfi Alam and Mariusz Andrzejczuk and Andrey Antipov and Theodora Asimakidis and Mikhail Astafev and Lukas Avilovas and Ahmad Azizimanesh and Amin Barzegar and Bela Bauer and Jonathan Becker and Umesh Kumar Bhaskar and Andrea G. Boa and Srini Boddapati and Nichlaus Bohac and Jouri Bommer and Jan Borovsky and Léo Bourdet and Samuel Boutin and Srivatsa Chakravarthi and Benjamin J. Chapman and Nikolaos Chatzaras and Tzu-Chiao Chien and Jason Cho and Patrick T. Codd and William Cole and Paul W. Cooper and Fabiano Corsetti and Ajuan Cui and Tareq El Dandachi and Konstantinos Divanis and Clayton Doyle and Andreas Ekefjard and Javier A. Falcon and Saeed Fallahi and Luca Galletti and Geoffrey C. Gardner and Haris Gavranovic and João Pedro Morais Gomes and Deshan Govender and Flavio Griggio and Ruben Grigoryan and Sebastian Grijalva and Sergei Gronin and Jan Gukelberger and Marzie Hamdast and Esben Bork Hansen and Sebastian Heedt and Samantha Ho and Laurens Holgaard and Kevin van Hoogdalem and Jinnapat Indrapiromkul and Henrik Ingerslev and Lovro Ivancevic and Max Jantos and Thomas Jensen and Jaspreet Singh Jhoja and Vidul R. Joshi and Konstantin V. Kalashnikov and Ray Kallaher and Rachpon Kalra and Farhad Karimi and Torsten Karzig and Maren Elisabeth Kloster and Christina Knapp and Jonathan Knoblauch and Jonne Koski and Anders Kringhøj and Tom Laeven and Jeffrey Lai and de Lange, Gijs  and Thorvald W. Larsen and Kyunghoon Lee and Kongyi Li and Shuang Liang and Tyler Lindemann and Luna Lochmatter and Marijn Lucas and Roman Lutchyn and Morten Hannibal Madsen and Nasiari Madulid and Ivan Maliyov and Yanick Mampaey and Michael Manfra and Signe Brynold Markussen and Esteban A. Martinez and J. R. Mattinson and Mónica Meira and Camille A. Mikolas and Sarang Mittal and Gopakumar Mohandas and Christian Mollgaard and de Moor, Michiel W. A.  and Chris Moore and George Moussa and Bhargav Nabar and Anirudh Narla and Ahmad Naseri and Chetan Nayak and Bjørn Funch Schrøder Nielsen and Jens Hedegaard Nielsen and Michael J. Nystrom and Eoin O'Farrell and Keita Ohtani and Theodore Rex Orth and Sara Di Paolo and Camille Papon and Luca Petit and Dima Pikulin and Mohana Rajpalke and Alejandro Alcaraz Ramirez and Katrine Rasmussen and David Razmadze and Yuan Ren and Mariya Romanova and Loïc Roure and Ivan Sadovskyy and Lauri Sainiemi and Juan Carlos Estrada Saldaña and Irene Sanlorenzo and dos Santos, Tatiane Pereira  and Carl Vincent C. Saruda and Simon Schaal and John Schack and Emma R. Schmidgall and Christina Sfetsou and Cristina Sfiligoj and Zahra Shekason and Sarat Shankar Sinha and Patrick Sohr and de Sousa, Maria João Lourenço  and Kasper Roed Spiegelhauer and Tomas Stankevic and Henri J. Suominen and Judith Suter and Attila Szénási and Samuel M. L. Teicher and Naganivetha Thiyagarajah and Raj Tholapi and Mason Thomas and Dennis Tom and Emily Toomey and Joshua Tracy and Michelle Turley and Matthew D. Turner and Ivan Urban and Aakash Valliappan and Dmitrii V. Viazmitinov and Anna Wulff Viazmitinova and Dominik Johannes Vogel and Wenbo Wang and Christopher A. Watson and John Watson and Alex Webster and Joseph Weston and Timothy Williamson and Georg W. Winkler and van Woerkom, David J.  and Brian Paquelet Wuetz and Cécile X. Yu and Emrah Yucelen and Jesús Herranz Zamorano and Roland Zeisel and Guoji Zheng and Andrew M. Zimmerman},
  year = 2026,
  month = jun,
  journal = {arXiv preprint},
  url = {https://arxiv.org/abs/2606.03884},
  eprint = {2606.03884},
  archiveprefix = {arXiv},
  primaryclass = {cond-mat.mes-hall},
}

@article{cao2026interferometric,
      title={Interferometric Signatures of Zero Modes in Fractional Quantum {Hall}-Superconductor Heterostructures}, 
      author={Junyi Cao and Ramanjit Sohal and Angela Kou and Eduardo Fradkin},
      year={2026},
  month = aug,
  journal = {arXiv preprint},
      eprint={2608.19312},
      archivePrefix={arXiv},
      primaryClass={cond-mat.str-el},
      url={https://arxiv.org/abs/2608.19312}, 
}

@book{cardy1996scaling,
  title = {Scaling and Renormalization in Statistical Physics},
  author = {Cardy, John},
  year = 1996,
  publisher = {Cambridge University Press},
  series = {Cambridge Lecture Notes in Physics},
  doi = {10.1017/CBO9781316036440},
  place = {Cambridge},
  collection = {Cambridge Lecture Notes in Physics},
}

@book{giamarchi2004onedim,
  title = {Quantum Physics in One Dimension},
  author = {Giamarchi, Thierry},
  year = 2004,
  publisher = {Oxford University Press},
  series = {International Series of Monographs on Physics},
  doi = {10.1093/acprof:oso/9780198525004.001.0001},
  isbn = 9780198525004,
  url = {https://books.google.co.il/books?id=1MwTDAAAQBAJ},
  lccn = 2004299020,
}

@book{breuer2002theory,
  title = {The Theory of Open Quantum Systems},
  author = {Breuer, Heinz-Peter and Petruccione, Francesco},
  year = 2007,
  month = jan,
  publisher = {Oxford University Press},
  doi = {10.1093/acprof:oso/9780199213900.001.0001},
  isbn = 9780199213900,
  url = {https://doi.org/10.1093/acprof:oso/9780199213900.001.0001},
}

@inproceedings{kupriyanov1975js,
  title = {The {$J_s(\phi)$} relationship, {Abrikosov} vortices and {Josephson} vortices in variable thickness bridges},
  author = {Kupriyanov, Mikhail {Yu}. and Likharev, Konstantin K. and Maslova, L. A.},
  year = 1975,
  month = jan,
  booktitle = {Proceedings of the 14th International Conference on Low Temperature Physics, Otaniemi, Finland},
  pages = {14--20},
  url = {https://www.researchgate.net/publication/230681071_The_Jsphi_relationship_Abrikosov_vortices_and_Josephson_vortices_in_variable_thickness_bridges},
}

@misc{SupplementaryMaterial,
  note = {See supplementary material},
}

\clearpage
\appendix

\begin{widetext}

\section{Full current calculation}
\label{app:full-current}
In this section, we derive
\cref{eq:diff}. We work in
natural units $\hbar=k_{\mathrm{B}}=1$ and restore the conductance
quantum when relating potential to physical current. The ultraviolet
length cutoff is $a$. The finger is divided into coherent segments
$[x_i,x_{i+1}]$ of length $\ell_{\mathrm{seg}}=x_{i+1}-x_i$; there are $N_{\rm seg}=L/\ell_{\mathrm{seg}}$ such segments. The currents
generated in different segments are assumed to add incoherently. In the
dissipation-dominated regime, $\ell_{\mathrm{seg}}$ is chosen as
$\ell_{\alpha,{\rm I}}=a/(2\pi\alpha_{\rm I})$, the bare dissipation length of the integer mode. We use $i$ for
a segment, $r$ for a perturbing process, and $\beta$ for a normal mode.
As in \cref{sec:currents}, bare $\Km$, $\Um$, $\vb*{\phi}$, and
$\vb{q}$ refer to the doubled top--bottom system. The reference velocity
is $u$, as in \cref{eq:action-tunneling}; mode velocities carry an index.

\subsection{Continuity equation and segment current}
The density of component $j$ is
$\hat{\rho}_j=-\partial_x\phi_j/(2\pi)$. The physical charge density is
$\hat{\rho}_Q=e\vb{q}^\transp\hat{\vb*{\rho}}$.
We first treat the coherent perturbations; for that, we
decompose the Hamiltonian as
\begin{align}
    H
    &=H_0+\sum_rH_r,
    \label{appeq:hamiltonian-decomposition}\\
    H_0
    &=\frac{1}{4\pi}\int\dd{x}\,
    \partial_x\vb*{\phi}^\transp\Um\partial_x\vb*{\phi}.
    \label{appeq:quadratic-hamiltonian}
\end{align}
The equal-time commutator is
\begin{align}
    \comm{\partial_x\phi_j(x)}{\phi_k(x')}
    &=2\pi i\qty(\Km^{-1})_{jk}\delta(x-x').
    \label{appeq:comm_phi}
\end{align}
For the vertex operator
$V_{\vb{m}_r}=\exp(i\vb{m}_r^\transp\vb*{\phi})$, define
$\Theta_r=\vb{m}_r^\transp\vb*{\phi}$.  The commutator in
\cref{appeq:comm_phi}  gives
\begin{align}
    \comm{\partial_x\vb*{\phi}(x)}
    {\exp(\pm i\Theta_r(x'))}
    &=\mp2\pi\Km^{-1}\vb{m}_r\delta(x-x')
    \exp(\pm i\Theta_r(x')).
    \label{appeq:vertex-commutator}
\end{align}
Consequently $V_{\vb{m}}$ creates mode charge $\Km^{-1}\vb{m}$.

We write the interaction Hamiltonian term as
\begin{align}
    H_r&=\int\dd{x}\,\mathcal{H}_r(x)
    \\
    \mathcal{H}_r(x)
    &=\frac{u}{2\pi a^2}\qty(
        y_r(x)\exp(i\Theta_r(x))
        +y_r^*(x)\exp(-i\Theta_r(x))
    ).
    \label{appeq:generic-perturbation}
\end{align}
Here $u$ is the same reference velocity for every process.
For the coherent perturbations in
\cref{eq:full-perturbation-action}, the process label $r$ runs over
$(T,jk)$, $(A,jk)$, $(E,s;jk)$, and $(\mathrm{N},X;jk)$,
with the corresponding amplitudes $y_r(x)$. 
For a real amplitude, the Hamiltonian term~\eqref{appeq:generic-perturbation} is given by
$u y_r(x)\cos(\Theta_r)/(\pi a^2)$.

We use the Heisenberg equation
$\partial_t\hat{O}=-i\comm{\hat{O}}{H}$. 
The flux is then given by
\begin{align}
    \hat{\vb{j}}(x,t)
    &\equiv
    \frac{i}{2\pi}
    \comm{H_0}{\vb*{\phi}(x,t)} =-\frac{1}{2\pi}
    \Km^{-1}\Um\partial_x\vb*{\phi}(x,t).
    \label{appeq:mode-current-commutator}
\end{align}
The second equality follows from
\cref{appeq:quadratic-hamiltonian,appeq:comm_phi}.
Equivalently, its spatial derivative is
\begin{align}
    \partial_x\hat{\vb{j}}(x,t)
    &=i\comm{\hat{\vb*{\rho}}(x,t)}{H_0}.
    \label{appeq:mode-current-density-commutator}
\end{align}

The part of the density evolution generated by $H_r$ is proportional
to $\Km^{-1}\vb{m}_r$. We use this fact to define the scalar $\hat{J}_r(x,t)$ by
\begin{align}
    \Km^{-1}\vb{m}_r\hat{J}_r(x,t)
    &\equiv
    i\comm{\hat{\vb*{\rho}}(x,t)}{H_r}.
    \label{appeq:transfer-current-commutator}
\end{align}
Using
\cref{appeq:vertex-commutator,appeq:generic-perturbation} gives
\begin{align}
    \hat{J}_r(x,t)
    &=-\frac{u}{\pi a^2}\Im\qty(
        y_r(x)\exp(i\Theta_r(x,t))
    ).
    \label{appeq:local-transfer-current}
\end{align}
Positive $\hat{J}_r$ removes mode density in the direction
$\Km^{-1}\vb{m}_r$; a negative component of that vector describes a gain in the corresponding mode. Thus $\hat{J}_r$ measures signed particle transfer per unit time and length.  
Its integral over segment $i$,
\begin{align}
    \hat{\mathcal{J}}_r[i](t)
    &=\int_{x_i}^{x_{i+1}}\dd{x}\,\hat{J}_r(x,t),
    \label{appeq:segment-transfer-current}
\end{align}
is the signed particle transfer per unit time for the entire segment. The corresponding
electrical transfer is obtained by multiplying by
$e\vb{q}^\transp\Km^{-1}\vb{m}_r$.  

Applying the Heisenberg equation to
\cref{appeq:hamiltonian-decomposition} gives the local continuity equation:
\begin{align}
    \partial_t\hat{\vb*{\rho}}(x,t)
    &=-i\comm{\hat{\vb*{\rho}}(x,t)}{H}
    \nonumber\\
    &=-i\comm{\hat{\vb*{\rho}}(x,t)}{H_0}
    -i\sum_r\comm{\hat{\vb*{\rho}}(x,t)}{H_r}
    \nonumber\\
    &=- \partial_x\hat{\vb{j}}(x,t)
    -\sum_r\Km^{-1}\vb{m}_r\hat{J}_r(x,t).
    \label{appeq:local-continuity}
\end{align}
Integration over segment $i$ gives
\begin{align}
    \partial_t
    \int_{x_i}^{x_{i+1}}\dd{x}\,
    \hat{\vb*{\rho}}(x,t)
    &=
    -\hat{\vb{j}}(x_{i+1},t)+\hat{\vb{j}}(x_i,t) -
    \sum_r\Km^{-1}\vb{m}_r\hat{\mathcal{J}}_r[i](t).
    \label{appeq:integrated-continuity}
\end{align}
In a steady state, the expectation value of the left-hand side of
\cref{appeq:integrated-continuity} vanishes, giving the
``Kirchhoff-like'' equation
\begin{align}
    -\qty(
        \expval{\hat{\vb{j}}(x_{i+1})}
        -\expval{\hat{\vb{j}}(x_i)}
    )
    &=\sum_r\Km^{-1}\vb{m}_r
    \expval{\hat{\mathcal{J}}_r[i]}.
    \label{appeq:segment-continuity}
\end{align}

\subsection{Kubo formula and the coherent-segment approximation}
The current operator in
\cref{appeq:segment-transfer-current} is measured in segment $i$, but
the perturbing Hamiltonian in \cref{appeq:generic-perturbation}
initially extends over the entire finger.  To leading nonvanishing
order in $y_r(x)$, the Kubo formula therefore gives
\begin{align}
    \expval{\hat{\mathcal{J}}_r[i](t)}
    &=-i\int_{x_i}^{x_{i+1}}\dd{x}
    \int_{0}^{L}\dd{x'}
    \int_{-\infty}^{t}\dd{t'}
    \expval{\comm{\hat{J}_r(x,t)}{\mathcal{H}_r(x',t')}}_0.
    \label{appeq:kubo-full-finger}
\end{align}
The subscript $0$ denotes the Gaussian theory: the expectation value and
the interaction-picture time evolution are evaluated with the quadratic
Hamiltonian.  

The coherent-segment approximation discards a vertex correlator when
$x$ and $x'$ belong to different segments.   Since $x$ in
\cref{appeq:kubo-full-finger} already lies in segment $i$, this
approximation replaces the full-finger integral over $x'$ by
\begin{align}
    \expval{\hat{\mathcal{J}}_r[i](t)}
    &\simeq-i\int_{x_i}^{x_{i+1}}\dd{x}
    \int_{x_i}^{x_{i+1}}\dd{x'}
    \int_{-\infty}^{t}\dd{t'}
    \expval{\comm{\hat{J}_r(x,t)}{\mathcal{H}_r(x',t')}}_0.
    \label{appeq:kubo-segment}
\end{align}

\subsection{Quadratic modes, potential phase, and thermal correlator}
For convenience, we repeat the quadratic Lagrangian in
\cref{eq:edge_action}:
\begin{align}
    S_0&=\int\dd{x}\dd{t}\,\mathcal{L}_0,
    \nonumber\\
    \mathcal{L}_0
    &=-\frac{1}{4\pi}\qty[
        \partial_x\vb*{\phi}^\transp\Km\partial_t\vb*{\phi}
        +\partial_x\vb*{\phi}^\transp\Um\partial_x\vb*{\phi}
    ].
    \label{appeq:quadratic-lagrangian}
\end{align}
We use the simultaneous diagonalization and transformed fields
of \cref{eq:diagonalizing-W}:
\begin{align}
    \vb*{\phi}&=W\tilde{\vb*{\phi}},
    \\
    \tilde{\Km}=W^\transp\Km W&=\mathrm{diag}(\sigma_\beta),
    \\
    \tilde{\Um}=W^\transp\Um W&=\mathrm{diag}(u_\beta),
    \label{appeq:current-diagonal-basis}
\end{align}
where $\sigma_\beta=\pm1$ specifies the propagation direction and
$u_\beta>0$ is the velocity of mode $\beta$.  
For nonsingular symmetric integer $\Km$ and symmetric positive-definite
$\Um$, this simultaneous diagonalization can be constructed from the
symmetric generalized eigenvalue problem
\begin{align}
    \Km\vb{v}_\beta&=\eta_\beta\Um\vb{v}_\beta \nonumber \\
    \vb{v}_\beta^\transp\Um\vb{v}_\gamma&=\delta_{\beta\gamma}.
    \label{appeq:generalized-eigenproblem}
\end{align}
The eigenvalues $\eta_\beta$ are real and non-zero. Choosing the columns of
$W$ and the diagonal entries as
\begin{align}
    W_\beta&=\frac{\vb{v}_\beta}{\sqrt{\abs{\eta_\beta}}}
    \nonumber  \\\sigma_\beta&=\operatorname{sgn}(\eta_\beta)\nonumber \\
    u_\beta&=\frac{1}{\abs{\eta_\beta}},
    \label{appeq:generalized-eigenbasis}
\end{align}
then gives \cref{appeq:current-diagonal-basis}.

A vertex vector and a potential
transform as
\begin{align}
    \vb*{\lambda}_r&=W^\transp\vb{m}_r,
    \\
    \tilde{\vb*{\mu}}_i&=W^\transp\vb*{\mu}_i.
    \label{appeq:transformed-vector-voltage}
\end{align}
The scaling dimension is given by \cref{eq:sd} 
\begin{align}
    d_r&=\frac{1}{2}\vb*{\lambda}_r\vdot\vb*{\lambda}_r
    =\frac{1}{2}\sum_\beta\lambda_{r\beta}^2.
    \label{appeq:current-scaling-dimension}
\end{align}

The chemical potential $\vb*{\mu}_i$ is constant within
segment $i$ and has energy units. Its coupling to the edge is
\begin{align}
    H_\mu[i]
    &=\frac{1}{2\pi}\int_{x_i}^{x_{i+1}}\dd{x}\,
    \vb*{\mu}_i^\transp\partial_x\vb*{\phi}
    =\frac{1}{2\pi}\int_{x_i}^{x_{i+1}}\dd{x}\,
    \tilde{\vb*{\mu}}_i^\transp\partial_x\tilde{\vb*{\phi}}.
    \label{appeq:bias-hamiltonian}
\end{align}
An external electrical voltage $V$ instead couples through
$H_V=-eV\int\dd{x}\,\vb{q}^\transp\hat{\vb*{\rho}}$, and therefore
sets $\vb*{\mu}=eV\vb{q}$ in an equilibrated incoming edge.

Within that segment, the quadratic energy density can be written as
\begin{align}
    \mathcal{H}_0+\mathcal{H}_\mu[i]
    &=\frac{1}{4\pi}
    \qty(\partial_x\tilde{\vb*{\phi}}
        +\tilde{\Um}^{-1}\tilde{\vb*{\mu}}_i)^\transp
    \tilde{\Um}
    \qty(\partial_x\tilde{\vb*{\phi}}
        +\tilde{\Um}^{-1}\tilde{\vb*{\mu}}_i) -\frac{1}{4\pi}
    \tilde{\vb*{\mu}}_i^\transp\tilde{\Um}^{-1}\tilde{\vb*{\mu}}_i.
    \label{appeq:completed-square}
\end{align}
The biased Gaussian state has the uniform mean field gradient
\begin{align}
    \expval{\partial_x\tilde{\phi}_\beta(x,0)}_{\vb*{\mu}}
    &=-\frac{\tilde{\mu}_{i,\beta}}{u_\beta}.
    \label{appeq:voltage-induced-gradient}
\end{align}

Thus, the potential is incorporated in the interaction-picture
field by the c-number shift
\begin{align}
    \vb*{\phi}(x,t)
    &\mapsto\vb*{\phi}_0(x,t)
    +\qty(-\Um^{-1}x+\Km^{-1}t)\vb*{\mu}_i,
    \label{appeq:biased-field}
\end{align}
where $\vb*{\phi}_0$ is the zero-bias Gaussian field and we used
$W\tilde{\Um}^{-1}W^\transp=\Um^{-1}$ and
$W\tilde{\Km}^{-1}W^\transp=\Km^{-1}$.

For two spacetime points, define the center and relative coordinates
\begin{align}
    x_+&=\frac{x+x'}{2},
    \\
    x_-&=x-x',
    \\
    t_-&=t-t'.
    \label{appeq:relative-coordinates}
\end{align}
The difference between the two potential-induced c-number parts of
$\Theta_r=\vb{m}_r^\transp\vb*{\phi}$ is the bias phase
\begin{align}
    \vartheta_{r,i}(x_-,t_-)
    &\equiv \Theta_r(x,t) - \Theta_r(x', t') =  \sum_\beta\lambda_{r\beta}
    \qty(-\frac{x_-}{u_\beta}+\sigma_\beta t_-)\tilde{\mu}_{i,\beta} =\vb{m}_r^\transp
    \qty(-\Um^{-1}x_-+\Km^{-1}t_-)\vb*{\mu}_i.
    \label{appeq:bias-phase}
\end{align}

For a single diagonal mode, the finite-temperature correlator is
\citep[][Eq.\ C.54]{giamarchi2004onedim} (see also \cite{ginsparg1988applied,delft1998bosonization})
\begin{align}
    \mathcal{C}_{r\beta}(x_-,t_-)
    &\equiv
    \expval{
        \exp(i\lambda_{r\beta}\tilde{\phi}_\beta(x,t))
        \exp(-i\lambda_{r\beta}\tilde{\phi}_\beta(x',t'))}_0
    \nonumber\\
    &=\qty(
        \frac{\pi T a/u_\beta}
        {i\sinh(\pi T\qty(t_--\sigma_\beta x_-/u_\beta-i0^+))}
    )^{\lambda_{r\beta}^2}.
    \label{appeq:single-mode-correlator}
\end{align}
 The $i0^+$ prescription fixes
the analytic continuation below the branch singularity. Since the modes are independent, the full vertex correlator factorizes:
\begin{align}
    \mathcal{C}_r(x_-,t_-)
    &\equiv
    \expval{
        \exp(i\vb*{\lambda}_r^\transp\tilde{\vb*{\phi}}(x,t))
        \exp(-i\vb*{\lambda}_r^\transp\tilde{\vb*{\phi}}(x',t'))}_0
    \nonumber\\
    &=\prod_\beta\mathcal{C}_{r\beta}(x_-,t_-) =\prod_\beta
    \qty(
        \frac{\pi T a/u_\beta}
        {i\sinh(\pi T\qty(t_--\sigma_\beta x_-/u_\beta-i0^+))}
    )^{\lambda_{r\beta}^2}.
    \label{appeq:finite-temperature-vertex}
\end{align}

We can now evaluate the commutator in
\cref{appeq:kubo-segment}.  Charge neutrality of the Gaussian
expectation value retains only a vertex paired with its Hermitian
conjugate.  The pairing
$\exp(i\Theta_r(x,t))\exp(-i\Theta_r(x',t'))$ carries the amplitude
$y_r(x)y_r^*(x')$ and the bias phase
$\exp(i\vartheta_{r,i})$.  Combining it with the conjugate pairing
produces
$\Im\qty[y_r(x)y_r^*(x')\exp(i\vartheta_{r,i})]$.  This factor reduces
to $\abs{y_r}^2\sin(\vartheta_{r,i})$ for a spatially uniform
amplitude. Exchanging the two spacetime points in the second term of
the commutator extends the relative-time integral from the causal
half-line to the whole real axis. The explicit result is
\begin{align}
    \expval{\hat{\mathcal{J}}_r[i]}
    &=\frac{i u^2}{2\qty(\pi a^2)^2}
    \int_{x_i}^{x_{i+1}}\dd{x}
    \int_{x_i}^{x_{i+1}}\dd{x'}
    \int_{-\infty}^{\infty}\dd{t_-}\,
    \Im\qty[
        y_r(x)y_r^*(x')
        \exp(i\vartheta_{r,i}(x_-,t_-))
    ]
    \nonumber\\
    &\quad\times\prod_\beta
    \qty(
        \frac{\pi T a/u_\beta}
        {i\sinh(\pi T\qty(t_--\sigma_\beta x_-/u_\beta-i0^+))}
    )^{\lambda_{r\beta}^2}.
    \label{appeq:kubo-correlator}
\end{align}

\subsection{Linear response and the dimensionless integral}

We use the  dimensionless center and relative coordinates
\begin{align}
    \widetilde{x}_+&=\frac{x_+-x_i}{\ell_{\mathrm{seg}}},
    \\
    \widetilde{x}_-&=\frac{x_-}{\ell_{\mathrm{seg}}},
    \\
    \widetilde{t}&=\frac{ut_-}{\ell_{\mathrm{seg}}},
    \\
    \widetilde{T}&=\frac{T\ell_{\mathrm{seg}}}{u}.
    \label{appeq:dimensionless-current-variables}
\end{align}
Their spatial integration domain is
$-1\leq\widetilde{x}_-\leq1$ and
$\abs{\widetilde{x}_-}/2\leq\widetilde{x}_+
\leq1-\abs{\widetilde{x}_-}/2$.
In these variables,
\cref{appeq:bias-phase} becomes
\begin{align}
    \vartheta_{r,i}
    &=-\widetilde{x}_-\,
    \ell_{\mathrm{seg}}\vb{m}_r^\transp\Um^{-1}\vb*{\mu}_i
    +\widetilde{t}\,
    \frac{\ell_{\mathrm{seg}}}{u}\vb{m}_r^\transp\Km^{-1}\vb*{\mu}_i.
    \label{appeq:dimensionless-bias-phase}
\end{align}
We now expand the integrand to the linear order in $\vb*{\mu}$:
\begin{align}
    \Im\qty[
        y_r(x)y_r^*(x')\exp(i\vartheta_{r,i})
    ]
    &=\Im\qty[y_r(x)y_r^*(x')] +\Re\qty[y_r(x)y_r^*(x')]\vartheta_{r,i}
    +\mathcal{O}\qty(\vartheta_{r,i}^2).
    \label{appeq:linearized-amplitude-phase}
\end{align}
Substituting the linear
term into \cref{appeq:kubo-correlator} gives
\begin{align}
    \expval{\hat{\mathcal{J}}_r[i]}_{\mathrm{lin}}
    &=\frac{i u}{2\pi^2a}
    \qty(\frac{\ell_{\mathrm{seg}}}{a})^{3-2d_r}
    \qty(\pi\widetilde{T})^{2d_r}
    \prod_\beta\qty(\frac{u}{u_\beta})^{\lambda_{r\beta}^2}
    \int_{-1}^{1}\dd{\widetilde{x}_-}
    \int_{-\infty}^{\infty}\dd{\widetilde{t}} 
    \frac{
        -\widetilde{x}_-\,
        \ell_{\mathrm{seg}}\vb{m}_r^\transp\Um^{-1}\vb*{\mu}_i
        +\widetilde{t}\,
        \frac{\ell_{\mathrm{seg}}}{u}\vb{m}_r^\transp\Km^{-1}\vb*{\mu}_i
    }{
        \displaystyle\prod_\beta
        \qty[
            i\sinh\qty(
                \pi\widetilde{T}
                \qty(
                    \widetilde{t}
                    -\sigma_\beta\frac{u}{u_\beta}\widetilde{x}_-
                    -i0^+
                )
            )
        ]^{\lambda_{r\beta}^2}
    }
    \nonumber\\
    &\quad\times
    \int_{\abs{\widetilde{x}_-}/2}^{1-\abs{\widetilde{x}_-}/2}
    \dd{\widetilde{x}_+}\,
    \Re\qty[
        y_r\qty(
            x_i+\ell_{\mathrm{seg}}
            \qty(\widetilde{x}_++\frac{\widetilde{x}_-}{2})
        )
        y_r^*\qty(
            x_i+\ell_{\mathrm{seg}}
            \qty(\widetilde{x}_+-\frac{\widetilde{x}_-}{2})
        )
    ].
    \label{appeq:dimensionless-linear-current}
\end{align}
We now assume uniform amplitude  $y_r(x)=y_r$. In this case, the $\widetilde{x}_+$
integral in
\cref{appeq:dimensionless-linear-current} becomes
$\abs{y_r}^2\qty(1-\abs{\widetilde{x}_-})$.
We then shift the time contour according to
\begin{align}
    t_{\rm c}
    &=\pi\widetilde{T}\qty(
        \widetilde{t}-i0^++\frac{i}{2\widetilde{T}}
    ).
    \label{appeq:current-contour-shift}
\end{align}
The shifted contour has $t_{\rm c}\in\mathbb{R}$; this is a dimensionless real integration coordinate. No singularity is crossed. For each mode, the contour shift gives
\begin{align}
    i\sinh\qty(
        t_{\rm c}
        -\sigma_\beta\pi T\frac{\ell_{\mathrm{seg}}}{u_\beta}
        \widetilde{x}_-
        -\frac{i\pi}{2}
    )
    &=\cosh\qty(
        t_{\rm c}
        -\sigma_\beta\pi T\frac{\ell_{\mathrm{seg}}}{u_\beta}
        \widetilde{x}_-
    ),
    \nonumber\\
    \mathcal{D}_r(\widetilde{x}_-,t_{\rm c})
    &=\prod_\beta
    \cosh(t_{\rm c}
        -\sigma_\beta\pi T\frac{\ell_{\mathrm{seg}}}{u_\beta}
        \widetilde{x}_-)^{\lambda_{r\beta}^2}.
    \label{appeq:shifted-denominator}
\end{align}
Since any Hamiltonian term has integer conformal spin,
\begin{align}
    s_r
    &=\frac{1}{2}\vb{m}_r^\transp\Km^{-1}\vb{m}_r
    =\frac{1}{2}\sum_\beta\sigma_\beta\lambda_{r\beta}^2,
    \label{appeq:zero-conformal-spin}
\end{align}
the two neutral vertex orderings have a relative exchange phase
$\exp(\pm i\pi\sum_\beta\sigma_\beta\lambda_{r\beta}^2)
=\exp(\pm2\pi i s_r)=1$.  The complete shifted Kubo kernel is therefore
the real function $1/\mathcal{D}_r$.  Both this
kernel and the uniform pair weight
$1-\abs{\widetilde{x}_-}$ are invariant under
$(\widetilde{x}_-,t_{\rm c})\mapsto
(-\widetilde{x}_-,-t_{\rm c})$.  Consequently, the terms proportional to
$\widetilde{x}_-$ and $t_{\rm c}$ integrate to zero, whereas the constant
imaginary displacement multiplies an even kernel and survives.
Thus the spatial bias term vanishes, and the temporal potential combination remains,
giving
\begin{align}
    \expval{\hat{\mathcal{J}}_r[i]}
    &=\ell_{\mathrm{seg}}\frac{1}{a}A_r(T)\abs{y_r}^2
    \vb{m}_r^\transp\Km^{-1}\vb*{\mu}_i,
    \label{appeq:current_final_2}\\
    A_r(T)
    &=\frac{1}{4\pi}\frac{\ell_{\mathrm{seg}}}{a}
    \qty(\frac{\pi Ta}{u})^{2d_r-2}
    \prod_\beta\qty(\frac{u}{u_\beta})^{\lambda_{r\beta}^2}\mathcal{I}_r(T),
    \label{appeq:current_final_3}\\
    \mathcal{I}_r(T)
    &=\int_{-1}^{1}\dd{\widetilde{x}_-}
    \int_{-\infty}^{\infty}\dd{t_{\rm c}}\,
    \frac{1-\abs{\widetilde{x}_-}}
    {\mathcal{D}_r(\widetilde{x}_-,t_{\rm c})}.
    \label{appeq:current-integral}
\end{align}
The integral and $A_r$ are real and positive. 
\subsubsection{Equal-velocity reduction}
When all participating modes have velocity $u_\beta=u$, the time
integral in \cref{appeq:current-integral} can be evaluated analytically.
Reflecting $t_c\mapsto-t_c$ changes its spatial signs to those in the following convenient representation without changing the integral.
Define the chiral exponents
$q_{r,\pm}=\sum_{\sigma_\beta=\pm1}\lambda_{r\beta}^2=d_r\pm s_r$
and their smaller value $q_r=\min(q_{r,+},q_{r,-})=d_r-\abs{s_r}$.
For $d_r>0$, $\abs{s_r}\leq d_r$, and $z\geq0$, the result is
\begin{align}
    \mathcal{T}_r(z)
    &\equiv\int_{-\infty}^{\infty}\dd{t_{\rm c}}\,
    \frac{1}{\cosh(t_{\rm c}+z)^{q_{r,+}}
             \cosh(t_{\rm c}-z)^{q_{r,-}}}
    \nonumber\\
    &=2^{2d_r-1}\mathcal{B}(d_r,d_r)e^{-2q_r z}
    {}_2F_1\qty(q_r,d_r;2d_r;1-e^{-4z}),
    \label{appeq:equal-velocity-time-integral}
\end{align}
where $\mathcal{B}$ is Euler's beta function and ${}_2F_1$ is the
Gauss hypergeometric function. To obtain this expression, first
reflect $t_{\rm c}\mapsto-t_{\rm c}$ if necessary so that the smaller
exponent multiplies $\cosh(t_{\rm c}-z)$. The substitution
$w=[1+\tanh(t_{\rm c}+z)]/2$ then gives Euler's integral representation,
\begin{align*}
    \mathcal{T}_r(z)
    &=2^{2d_r-1}e^{-2q_r z}
    \int_0^1\dd{w}\,
    w^{d_r-1}(1-w)^{d_r-1}
    \qty[1-(1-e^{-4z})w]^{-q_r}.
\end{align*}
The time integral is even in the signed spatial separation. Thus, with
$\zeta=\pi T\ell_{\rm seg}/u$, the full response integral reduces to
\begin{align}
    \mathcal{I}_r(T)
    &=2\int_0^1\dd{v}\,(1-v)\mathcal{T}_r(\zeta v).
    \label{appeq:equal-velocity-spatial-integral}
\end{align}
This leaves one spatial integral for numerical integration. At large $z$, the time integral can be solved numerically instead of the hypergeometric expression
to avoid loss of floating-point accuracy as $1-e^{-4z}$ approaches one.
For a fully chiral vertex, $q_r=0$ and
$\mathcal{T}_r(z)=2^{2d_r-1}\mathcal{B}(d_r,d_r)$ is independent of $z$;
the spatial integral is then also elementary.

\subsubsection{Short- and long-segment limits}
For $d>0$, we use the integral
\begin{align}
    \int_{-\infty}^{\infty}\dd{z}\,\cosh(z)^{-2d}
    &=2^{2d-1}\mathcal{B}\qty(d,d).
    \label{appeq:intcosh}
\end{align}
Mode $\beta$ has thermal
coherence length $\ell_{T,\beta}=u_\beta/(\pi T)$; we use $\ell_T$ for the
corresponding characteristic scale when the mode velocities are
comparable.

If $\pi T\ell_{\mathrm{seg}}/u_\beta\ll1$ for every participating mode,
the spatial arguments in \cref{appeq:shifted-denominator} may be
dropped.  The $\widetilde{x}_-$ integral is then one and
\cref{appeq:intcosh} with $d=d_r$ gives
\begin{align}
    A_r^{\mathrm{short}}(T)
    &=\frac{1}{2\pi}\frac{\ell_{\mathrm{seg}}}{a}
    \qty(\frac{2\pi Ta}{u})^{2d_r-2}
    \prod_\beta\qty(\frac{u}{u_\beta})^{\lambda_{r\beta}^2}
    \mathcal{B}\qty(d_r,d_r).
    \label{appeq:current-short-segment}
\end{align}

For an operator with non-zero weight in both propagation directions,
in the opposite limit $\ell_{\mathrm{seg}}\gg\ell_T$, only
$\abs{\widetilde{x}_-}\lesssim u/(T\ell_{\mathrm{seg}})$ contributes,
so $1-\abs{\widetilde{x}_-}$ may be replaced by one and the spatial range
extended to the real line. If all participating modes have the same
velocity $u_\beta=u$, define
\begin{align}
    d_{r,+}&=\frac{1}{2}\sum_{\sigma_\beta=+1}\lambda_{r\beta}^2,
    \\
    d_{r,-}&=\frac{1}{2}\sum_{\sigma_\beta=-1}\lambda_{r\beta}^2,
    \\
    d_r&=d_{r,+}+d_{r,-}.
    \label{appeq:chiral-dimensions}
\end{align}
The light-cone variables
$z_\pm=t_{\rm c}\mp
\pi T\ell_{\mathrm{seg}}\widetilde{x}_-/u$ have Jacobian
$\dd{z_+}\dd{z_-}=2\pi T\ell_{\mathrm{seg}}
\dd{\widetilde{x}_-}\dd{t_{\rm c}}/u$.  The two integrals therefore
factorize and yield
\begin{align}
    A_r^{\mathrm{long}}(T)
    &=\frac{1}{4\pi}
    \qty(\frac{2\pi Ta}{u})^{2d_r-3}
    \mathcal{B}\qty(d_{r,+},d_{r,+})
    \mathcal{B}\qty(d_{r,-},d_{r,-}).
    \label{appeq:current-long-segment}
\end{align}
\cref{appeq:current-long-segment} requires
$d_{r,+}>0$ and $d_{r,-}>0$. 
For the fully chiral operator, if all velocities are equal, the change
of variable $t'_{\rm c}=t_{\rm c}- \sigma\pi T\ell_{\mathrm{seg}}\widetilde{x}_-/u$ instead gives
\cref{appeq:current-integral} independent of the segment length. In that special case
$A_r=A_r^{\mathrm{short}}$ even for a long segment: the two spatial
integrals remain coherent along the chiral trajectory. This includes
the same-side $\nu=4/3$ IF operator at $g_{\mathrm{I}}=g=1$, where
$g\equiv g_{\rm C}$ is the Luttinger parameter for the pair of fractional modes.

Outside the limits in
\cref{appeq:current-short-segment,appeq:current-long-segment}, or for
unequal velocities, the integral in \cref{appeq:current-integral} can
be evaluated numerically, as illustrated in \cref{fig:response-crossover}.  

There is also a simple physical way to
understand the different temperature powers.  In a short segment, an
event at $x$ can interfere coherently with an event anywhere in the
same segment, so the two spatial integrations sample an area of order
$\ell_{\mathrm{seg}}^2$.  In a long segment, thermal fluctuations
restrict the second event to a window of width
$\ell_T\propto T^{-1}$ around the first.  The coherent spatial phase
space is then of order $\ell_{\mathrm{seg}}\ell_T$, and the factor
$\ell_T$ accounts for the additional inverse power of $T$ in
\cref{appeq:current-long-segment}.

\begin{figure*}[t]
 \centering
 \includegraphics[width=\textwidth]{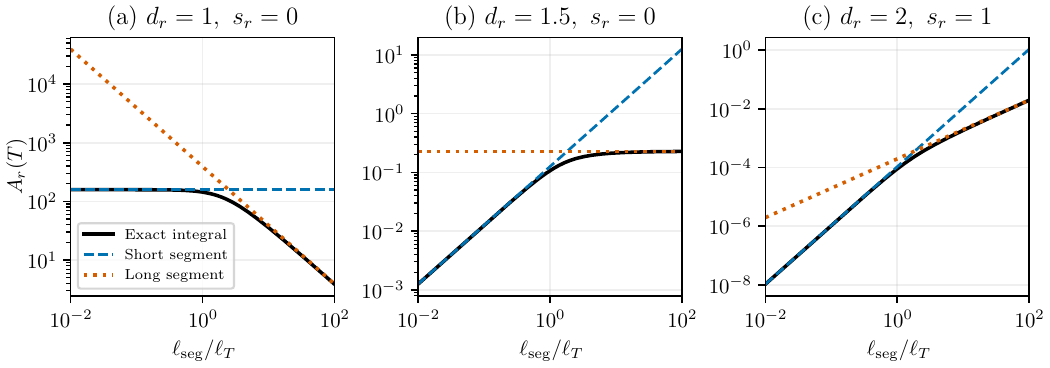}
 \caption{Crossover of the dimensionless response $A_r(T)$ between short
 and long coherent segments. Black curves evaluate the full integral in
 \cref{appeq:current-integral,appeq:current_final_3}; blue dashed and orange
 dotted curves show \cref{appeq:current-short-segment} and
 \cref{appeq:current-long-segment}, respectively. All participating modes
 have velocity $u$, $\ell_T=u/(\pi T)$, and
 $\ell_{\rm seg}/a=10^3$ is fixed while temperature varies.
 The panels use $(d_r,s_r)=(1,0),(3/2,0),(2,1)$, with
 $d_{r,\pm}=(d_r\pm s_r)/2$. The short-segment approximation applies
 at $\ell_{\rm seg}/\ell_T\ll1$, and the long-segment approximation at
 $\ell_{\rm seg}/\ell_T\gg1$. All three examples have weight in both
 chiralities; for a purely chiral operator at equal velocities,
 $A_r=A_r^{\rm short}$ at every segment length.}
 \label{fig:response-crossover}
\end{figure*}

\subsubsection{Disordered amplitudes}
We use the random-coupling notation of
\cref{eq:random-coupling-flow}.  For $X,Y\in\{A,T\}$, with the
mode-pair indices suppressed, $\eta_X=\delta y_X$ is the random
part of the corresponding main-text coupling. Let the overline denote
the disorder average and take independent complex Gaussian amplitudes satisfying
\begin{align}
    \overline{\eta_X(x)}&=0,
    \\
    \overline{\eta_X(x)\eta_Y^*(x')}
    &=\delta_{XY}aD_X\delta(x-x').
    \label{appeq:delta-disorder-variance}
\end{align}
In the dimensionless coordinates of
\cref{appeq:dimensionless-current-variables}, \cref{appeq:delta-disorder-variance} gives
\begin{align}
    \overline{
        \Re\qty[
            \eta_X\qty(
                x_i+\ell_{\mathrm{seg}}
                \qty(\widetilde{x}_++\frac{\widetilde{x}_-}{2})
            )
            \eta_X^*\qty(
                x_i+\ell_{\mathrm{seg}}
                \qty(\widetilde{x}_+-\frac{\widetilde{x}_-}{2})
            )
        ]
    }
    &=\frac{aD_X}{\ell_{\mathrm{seg}}}\delta(\widetilde{x}_-).
    \label{appeq:dimensionless-disorder-variance}
\end{align}
 Both $\eta_X$ and $D_X$ are dimensionless; the factor $a$ in the covariance carries the length unit. Substituting
\cref{appeq:dimensionless-disorder-variance} directly into
\cref{appeq:dimensionless-linear-current} collapses the
$\widetilde{x}_-$ integral to $\widetilde{x}_-=0$.   The spatial potential term vanishes, and the
remaining time integral gives
\begin{align}
    \overline{\expval{\hat{\mathcal{J}}_{X,\mathrm{dis}}[i]}}
    &=\ell_{\mathrm{seg}}\frac{1}{a}
    A_{X,\mathrm{dis}}(T)D_X
    \vb{m}_X^\transp\Km^{-1}\vb*{\mu}_i,
    \label{appeq:delta-disorder-current}\\
    A_{X,\mathrm{dis}}(T)
    &=\frac{1}{4\pi}
    \qty(\frac{\pi Ta}{u})^{2d_X-2}
    \prod_\beta\qty(\frac{u}{u_\beta})^{\lambda_{X\beta}^2}
    \mathcal{I}_{X,\mathrm{dis}},
    \label{appeq:delta-disorder-prefactor}\\
    \mathcal{I}_{X,\mathrm{dis}}
    &=\int_{-\infty}^{\infty}\dd{t_{\rm c}}\,
    \frac{1}{\cosh(t_{\rm c})^{2d_X}}.
    \label{appeq:delta-disorder-integral}
\end{align}
Using \cref{appeq:intcosh} in
\cref{appeq:delta-disorder-integral} evaluates the coefficient as
\begin{align}
    A_{X,\mathrm{dis}}(T)
    &=\frac{1}{2\pi}
    \qty(\frac{2\pi Ta}{u})^{2d_X-2}
    \prod_\beta\qty(\frac{u}{u_\beta})^{\lambda_{X\beta}^2}
    \mathcal{B}\qty(d_X,d_X).
    \label{appeq:delta-disorder-coefficient}
\end{align}
\cref{appeq:delta-disorder-current,appeq:delta-disorder-prefactor,appeq:delta-disorder-integral,appeq:delta-disorder-coefficient}
are valid for every value of $\ell_{\mathrm{seg}}/\ell_T$.  A disordered
coupling conductance therefore scales as $T^{2d_X-2}$.  Its temperature dependence
is the same as that of a clean short segment, but the factor
$\abs{y_X}^2\ell_{\mathrm{seg}}/a$ is replaced by $D_X$.

\subsubsection{Current into a local dissipative bath}
The remaining perturbation in \cref{eq:full-perturbation-action} is the
dissipative term in \cref{eq:diss_action}.  Before integrating out the
bath, the Hamiltonian corresponding to \cref{eq:diss_action} is
\begin{align}
    H_{\alpha,s,j}
    &=\int\dd{x}\,\qty(
        \frac{w_{s,j}}{\sqrt{a}}
        \zeta_{s,j,x}^\dagger\Psi_{s,j}(x)+\hc
    ).
    \label{appeq:edge-bath-hamiltonian}
\end{align}
Bath modes attached at different $x$ are uncorrelated.  Their
two-point function therefore contains $\delta(x-x')$, while the local
bath fermion contributes one additional inverse hyperbolic sine to the
time correlator.  The embedded electron vector $\vb{m}^{\rm tot}_{s,j}$ is defined in \cref{eq:doubled-electron-vectors}. Its
diagonal-basis vector and edge scaling dimension are
\begin{align}
    \vb*{\lambda}_{e,s,j}
    &=W^\transp\vb{m}^{\rm tot}_{s,j},
    \\
    d_{e,s,j}
    &=\frac{1}{2}\vb*{\lambda}_{e,s,j}
    \vdot\vb*{\lambda}_{e,s,j}
    =\frac{1}{2}\sum_\beta\lambda_{e,s,j,\beta}^2.
    \label{appeq:edge-electron-dimension}
\end{align}
The two edge-electron vertices in the current correlator contribute
the exponent $2d_{e,s,j}$, and the bath-fermion correlator contributes
one more.  With a zero-potential bath and
\begin{align}
    \alpha_{s,j}
    &=\frac{\abs{w_{s,j}}^2}{2\pi u v_{b,s,j}},
    \label{appeq:dissipation-strength}
\end{align}
where $v_{b,s,j}$ is the bath velocity, the dissipative particle
transfer per unit time from segment $i$ in linear response is
\begin{align}
    \expval{\hat{\mathcal{J}}_{\alpha,s,j}[i]}
    &=\ell_{\mathrm{seg}}\frac{1}{a}
    A_{\alpha,s,j}(T)\alpha_{s,j}
    (\vb{m}^{\rm tot}_{s,j})^\transp\Km^{-1}\vb*{\mu}_i,
    \label{appeq:dissipative-current}\\
    A_{\alpha,s,j}(T)
    &=\qty(\frac{2\pi Ta}{u})^{2d_{e,s,j}-1}
    \prod_\beta\qty(\frac{u}{u_\beta})^{\lambda_{e,s,j,\beta}^2}
    \mathcal{B}\qty(
        d_{e,s,j}+\frac{1}{2},d_{e,s,j}+\frac{1}{2}
    ).
    \label{appeq:dissipative-coefficient}
\end{align}

\subsection{From segment currents to the potential recurrence}
We now restore the conductance quantum
\begin{align}
    G_{\mathrm{Q}}&=\frac{e^2}{h}.
    \label{appeq:conductance-quantum}
\end{align}
In any basis, the relation between the generalized flux and the
chemical potentials is
\begin{align}
    \expval{\hat{\vb{j}}(x_i)}
    &=\frac{1}{h}\Km^{-1}\vb*{\mu}_i.
    \label{appeq:current-voltage-equation}
\end{align}
Substitution into \cref{appeq:segment-continuity} gives
\begin{align}
    -\frac{1}{h}\Km^{-1}
    \qty(\vb*{\mu}_{i+1}-\vb*{\mu}_i)
    &=\sum_r\Km^{-1}\vb{m}_r
    \expval{\hat{\mathcal{J}}_r[i]}.
    \label{appeq:voltage-balance}
\end{align}

For a coherent process in a uniform segment, define 
\begin{align}
    \gamma_r(T)
    &=\frac{e^2}{a}A_r(T)\abs{y_r}^2,
   \\
    \mathcal{M}_r
    &=\vb{m}_r
    \qty(\Km^{-1}\vb{m}_r)^\transp.
    \label{appeq:update-matrix}
\end{align}
The particle-transfer response is
$\expval{\hat{\mathcal{J}}_r[i]}=
\ell_{\mathrm{seg}}\gamma_r\vb{m}_r^\transp\Km^{-1}\vb*{\mu}_i/e^2$.
Multiplying \cref{appeq:voltage-balance} by $\Km$ yields the recurrence
\begin{align}
    \vb*{\mu}_{i+1}
    &=\qty(
        \mathds{1}-\frac{\ell_{\mathrm{seg}}}{G_{\mathrm{Q}}}
        \sum_r\gamma_r(T)\mathcal{M}_r
    )\vb*{\mu}_i.
    \label{appeq:voltage-recurrence}
\end{align}
A random process $X$ is
included by replacing $\gamma_r$ with
$e^2A_{X,\mathrm{dis}}D_X/a$; a bath process uses
$\gamma_{\alpha,s,j}=e^2A_{\alpha,s,j}\alpha_{s,j}/a$ and the electron
vector $\vb{m}^{\rm tot}_{s,j}$ in
\cref{eq:doubled-electron-vectors}.

We finally make explicit how the potential solution determines the
measured differential conductance.  Let
$\vb*{\mu}(x;V_{\rm in})$ denote the solution after imposing the incoming
boundary conditions and continuity around the tip, and $\hat{\vb{j}}_{\rm b}$ and $\vb*{\mu}_{\rm b}$ be the
bottom-side blocks of the current and potential vectors.

The positive physical direction of the outgoing bottom-edge current is
toward decreasing $x$.  Its total charge current is consequently
\begin{align}
    I_{\rm out}(V_{\rm in})
    &=-e\vb{q}_{\rm edge}^{\transp}
    \expval{\hat{\vb{j}}_{\rm b}(0;V_{\rm in})}
    \nonumber\\
    &=\frac{e}{h}\vb{q}_{\rm edge}^{\transp}
    \Km_{\rm edge}^{-1}\vb*{\mu}_{\rm b}(0;V_{\rm in}).
    \label{appeq:outgoing-current}
\end{align}

As our potential solution is linearized, the differential
conductance is 
\begin{align}
    G
    &\equiv
    \eval{\dv{I_{\rm out}}{V_{\rm in}}}_{V_{\rm in}=0}
    \nonumber\\
    &=\frac{e}{h}\vb{q}_{\rm edge}^{\transp}
    \Km_{\rm edge}^{-1}
    \eval{
        \pdv{\vb*{\mu}_{\rm b}(0;V_{\rm in})}{V_{\rm in}}
    }_{V_{\rm in}=0}.
\label{appeq:differential-conductance}
\end{align}

\subsection{Examples}
\subsubsection{Laughlin edge and the two-mode recurrence}
As a check, consider one $\nu=1/m$ Laughlin mode on each side of the
finger, ordered as $({\rm t},{\rm b})$ using the orientation convention
in \cref{eq:side-K-convention}:
\begin{align}
    \Km&=\begin{pmatrix}m&0\\0&-m\end{pmatrix},
    \\
    \vb*{\mu}_i&=\begin{pmatrix}\mu_{\rm t}(x_i)\\\mu_{\rm b}(x_i)\end{pmatrix},
    \\
    \sigma_{\mathrm{xy}}&=\frac{G_{\mathrm{Q}}}{m}.
    \label{appeq:laughlin-data}
\end{align}
We denote the scalar Luttinger parameter by $g\equiv g_{\rm C}$ and choose $u=u_{\rm C}$,
the common velocity of the two normalized modes. Normal tunneling and CAR, given in
\cref{eq:action-tunneling,eq:action-andreev}, are represented by
\begin{align}
    \vb{m}_T
    &=\begin{pmatrix}m\\m\end{pmatrix},
    \\
    \vb{m}_A
    &=\begin{pmatrix}m\\-m\end{pmatrix}.
    \label{appeq:laughlin-vertex-vectors}
\end{align}
meaning
\begin{align}
    \vb{m}_T^\transp\Km^{-1}\vb*{\mu}_i
    &=\mu_{\rm t}(x_i)-\mu_{\rm b}(x_i),
    \\
    \vb{m}_A^\transp\Km^{-1}\vb*{\mu}_i
    &=\mu_{\rm t}(x_i)+\mu_{\rm b}(x_i).
    \label{appeq:laughlin-voltage-combinations}
\end{align}
In the diagonal basis
\begin{align}
    \vb*{\lambda}_T
    &=\begin{pmatrix}\sqrt{mg}\\\sqrt{mg}\end{pmatrix},
    \\
    \vb*{\lambda}_A
    &=\begin{pmatrix}\sqrt{m/g}\\-\sqrt{m/g}\end{pmatrix},
    \label{appeq:laughlin-diagonal-vectors}
\end{align}
and hence the dimensions denoted by $d_X$ in
\cref{eq:general-dim-A,eq:general-dim-T} are
\begin{align}
    d_T&=mg,
    \\
    d_A&=\frac{m}{g}.
    \label{appeq:laughlin-dimensions}
\end{align}

For the dissipative term in \cref{eq:diss_action}, the edge-electron
vectors are
\begin{align}
    \vb{m}^{\rm tot}_{{\rm t},1}
    &=\begin{pmatrix}m\\0\end{pmatrix},
    \\
    \vb{m}^{\rm tot}_{{\rm b},1}
    &=\begin{pmatrix}0\\-m\end{pmatrix}.
    \label{appeq:laughlin-electron-vectors}
\end{align}
Either electron has scaling
dimension
\begin{align}
    d_e&=\frac{m}{4}\qty(g+g^{-1}).
    \label{appeq:laughlin-electron-dimension}
\end{align}
Therefore \cref{appeq:dissipative-coefficient} becomes
$A_\alpha=(2\pi Ta/u)^{2d_e-1}
\mathcal{B}\qty(d_e+1/2,d_e+1/2)$, in agreement
with the single-mode result of \citet{schiller2022interplay}.

For compactness, abbreviate $y_T=y_{T,11}$,
$y_A=y_{A,11}$, take the two dissipative strengths to be equal, and
define
\begin{align}
    \gamma_T
    &=\frac{e^2}{a}A_T\abs{y_T}^2,
    \\
    \gamma_A
    &=\frac{e^2}{a}A_A\abs{y_A}^2,
    \\
    \gamma_\alpha
    &=\frac{e^2}{a}A_\alpha\alpha.
    \label{appeq:laughlin-rates}
\end{align}
The segment currents are then simply
\begin{align}
    \expval{\hat{\mathcal{J}}_T[i]}
    &=\frac{\ell_{\mathrm{seg}}}{e^2}\gamma_T
    \qty(\mu_{\rm t}-\mu_{\rm b}),
    \\
    \expval{\hat{\mathcal{J}}_A[i]}
    &=\frac{\ell_{\mathrm{seg}}}{e^2}\gamma_A
    \qty(\mu_{\rm t}+\mu_{\rm b}),
    \\
    \expval{\hat{\mathcal{J}}_{\alpha,{\rm t}/{\rm b}}[i]}
    &=\frac{\ell_{\mathrm{seg}}}{e^2}\gamma_\alpha \mu_{{\rm t}/{\rm b}},
    \label{appeq:laughlin-segment-currents}
\end{align}
where all potentials on the right-hand side are evaluated at $x_i$.
Substitution into \cref{appeq:voltage-balance} gives
\begin{align}
    \frac{\vb*{\mu}_{i+1}-\vb*{\mu}_i}{\ell_{\mathrm{seg}}}
    &=\mathcal{A}\vb*{\mu}_i,
    \label{appeq:laughlin-difference-equation}\\
    \mathcal{A}
    &=\frac{1}{\sigma_{\mathrm{xy}}}
    \begin{pmatrix}
        -\gamma_T-\gamma_A-\gamma_\alpha
        &\gamma_T-\gamma_A\\
        -\gamma_T+\gamma_A
        &\gamma_T+\gamma_A+\gamma_\alpha
    \end{pmatrix}.
    \label{appeq:laughlin-recurrence-matrix}
\end{align}

For $N_{\rm seg}$ identical segments, the discrete solution is
\begin{align}
    \vb*{\mu}_{N_{\rm seg}}
    &=\mathcal{P}^{\rm disc}(L)\vb*{\mu}_0\\\mathcal{P}^{\rm disc}(L)&=\qty(\mathds{1}+\ell_{\mathrm{seg}}\mathcal{A})^{N_{\rm seg}}.
    \label{appeq:discrete-voltage-solution}
\end{align}
If the potential varies weakly over one segment,
$\ell_{\mathrm{seg}}\norm{\mathcal{A}}\ll1$, the continuum equation and its
solution are
\begin{align}
    \partial_x\vb*{\mu}(x)&=\mathcal{A}\vb*{\mu}(x),
    \\
    \vb*{\mu}(x)&=\mathcal{P}^{\rm cont}(x)\vb*{\mu}(0),
    \\\mathcal{P}^{\rm cont}(x)&=\exp(\mathcal{A} x).
    \label{appeq:continuum-voltage-solution}
\end{align}
Equivalently, if $\mathcal{A}$ is diagonalizable,
$\vb*{\mu}(x)=\sum_{\omega\in\mathrm{spec}(\mathcal{A})}
b_\omega\exp(\omega x)\vb{v}_\omega$, where $\vb{v}_\omega$ is the
eigenvector with eigenvalue $\omega$. The constants $b_\omega$ are fixed
by the potentials of the incoming modes
at the two ends.

\subsection{Analytic solutions for \texorpdfstring{$\nu=2/3$ and $\nu=4/3$}{ν=2/3 and ν=4/3}}
\label{appsec:23and43sol}
Both fillings contain one integer mode and one fractional mode on each
side of the finger.  To keep the two top--bottom pairs adjacent, we
permute the top--bottom ordering implied by
\cref{eq:side-K-convention} and write
\begin{align}
    \vb*{\mu}_{\rm pair}
    &=
    \begin{pmatrix}
        \mu_{{\rm t},{\rm I}}&
        \mu_{{\rm b},{\rm I}}&
        \mu_{{\rm t},{\rm f}}&
        \mu_{{\rm b},{\rm f}}
    \end{pmatrix}^{\transp},
    \\
    \Km_{\rm pair}
    &=
    \mathrm{diag}\qty(
        1,-1,3\chi_{\rm f},-3\chi_{\rm f}
    ),
    \label{appeq:integer-fractional-paired-basis}
\end{align}
where $\chi_{\rm f}=-1$ for $\nu=2/3$ and
$\chi_{\rm f}=+1$ for $\nu=4/3$.  

For $c\in\{{\rm I},{\rm f}\}$ and $X\in\{T,A\}$, let
$y_{X,c}$ denote the same-mode amplitude $y_{X,cc}$ and define
\begin{align}
    \gamma_{X,c}
    &=
    \frac{e^2}{a}A_{X,c}\abs{y_{X,c}}^2,
    \\
    \gamma_{\alpha,c}
    &=
    \frac{e^2}{a}A_{\alpha,c}\alpha_c,
    \\
    \alpha_c
    &=
    \frac{\abs{w_c}^2}{2\pi u v_{b,c}}.
    \label{appeq:integer-fractional-rates}
\end{align}
The top and bottom conductance coefficients are taken equal, as in
\cref{appeq:laughlin-rates}.
The coefficients in \cref{appeq:integer-fractional-rates} describe normal
inter-side tunneling $y_T$ and CAR $y_A$ in
\cref{eq:full-perturbation-action}, and the local bath coupling $w$
defined in \cref{appeq:edge-bath-hamiltonian}.  A delta-correlated
part of $y_T$ or $y_A$ is included by using the disorder-averaged coefficient
stated after \cref{appeq:voltage-recurrence}; the potential algebra is
unchanged.

For completeness, collect every coherent process that mixes the
sectors of integer and fractional modes into
\begin{align}
    \mathfrak{C}_{\rm mix}
    &=
    \qty{
        (T,{\rm I}{\rm f}),(T,{\rm f}{\rm I}),
        (A,{\rm I}{\rm f}),(A,{\rm f}{\rm I}),
        (E,{\rm t};{\rm I}{\rm f}),(E,{\rm b};{\rm I}{\rm f})
    }.
    \label{appeq:mixing-processes}
\end{align}
This set contains both the off-diagonal $y_T,y_A$ amplitudes and the
same-side equilibration amplitude $y_E$ from
\cref{eq:same-side-equilibration}. Applying \cref{appeq:current_final_2,appeq:voltage-balance} gives
the conductance coefficient of each process and the total mixing
contribution to the potential generator:
\begin{align}
    \gamma_r(T)
    &=
    \frac{e^2}{a}A_r(T)\abs{y_r}^2,
    \nonumber\\
    \mathcal{A}_{\rm mix}
    &=
    -\frac{1}{G_{\rm Q}}
    \sum_{r\in\mathfrak{C}_{\rm mix}}
    \gamma_r(T)\vb{m}_r
    \qty(\Km_{\rm pair}^{-1}\vb{m}_r)^\transp.
    \label{appeq:mixing-generator}
\end{align}

Strong dissipation of the integer mode keeps its Luttinger parameter near $g_{\mathrm{I}}=1$ according to \cref{eq:integer-g-flow}, requiring the additional limit
\begin{align}
    \varepsilon_{\rm mix}
    &\equiv
    \max\qty(
        \frac{G_{\rm Q}\norm{\mathcal{A}_{\rm mix}}}
        {\gamma_{\alpha,{\rm I}}},
        L\norm{\mathcal{A}_{\rm mix}}
    )
    \ll1.
    \label{appeq:strong-dissipation-assumption}
\end{align}
Before taking this limit, the four-mode potential generator is
\begin{align}
    \partial_x\vb*{\mu}_{\rm pair}
    &=
    \mathcal{A}_{\chi_{\rm f}}\vb*{\mu}_{\rm pair},
    \\
    \mathcal{A}_{\chi_{\rm f}}
    &=
    \mathcal{A}_{\rm I}\oplus
    \chi_{\rm f}\mathcal{A}_{\rm f}
    +\mathcal{A}_{\rm mix}.
    \label{appeq:integer-fractional-generator}
\end{align}
Writing
$\mathcal{A}_0=\mathcal{A}_{\rm I}\oplus\chi_{\rm f}\mathcal{A}_{\rm f}$,
the first correction in $\varepsilon_{\rm mix}$ to the continuum propagator is explicitly
\begin{align}
 \mathcal{P}(L)&=e^{L\mathcal{A}_0}
 +\int_0^L\dd{x}\,e^{(L-x)\mathcal{A}_0}\mathcal{A}_{\rm mix}
 e^{x\mathcal{A}_0}+O(\mathcal{A}_{\rm mix}^2).
\end{align}
The discrete recurrence has the corresponding sum over insertions
between unperturbed segment propagators.
The second entry of the maximum in
\cref{appeq:strong-dissipation-assumption} also specifies the order of
limits: small mixing coefficients must not produce order-one
conversion over the finite finger. 

It is useful to solve a downstream block once. With
$\sigma_{\mathrm{xy},{\rm I}}=G_{\rm Q}$ and
$\sigma_{\mathrm{xy},{\rm f}}=G_{\rm Q}/3$, \cref{appeq:laughlin-recurrence-matrix} gives
\begin{align}
    \mathcal{A}_c
    &=
    \frac{1}{\sigma_{\mathrm{xy},c}}
    \begin{pmatrix}
        -\gamma_{T,c}-\gamma_{A,c}-\gamma_{\alpha,c}
        &\gamma_{T,c}-\gamma_{A,c}\\
        -\gamma_{T,c}+\gamma_{A,c}
        &\gamma_{T,c}+\gamma_{A,c}+\gamma_{\alpha,c}
    \end{pmatrix}.
    \label{appeq:sector-voltage-generator}
\end{align}
This traceless matrix obeys
\begin{align}
    \mathcal{A}_c^2
    &=
    \frac{
        \qty(2\gamma_{T,c}+\gamma_{\alpha,c})
        \qty(2\gamma_{A,c}+\gamma_{\alpha,c})
    }{\sigma_{\mathrm{xy},c}^2}\mathds{1}_2
    =
    \frac{1}{L_{{\rm rel},c}^2}\mathds{1}_2,
    \nonumber\\
    L_{{\rm rel},c}
    &=
    \frac{\sigma_{\mathrm{xy},c}}{
        \sqrt{
            \qty(2\gamma_{T,c}+\gamma_{\alpha,c})
            \qty(2\gamma_{A,c}+\gamma_{\alpha,c})
        }
    }.
    \label{appeq:sector-relaxation-length}
\end{align}
Consequently, its transfer matrix is
\begin{align}
    \mathcal{P}^{\rm cont}_c(L)
    &\equiv
    \exp(\mathcal{A}_c L)
    \nonumber\\
    &=
    \cosh(\frac{L}{L_{{\rm rel},c}})\mathds{1}_2
    +
    L_{{\rm rel},c}
    \sinh(\frac{L}{L_{{\rm rel},c}})
    \mathcal{A}_c.
    \label{appeq:sector-transfer-matrix}
\end{align}

For a downstream pair incident from the biased source, the boundary
conditions and the isolated-pair response are
\begin{align}
    \mu_{{\rm t},c}(0)&=eV_{\rm in},
    \\
    \mu_{{\rm t},c}(L)&=\mu_{{\rm b},c}(L),
    \\
    r_c(L)
    &=
    \frac{\mu_{{\rm b},c}(0)}{eV_{\rm in}}.
    \label{appeq:sector-boundary-conditions}
\end{align}
The first equation fixes the incoming potential.  The second expresses
continuity where the edge turns around the tip of the finger.  If
$\mathcal{P}^{\rm cont}_{c,jk}$, with $j,k\in\{1,2\}$, denotes an entry of
\cref{appeq:sector-transfer-matrix}, these equations give
\begin{align}
    r_c(L)
    &=
    \frac{
        \mathcal{P}^{\rm cont}_{c,11}-\mathcal{P}^{\rm cont}_{c,21}
    }{
        \mathcal{P}^{\rm cont}_{c,22}-\mathcal{P}^{\rm cont}_{c,12}
    }
    \nonumber\\
    &=
    \frac{
        1-
        \sqrt{
            \frac{2\gamma_{A,c}+\gamma_{\alpha,c}}
            {2\gamma_{T,c}+\gamma_{\alpha,c}}
        }
        \tanh(\frac{L}{L_{{\rm rel},c}})
    }{
        1+
        \sqrt{
            \frac{2\gamma_{A,c}+\gamma_{\alpha,c}}
            {2\gamma_{T,c}+\gamma_{\alpha,c}}
        }
        \tanh(\frac{L}{L_{{\rm rel},c}})
    }.
    \label{appeq:sector-conductance-finite}
\end{align}
For a finger long
compared with $L_{{\rm rel},c}$,
\begin{align}
    r_c^{(\infty)}
    &=
    \frac{
        \sqrt{2\gamma_{T,c}+\gamma_{\alpha,c}}
        -
        \sqrt{2\gamma_{A,c}+\gamma_{\alpha,c}}
    }{
        \sqrt{2\gamma_{T,c}+\gamma_{\alpha,c}}
        +
        \sqrt{2\gamma_{A,c}+\gamma_{\alpha,c}}
    }.
    \label{appeq:sector-conductance-long}
\end{align}
Normal tunneling therefore contributes with a positive sign, CAR with
negative sign, and dissipation reduces the magnitude.  \cref{appeq:sector-conductance-long} is the
single-mode result of \citet{schiller2022interplay}.

For the pair of integer modes, the hierarchy suggested by the dissipation-dominated limit gives
\begin{align}
    \gamma_{\alpha,{\rm I}}
    &\gg
    \gamma_{T,{\rm I}},\gamma_{A,{\rm I}},
    \\
    \frac{L\gamma_{\alpha,{\rm I}}}{G_{\rm Q}}
    &\gg1,
    \nonumber\\
    r_{\rm I}^{(\infty)}
    &=
    \frac{
        \gamma_{T,{\rm I}}-\gamma_{A,{\rm I}}
    }{
        2\gamma_{\alpha,{\rm I}}
    }
    +
    \mathcal{O}\qty(
        \frac{
            \qty(\gamma_{T,{\rm I}}+\gamma_{A,{\rm I}})^2
        }{
            \gamma_{\alpha,{\rm I}}^2
        }
    ).
    \label{appeq:integer-strong-dissipation}
\end{align}

\subsubsection{\texorpdfstring{$\nu=2/3$}{ν=2/3} with KFP contacts}
For $\chi_{\rm f}=-1$, the microscopic integer mode is downstream and
fractional mode is upstream inside the finger. We assume that the exterior
edges reach the KFP fixed point, and that matching at $x=0$ introduces no
additional local transfer. Reservoirs therefore fix the incoming charge
and neutral modes of the exterior edge. In terms of the microscopic
potentials, the normalized charge and neutral potentials are
\begin{align}
    \mu_{s,\rho}
    &=\frac{3\mu_{s,{\rm I}}-\mu_{s,{\rm f}}}{\sqrt{6}},
    &\mu_{s,\sigma}
    &=\frac{\mu_{s,{\rm f}}-\mu_{s,{\rm I}}}{\sqrt{2}}.
    \label{appeq:kfp-potentials}
\end{align}
The corresponding normalized charge vector is
$(q_\rho,q_\sigma)=(\sqrt{2/3},0)$. Thus the source fixes
$\mu_{{\rm t},\rho}(0)=\sqrt{2/3}\,eV_{\rm in}$ and the  incoming neutral mode has $\mu_{{\rm b},\sigma}(0)=0$.
Together with physical-sector continuity at the tip, the boundary
conditions become
\begin{equation}
    \begin{aligned}
        3\mu_{{\rm t},{\rm I}}(0)-\mu_{{\rm t},{\rm f}}(0)
        &=2eV_{\rm in},\\
        \mu_{{\rm b},{\rm f}}(0)-\mu_{{\rm b},{\rm I}}(0)&=0,\\
        \mu_{{\rm t},{\rm I}}(L)&=\mu_{{\rm b},{\rm I}}(L),\\
        \mu_{{\rm t},{\rm f}}(L)&=\mu_{{\rm b},{\rm f}}(L).
    \end{aligned}
    \label{appeq:nu23-boundary-conditions}
\end{equation}
In the
decoupled-sector limit, the pair of integer modes evolves with
$\exp(\mathcal{A}_{\rm I}L)$ and the pair of upstream fractional modes with
$\exp(-\mathcal{A}_{\rm f}L)$. Their tip conditions imply
\begin{align}
    \mu_{{\rm b},{\rm I}}(0)
    &=r_{\rm I}(L)\mu_{{\rm t},{\rm I}}(0),\\
    \mu_{{\rm t},{\rm f}}(0)
    &=r_{\rm f}(L)\mu_{{\rm b},{\rm f}}(0),
    \label{appeq:nu23-pair-responses}
\end{align}
where $r_c$ is the auxiliary response in
\cref{appeq:sector-conductance-finite}. 
Combining these relations with \cref{appeq:nu23-boundary-conditions},
and suppressing the argument $L$ on $r_c$, gives
\begin{align}
    \mu_{{\rm t},{\rm I}}(0)
    &=\frac{2eV_{\rm in}}{3-r_{\rm I}r_{\rm f}},\\
    \mu_{{\rm b},{\rm I}}(0)=\mu_{{\rm b},{\rm f}}(0)
    &=\frac{2r_{\rm I}eV_{\rm in}}
    {3-r_{\rm I}r_{\rm f}},\\
    \mu_{{\rm t},{\rm f}}(0)
    &=\frac{2r_{\rm I}r_{\rm f}eV_{\rm in}}
    {3-r_{\rm I}r_{\rm f}}.
    \label{appeq:nu23-matched-potentials}
\end{align}
The outgoing
charge current is
\begin{align}
    I_{\rm out}
    &=\frac{e}{h}\qty[\mu_{{\rm b},{\rm I}}(0)
       -\frac{\mu_{{\rm b},{\rm f}}(0)}{3}]
    =\frac{2e}{3h}\mu_{{\rm b},{\rm I}}(0).
    \label{appeq:nu23-kfp-current}
\end{align}
Consequently,
\begin{align}
    \frac{G_{2/3}}{G_{\rm Q}}
    &=\frac{4r_{\rm I}}
      {3\qty(3-r_{\rm I}r_{\rm f})}
      +\mathcal{O}(\varepsilon_{\rm mix}).
    \label{appeq:nu23-conductance}
\end{align}
The mixing correction is controlled by the smallness condition in
\cref{appeq:strong-dissipation-assumption}.
For a small response of the integer modes and a bounded response of the fractional modes, the decoupled expression expands as
\begin{align}
    \frac{G_{2/3}}{G_{\rm Q}}
    &=\frac{4}{9}r_{\rm I}
      +\frac{4}{27}r_{\rm I}^2r_{\rm f}
      +\mathcal{O}(r_{\rm I}^3,\varepsilon_{\rm mix}).
    \label{appeq:nu23-small-integer-response}
\end{align}
Thus the response of the fractional modes affects the total signal only at second
order in the small response of the integer modes. Both contributions are small under strong dissipation of the integer mode.

In the long-finger hierarchy of
\cref{appeq:integer-strong-dissipation}, this gives
\begin{align}
    \frac{G_{2/3}}{G_{\rm Q}}
    &=\frac{2\qty(\gamma_{T,{\rm I}}-\gamma_{A,{\rm I}})}
      {9\gamma_{\alpha,{\rm I}}}
    +\mathcal{O}\qty(
        \varepsilon_{\rm mix},
        \frac{(\gamma_{T,{\rm I}}+\gamma_{A,{\rm I}})^2}
             {\gamma_{\alpha,{\rm I}}^2}
    ).
    \label{appeq:nu23-strong-dissipation}
\end{align}
At $g_{\mathrm{I}}=1$, the three conductance coefficients involving the integer mode have the same temperature
scaling by \cref{eq:integer-equal-dimensions}. Their ratio, and hence the
leading conductance, is temperature independent within this regime,
with a non-universal value and subleading corrections from the response of the fractional modes.

\subsubsection{\texorpdfstring{$\nu=4/3$}{ν=4/3}}
For $\chi_{\rm f}=+1$, both sectors propagate downstream.  The source
therefore fixes both incoming top potentials, and each mode is
continuous around the tip:
\begin{align}
    \mu_{{\rm t},{\rm I}}(0)
    &=
    \mu_{{\rm t},{\rm f}}(0)
    =
    eV_{\rm in},
    \nonumber\\
    \mu_{{\rm t},{\rm I}}(L)
    &=
    \mu_{{\rm b},{\rm I}}(L),
    \\
    \mu_{{\rm t},{\rm f}}(L)
    &=
    \mu_{{\rm b},{\rm f}}(L).
    \label{appeq:nu43-boundary-conditions}
\end{align}

We treat every other entry of the $4\times4$ potential generator
perturbatively relative to the bath contribution involving the integer mode.  In the paired
ordering of \cref{appeq:integer-fractional-paired-basis}, define
\begin{align}
    \vb*{\mu}_c
    &=
    \begin{pmatrix}
        \mu_{{\rm t},c}\\
        \mu_{{\rm b},c}
    \end{pmatrix},
    \\
    \partial_x
    \begin{pmatrix}
        \vb*{\mu}_{\rm I}\\
        \vb*{\mu}_{\rm f}
    \end{pmatrix}
    &=
    \mathcal{A}_{4/3}
    \begin{pmatrix}
        \vb*{\mu}_{\rm I}\\
        \vb*{\mu}_{\rm f}
    \end{pmatrix},
    \nonumber\\
    \mathcal{A}_{4/3}
    &\equiv
    \mathcal{A}_{\chi_{\rm f}=+1}
    =
    \mathcal{A}_{\rm I}\oplus\mathcal{A}_{\rm f}
    +\mathcal{A}_{\rm mix}
    =
    \begin{pmatrix}
        \mathcal{A}_{{\rm I}{\rm I}}
        &\mathcal{A}_{{\rm I}{\rm f}}\\
        \mathcal{A}_{{\rm f}{\rm I}}
        &\mathcal{A}_{{\rm f}{\rm f}}
    \end{pmatrix}.
    \label{appeq:nu43-block-generator}
\end{align}
Thus all same-side IF terms and all other cross-sector coefficients in
\cref{appeq:mixing-generator} are retained in the four blocks of
\cref{appeq:nu43-block-generator}.  In particular,
$\mathcal{A}_{{\rm I}{\rm f}}$ and
$\mathcal{A}_{{\rm f}{\rm I}}$ are the off-block-diagonal pieces that
are lost in the decoupled-sector approximation.  Separate only the
dominant part,
\begin{align}
    \mathcal{A}_0
    &=
    \frac{\gamma_{\alpha,{\rm I}}}{G_{\rm Q}}
    \begin{pmatrix}
        -1&0&0&0\\
        0&1&0&0\\
        0&0&0&0\\
        0&0&0&0
    \end{pmatrix},
    \\
    \mathcal{A}_1
    &=
    \mathcal{A}_{4/3}-\mathcal{A}_0,
    \nonumber\\
    \varepsilon
    &\equiv
    \frac{G_{\rm Q}\norm{\mathcal{A}_1}}
    {\gamma_{\alpha,{\rm I}}}
    \ll1,
    \\
    \frac{L\gamma_{\alpha,{\rm I}}}{G_{\rm Q}}
    &\gg1,
    \\
    L\norm{\mathcal{A}_1}
    &=\mathcal{O}(1).
    \label{appeq:nu43-strong-matrix-hierarchy}
\end{align}
The last condition keeps the accumulated evolution of the slow modes
finite.

The two zero eigenvalues of $\mathcal{A}_0$ are degenerate, so the slow
subspace must be diagonalized as a block. For a slow eigenvalue
$\omega=\mathcal{O}(\norm{\mathcal{A}_1})$, the components of its
eigenvector associated with the integer modes satisfy
\begin{align}
    \vb{v}_{\rm I}
    &=
    -\qty(
        \mathcal{A}_{{\rm I}{\rm I}}
        -\omega\mathds{1}_2
    )^{-1}
    \mathcal{A}_{{\rm I}{\rm f}}\vb{v}_{\rm f}
    \nonumber\\
    &=
    -\mathcal{A}_{{\rm I}{\rm I}}^{-1}
    \mathcal{A}_{{\rm I}{\rm f}}\vb{v}_{\rm f}
    +\mathcal{O}(\varepsilon^2)\norm{\vb{v}_{\rm f}}.
    \label{appeq:nu43-slow-eigenvector}
\end{align}
Substitution into the block of fractional modes gives the degenerate
perturbation matrix
\begin{align}
    \mathcal{A}_{\rm slow}
    &=
    \mathcal{A}_{{\rm f}{\rm f}}
    -
    \mathcal{A}_{{\rm f}{\rm I}}
    \mathcal{A}_{{\rm I}{\rm I}}^{-1}
    \mathcal{A}_{{\rm I}{\rm f}}
    +
    \mathcal{O}\qty(
        \frac{
            \varepsilon^3\gamma_{\alpha,{\rm I}}
        }{
            G_{\rm Q}
        }
    ).
    \label{appeq:nu43-schur-generator}
\end{align}
The IF blocks therefore rotate the slow eigenvectors at first order in
$\varepsilon$ and change the slow eigenvalues through the second-order
term in \cref{appeq:nu43-schur-generator}.  Equivalently, to first
order, the physical and block-diagonal potentials are connected by
\begin{align}
    \begin{pmatrix}
        \vb*{\mu}_{\rm I}\\
        \vb*{\mu}_{\rm f}
    \end{pmatrix}
    &=
    \begin{pmatrix}
        \mathds{1}_2
        &-\mathcal{A}_{{\rm I}{\rm I}}^{-1}
        \mathcal{A}_{{\rm I}{\rm f}}\\
        \mathcal{A}_{{\rm f}{\rm I}}
        \mathcal{A}_{{\rm I}{\rm I}}^{-1}
        &\mathds{1}_2
    \end{pmatrix}
    \begin{pmatrix}
        \vb*{\mu}'_{\rm I}\\
        \vb*{\mu}'_{\rm f}
    \end{pmatrix}
    +
    \mathcal{O}(\varepsilon^2)eV_{\rm in}.
    \label{appeq:nu43-block-transformation}
\end{align}
For an explicit closed expression, consider the spatially uniform,
same-side, and side-symmetric IF current. Side symmetry makes $y_{E,{\rm IF}}$, $A_{E,{\rm IF}}$, and $\gamma_{E,{\rm IF}}$ equal on the two sides, so their side label is suppressed. Event operators and vectors retain $s$. The same-side current is
\begin{align}
    \expval{\hat{\mathcal{J}}_{E,{\rm IF},s}[i]}
    &=
    \frac{\ell_{\mathrm{seg}}}{e^2}\gamma_{E,{\rm IF}}
    \qty(
        \mu_{s,{\rm I}}-\mu_{s,{\rm f}}
    ).
    \label{appeq:nu43-if-current}
\end{align}
For this uniform segment,
$\gamma_{E,{\rm IF}}=e^2 A_{E,{\rm IF}}\abs{y_{E,{\rm IF}}}^2/a$, with the full
finite-segment integral in \cref{appeq:current-integral}. In the paired
ordering, the process vectors are
\begin{align}
    \vb{m}_{E,{\rm IF},{\rm t}}
    &=
    \begin{pmatrix}1\\0\\-3\\0\end{pmatrix},
    \\
    \vb{m}_{E,{\rm IF},{\rm b}}
    &=
    \begin{pmatrix}0\\-1\\0\\3\end{pmatrix}.
    \label{appeq:nu43-if-vectors}
\end{align}
\cref{appeq:voltage-balance,appeq:nu43-if-current} give
\begin{align}
    \mathcal{A}_{E,{\rm IF}}
    &=
    \frac{\gamma_{E,{\rm IF}}}{G_{\rm Q}}
    \begin{pmatrix}
        -1&0&1&0\\
        0&1&0&-1\\
        3&0&-3&0\\
        0&-3&0&3
    \end{pmatrix}.
    \label{appeq:nu43-if-matrix}
\end{align}
The factors of three follow from
$\sigma_{\mathrm{xy},{\rm f}}=G_{\rm Q}/3$.  Retaining the within-sector normal
tunneling, CAR, and bath contributions gives the four blocks
\begin{align}
    \mathcal{A}_{{\rm I}{\rm I}}
    &=
    \frac{1}{G_{\rm Q}}
    \begin{pmatrix}
        -\gamma_{T,{\rm I}}-\gamma_{A,{\rm I}}
        -\gamma_{\alpha,{\rm I}}-\gamma_{E,{\rm IF}}
        &\gamma_{T,{\rm I}}-\gamma_{A,{\rm I}}\\
        -\gamma_{T,{\rm I}}+\gamma_{A,{\rm I}}
        &\gamma_{T,{\rm I}}+\gamma_{A,{\rm I}}
        +\gamma_{\alpha,{\rm I}}+\gamma_{E,{\rm IF}}
    \end{pmatrix},
    \nonumber\\
    \mathcal{A}_{{\rm I}{\rm f}}
    &=
    \frac{\gamma_{E,{\rm IF}}}{G_{\rm Q}}
    \begin{pmatrix}1&0\\0&-1\end{pmatrix},
    \\
    \mathcal{A}_{{\rm f}{\rm I}}
    &=
    3\mathcal{A}_{{\rm I}{\rm f}},
    \nonumber\\
    \mathcal{A}_{{\rm f}{\rm f}}
    &=
    \frac{3}{G_{\rm Q}}
    \begin{pmatrix}
        -\gamma_{T,{\rm f}}-\gamma_{A,{\rm f}}
        -\gamma_{\alpha,{\rm f}}-\gamma_{E,{\rm IF}}
        &\gamma_{T,{\rm f}}-\gamma_{A,{\rm f}}\\
        -\gamma_{T,{\rm f}}+\gamma_{A,{\rm f}}
        &\gamma_{T,{\rm f}}+\gamma_{A,{\rm f}}
        +\gamma_{\alpha,{\rm f}}+\gamma_{E,{\rm IF}}
    \end{pmatrix}.
    \label{appeq:nu43-if-blocks}
\end{align}

Using
\begin{align}
    \mathcal{A}_{{\rm I}{\rm I}}^{-1}
    &=
    \frac{G_{\rm Q}}{\gamma_{\alpha,{\rm I}}}
    \begin{pmatrix}-1&0\\0&1\end{pmatrix}
    +
    \mathcal{O}\qty(
        \frac{
            \varepsilon G_{\rm Q}
        }{
            \gamma_{\alpha,{\rm I}}
        }
    )
    \label{appeq:nu43-fast-inverse}
\end{align}
in \cref{appeq:nu43-schur-generator} gives
\begin{align}
    \mathcal{A}_{\rm slow}
    &=
    \frac{3}{G_{\rm Q}}
    \begin{pmatrix}
        -\gamma_{T,{\rm f}}-\gamma_{A,{\rm f}}
        -\gamma_{\alpha,{\rm f}}^{({\rm eff})}
        &\gamma_{T,{\rm f}}-\gamma_{A,{\rm f}}\\
        -\gamma_{T,{\rm f}}+\gamma_{A,{\rm f}}
        &\gamma_{T,{\rm f}}+\gamma_{A,{\rm f}}
        +\gamma_{\alpha,{\rm f}}^{({\rm eff})}
    \end{pmatrix}
    +
    \mathcal{O}\qty(
        \frac{
            \varepsilon^3\gamma_{\alpha,{\rm I}}
        }{
            G_{\rm Q}
        }
    ),
    \nonumber\\
    \gamma_{\alpha,{\rm f}}^{({\rm eff})}
    &=
    \gamma_{\alpha,{\rm f}}
    +\gamma_{E,{\rm IF}}
    -\frac{\gamma_{E,{\rm IF}}^2}{\gamma_{\alpha,{\rm I}}}.
    \label{appeq:nu43-effective-fractional-dissipation}
\end{align}
Thus the part of \cref{appeq:nu43-if-matrix} acting directly on the fractional modes contributes $\gamma_{E,{\rm IF}}$, while the
off-block-diagonal return through the dissipative integer mode produces
the correction $-\gamma_{E,{\rm IF}}^2/\gamma_{\alpha,{\rm I}}$.

For this IF matrix, \cref{appeq:nu43-block-transformation} becomes
\begin{align}
    \vb*{\mu}_{\rm I}
    &=
    \vb*{\mu}'_{\rm I}
    +
    \frac{\gamma_{E,{\rm IF}}}{\gamma_{\alpha,{\rm I}}}
    \vb*{\mu}'_{\rm f}
    +
    \mathcal{O}(\varepsilon^2)eV_{\rm in},
    \nonumber\\
    \vb*{\mu}_{\rm f}
    &=
    \vb*{\mu}'_{\rm f}
    -
    \frac{3\gamma_{E,{\rm IF}}}{\gamma_{\alpha,{\rm I}}}
    \vb*{\mu}'_{\rm I}
    +
    \mathcal{O}(\varepsilon^2)eV_{\rm in}.
    \label{appeq:nu43-specialized-transformation}
\end{align}
The fast component associated with the top integer mode is confined to a source boundary layer
of width $G_{\rm Q}/\gamma_{\alpha,{\rm I}}$, while the fast component associated with the bottom
integer mode is confined to a tip boundary layer of the same
width.  Their overlap is exponentially small under
\cref{appeq:nu43-strong-matrix-hierarchy}.
Matching the source and tip conditions in
\cref{appeq:nu43-boundary-conditions} then gives
\begin{align}
    \mu'_{{\rm t},{\rm f}}(0)
    &=
    \qty(
        1+
        \frac{3\gamma_{E,{\rm IF}}}{\gamma_{\alpha,{\rm I}}}
    )eV_{\rm in}
    +
    \mathcal{O}(\varepsilon^2)eV_{\rm in},
    \nonumber\\
    \mu'_{{\rm t},{\rm f}}(L)
    &=
    \mu'_{{\rm b},{\rm f}}(L)
    +
    \mathcal{O}(\varepsilon^2)eV_{\rm in},
    \nonumber\\
    \mu'_{{\rm b},{\rm I}}(0)
    &=
    \frac{
        \gamma_{T,{\rm I}}-\gamma_{A,{\rm I}}
    }{
        2\gamma_{\alpha,{\rm I}}
    }eV_{\rm in}
    +
    \mathcal{O}\qty(
        \varepsilon^2,
        \exp(-\frac{L\gamma_{\alpha,{\rm I}}}{G_{\rm Q}})
    )eV_{\rm in}.
    \label{appeq:nu43-transformed-boundary-conditions}
\end{align}
The first equation includes the source boundary layer generated by the
rotation in \cref{appeq:nu43-specialized-transformation}.  The third
equation is the fast response of the integer mode given in
\cref{appeq:integer-strong-dissipation}.

Define
\begin{align}
    L_{{\rm rel},{\rm f}}^{({\rm eff})}
    &=
    \frac{G_{\rm Q}/3}{
        \sqrt{
            \qty(
                2\gamma_{T,{\rm f}}
                +\gamma_{\alpha,{\rm f}}^{({\rm eff})}
            )
            \qty(
                2\gamma_{A,{\rm f}}
                +\gamma_{\alpha,{\rm f}}^{({\rm eff})}
            )
        }
    },
    \nonumber\\
    r_{\rm f}^{({\rm eff})}(L)
    &\equiv
    \frac{\mu'_{{\rm b},{\rm f}}(0)}{\mu'_{{\rm t},{\rm f}}(0)}
    \nonumber\\
    &=
    \frac{
        1-
        \sqrt{
            \frac{
                2\gamma_{A,{\rm f}}
                +\gamma_{\alpha,{\rm f}}^{({\rm eff})}
            }{
                2\gamma_{T,{\rm f}}
                +\gamma_{\alpha,{\rm f}}^{({\rm eff})}
            }
        }
        \tanh(
            \frac{L}{
                L_{{\rm rel},{\rm f}}^{({\rm eff})}
            }
        )
    }{
        1+
        \sqrt{
            \frac{
                2\gamma_{A,{\rm f}}
                +\gamma_{\alpha,{\rm f}}^{({\rm eff})}
            }{
                2\gamma_{T,{\rm f}}
                +\gamma_{\alpha,{\rm f}}^{({\rm eff})}
            }
        }
        \tanh(
            \frac{L}{
                L_{{\rm rel},{\rm f}}^{({\rm eff})}
            }
        )
    }.
    \label{appeq:nu43-effective-fractional-response}
\end{align}
This is \cref{appeq:sector-conductance-finite} evaluated with
the effective slow matrix in
\cref{appeq:nu43-effective-fractional-dissipation}.

\cref{appeq:nu43-specialized-transformation,appeq:nu43-transformed-boundary-conditions} give the physical
outgoing potentials
\begin{align}
    \frac{\mu_{{\rm b},{\rm f}}(0)}{eV_{\rm in}}
    &=
    \qty(
        1+
        \frac{3\gamma_{E,{\rm IF}}}{\gamma_{\alpha,{\rm I}}}
    )
    r_{\rm f}^{({\rm eff})}(L)
    +
    \mathcal{O}\qty(
        \varepsilon^2,
        \exp(-\frac{L\gamma_{\alpha,{\rm I}}}{G_{\rm Q}})
    ),
    \nonumber\\
    \frac{\mu_{{\rm b},{\rm I}}(0)}{eV_{\rm in}}
    &=
    \frac{
        \gamma_{T,{\rm I}}-\gamma_{A,{\rm I}}
    }{
        2\gamma_{\alpha,{\rm I}}
    }
    +
    \frac{\gamma_{E,{\rm IF}}}{\gamma_{\alpha,{\rm I}}}
    r_{\rm f}^{({\rm eff})}(L)
    +
    \mathcal{O}\qty(
        \varepsilon^2,
        \exp(-\frac{L\gamma_{\alpha,{\rm I}}}{G_{\rm Q}})
    ).
    \label{appeq:nu43-output-potentials}
\end{align}
Using \cref{appeq:current-voltage-equation}, the total conductance is
therefore
\begin{align}
    \frac{G_{4/3}}{G_{\rm Q}}
    &=
    \frac{
        \gamma_{T,{\rm I}}-\gamma_{A,{\rm I}}
    }{
        2\gamma_{\alpha,{\rm I}}
    }
    +
    \qty(
        \frac{1}{3}
        +
        \frac{2\gamma_{E,{\rm IF}}}{\gamma_{\alpha,{\rm I}}}
    )
    r_{\rm f}^{({\rm eff})}(L)
    +
    \mathcal{O}\qty(
        \varepsilon^2,
        \exp(-\frac{L\gamma_{\alpha,{\rm I}}}{G_{\rm Q}})
    ).
    \label{appeq:nu43-conductance}
\end{align}
One factor $\gamma_{E,{\rm IF}}/\gamma_{\alpha,{\rm I}}$ in
\cref{appeq:nu43-conductance} is the component of the outgoing
slow eigenvector associated with the integer mode. The other comes from
matching the physical source potential to the rotated eigenvector
associated with the fractional modes. This is why the IF
off-block-diagonal entries influence the conductance already at first
order, even though their correction to the slow eigenvalues is second
order.

For $\gamma_{E,{\rm IF}}=0$, \cref{appeq:nu43-conductance} reduces to
\begin{align}
    \left.
        \frac{G_{4/3}}{G_{\rm Q}}
    \right|_{\gamma_{E,{\rm IF}}=0}
    &=
    \frac{
        \gamma_{T,{\rm I}}-\gamma_{A,{\rm I}}
    }{
        2\gamma_{\alpha,{\rm I}}
    }
    +
    \frac{1}{3}r_{\rm f}(L)
    +
    \mathcal{O}\qty(
        \varepsilon^2,
        \exp(-\frac{L\gamma_{\alpha,{\rm I}}}{G_{\rm Q}})
    ).
    \label{appeq:nu43-zero-if-limit}
\end{align}
The contribution of the fractional modes approaches $\pm G_{\rm Q}/3$ when their response is saturated and normal tunneling or CAR, respectively, dominates. 

For a long finger with
$\gamma_{E,{\rm IF}}\gg
\gamma_{T,{\rm f}},\gamma_{A,{\rm f}},\gamma_{\alpha,{\rm f}}$
while $\gamma_{E,{\rm IF}}/\gamma_{\alpha,{\rm I}}\ll1$,
\cref{appeq:nu43-effective-fractional-response} gives
\begin{align}
    r_{\rm f}^{({\rm eff},\infty)}
    &=
    \frac{
        \gamma_{T,{\rm f}}-\gamma_{A,{\rm f}}
    }{
        2\gamma_{\alpha,{\rm f}}^{({\rm eff})}
    }
    +
    \mathcal{O}\qty(
        \frac{
            \qty(
                \gamma_{T,{\rm f}}+\gamma_{A,{\rm f}}
            )^2
        }{
            \qty(
                \gamma_{\alpha,{\rm f}}^{({\rm eff})}
            )^2
        }
    )
    \longrightarrow0.
    \label{appeq:nu43-if-suppression}
\end{align}

\subsection{Dominant conductance coefficients and the boundary response}
\label{app:dominant-rate-proof}

We give the full derivation of the result in \cref{sec:plateau_in_general}. The in/out subscripts follow \cref{eq:formal_solution}. The prescribed data are
$\vb*{\mu}_{\text{in}}(0)$, so 
we must invert $\mathsf{B}=\mathcal{P}_{\text{out},\text{out}}-\mathcal{P}_{\text{in},\text{out}}$. A well-posed boundary problem has a unique
outgoing solution for every allowed incoming bias; on this space,
$\mathsf{B}$ is invertible. 

For a single dominant process, suppress its label $r$:
$\vb{m}=\vb{m}_r$, $\mathcal{M}=\mathcal{M}_r$, $\gamma=\gamma_r$,
and $s=s_r$. The matrix associated with its vertex obeys
\begin{align}
    \mathcal{M}&=\vb{m}(\Km^{-1}\vb{m})^\transp,\\
    s&=\frac{1}{2}\vb{m}^\transp\Km^{-1}\vb{m}\\
    \mathcal{M}^2&=2s\mathcal{M}.
    \label{eq:transport-spin-identity}
\end{align}
Consequently, for the discrete propagator $\mathcal{P}=\mathcal{P}^{\rm disc}$~\eqref{appeq:discrete-voltage-solution}
with $z=L\gamma/G_{\rm Q}$,
\begin{align}
    \mathcal{P}&=\mathds{1}+f_s(z)\mathcal{M},\\
    f_s(z)&=
    \begin{cases}
        \qty((1-2sz/N_{\rm seg})^{N_{\rm seg}}-1)/(2s),&s\ne0,\\
        -z,&s=0.
    \end{cases}
    \label{eq:one-term-transfer}
\end{align}

Write $\vb{m}=(\vb{m}_{\text{in}},\vb{m}_{\text{out}})^\transp$ and
$\vb{h}=\Km^{-1}\vb{m}=(\vb{h}_{\text{in}},\vb{h}_{\text{out}})^\transp$,
and define $\vb{d}=\vb{m}_{\text{in}}-\vb{m}_{\text{out}}$. 
Define the boundary-potential map by $\vb*{\mu}_{\text{out}}(0)=\mathsf{T}_{\rm bdry}\vb*{\mu}_{\text{in}}(0)$. It maps incoming to outgoing potentials at the same endpoint $x=0$. The rank-one inverse gives
\begin{align}
    \mathsf{T}_{\rm bdry}
    &=\mathds{1}
      +\frac{f_s}{1-f_s\vb{h}_{\text{out}}^\transp\vb{d}}
       \vb{d}(\vb{h}_{\text{in}}+\vb{h}_{\text{out}})^\transp.
    \label{appeq:one-term-block-response}
\end{align}
Thus the electrical conductance is a ratio of affine functions of
$f_s$. The saturated response is the limit
$z=L\gamma/G_{\rm Q}\to\infty$, while the local segment update
remains weak. 
For example, increase $L$ at fixed $\ell_{\rm seg}$ and $\gamma$, with $\ell_{\rm seg}\gamma\norm{\mathcal{M}}/G_{\rm Q}\ll1$.
In this limit,
$f_s\to-1/(2s)$ for $s>0$, whereas $\abs{f_s}\to\infty$ for
$s\leq0$. If the relevant denominator is non-zero,
\begin{align}
 \mathsf T_{\rm bdry}^{\rm sat}
 &=\mathds{1}-\frac{\vb d(\vb h_{\rm in}+\vb h_{\rm out})^\transp}
 {2s+\vb h_{\rm out}^\transp\vb d}, &&s>0,
 \label{appeq:saturated-boundary-map-pos}\\
 \mathsf T_{\rm bdry}^{\rm sat}
 &=\mathds{1}-\frac{\vb d(\vb h_{\rm in}+\vb h_{\rm out})^\transp}
 {\vb h_{\rm out}^\transp\vb d}, &&s\leq0.
 \label{appeq:saturated-boundary-map-neg}
\end{align}
The distinction between the two branches comes from spatial
propagation. The conformal spin
$s=\vb{m}^{\transp}\Km^{-1}\vb{m}/2$ measures the signed imbalance
between the right- and left-moving contributions to the vertex.
For $s>0$, this driving imbalance decays toward increasing $x$,
and a sufficiently long finger removes it. The propagator approaches
$\mathds{1}-\mathcal{M}/(2s)$, giving the finite limit
$f_s\to-1/(2s)$ in \cref{appeq:saturated-boundary-map-pos}.
For $s<0$, the same combination relaxes in the opposite spatial
direction. The growing factor occurs
in both the numerator and denominator of
\cref{appeq:one-term-block-response}, and cancels in their ratio.
For $s=0$, the driving imbalance stays constant and the potential
change accumulates linearly with length because
$\mathcal{M}^2=0$. Thus both $s<0$ and $s=0$ have
$\abs{f_s}\to\infty$ and yield \cref{appeq:saturated-boundary-map-neg}.

These limits no longer depend on the magnitude of $\gamma$.
A zero denominator signals that the dominant process alone does not
fix this boundary response; subleading processes or additional
boundary data must then be retained. For ideal contacts specified by
the integer charge lattice and rational connectivity, a finite
saturated conductance is a rational multiple of $G_{\rm Q}$.

For clarity, consider one Laughlin pair with
$\Km=\mathrm{diag}(m,-m)$, $m>0$, and incoming potential
$\mu_{\rm t}(0)=eV_{\rm in}$. The tip condition is
$\mu_{\rm t}(L)=\mu_{\rm b}(L)$ and
$G=(G_{\rm Q}/m)\mu_{\rm b}(0)/(eV_{\rm in})$.
Normal tunneling and CAR have, respectively,
\begin{align}
 \vb{m}_T&=(m,m)^\transp,\\
 \mathcal{M}_T&=m\begin{pmatrix}1&-1\\1&-1\end{pmatrix},\\
 \vb{m}_A&=(m,-m)^\transp,\\
 \mathcal{M}_A&=m\begin{pmatrix}1&1\\-1&-1\end{pmatrix}.
\end{align}
Both have $s=0$, so $\mathcal{P}=\mathds{1}-z\mathcal{M}$ exactly
within the recurrence. Substitution into the tip condition gives
\begin{align}
 G_T&=\frac{G_{\rm Q}}{m},\\
 G_A&=\frac{G_{\rm Q}}{m}\frac{1-2mz}{1+2mz}
 \underset{z\to\infty}{\longrightarrow}-\frac{G_{\rm Q}}{m}.
 \label{appeq:normal-car-plateaus}
\end{align}
For a fully downstream multicomponent edge, the same signs give
$G=\pm\nu G_{\rm Q}$ when the dominant normal or pairing processes
fix the entire electrically driven sector.

\paragraph{Complete conversion of a charge subspace.}

For the downstream Jain block $\Km_{\rm J}$, define the normalized
charge vector
\begin{align}
 \vb{c}&=\Km_{\rm J}^{-1/2}\vb{q}_{\rm f},
 \label{appeq:normalized-charge-vector}
\end{align}
where $\vb{q}_{\rm f}=(1,\ldots,1)^{\transp}$. The incoming bias
$\vb*{\mu}_{\rm t}(0)=eV_{\rm in}\vb{q}_{\rm f}$ therefore has
normalized direction $\vb{c}$. Define the boundary-response matrix
$\mathsf{R}$ in these coordinates by
\begin{align}
 \Km_{\rm J}^{-1/2}\vb*{\mu}_{\rm b}(0)
 &=\mathsf{R}\Km_{\rm J}^{-1/2}\vb*{\mu}_{\rm t}(0).
 \label{appeq:normalized-boundary-response}
\end{align}
Substitution in \cref{appeq:outgoing-current} gives
\begin{align}
 \frac{G}{G_{\rm Q}}&=\vb{c}^{\transp}\mathsf{R}\vb{c}.
 \label{appeq:normalized-charge-response}
\end{align}

Let $\mathcal{I}$ be a subset of $n$ electron indices, and let $E_{\mathcal{I}}$ be the
$N\times n$ matrix whose columns are the corresponding unit vectors
$\vb{e}_j$. An electron vector $\Km_{\rm J}\vb{e}_j$ becomes
$\Km_{\rm J}^{1/2}\vb{e}_j$ in normalized coordinates. The columns of
$V_{\mathcal{I}}=\Km_{\rm J}^{1/2}E_{\mathcal{I}}$ therefore span the subspace affected by
CAR on these indices. Suppose CAR is perfect in this subspace, while its orthogonal complement has the normal
response. 

To evaluate this response, decompose any normalized incoming vector
into its projection onto $\operatorname{span}(V_{\mathcal{I}})$ and the remaining
orthogonal component. The orthogonal projector is
\begin{align}
 \Pi_{\mathcal{I}}&=V_{\mathcal{I}}(V_{\mathcal{I}}^{\transp}V_{\mathcal{I}})^{-1}V_{\mathcal{I}}^{\transp}.
 \label{appeq:converted-subspace-projector}
\end{align}
The transmission coefficient is $-1$ on the projected component and $+1$ on the remaining component, giving
\begin{align}
 \mathsf{R}&=-\Pi_{\mathcal{I}}+(\mathds{1}-\Pi_{\mathcal{I}})
 =\mathds{1}-2\Pi_{\mathcal{I}}.
 \label{appeq:complete-conversion-response}
\end{align}
The conductance is consequently the normal charge weight
$\vb{c}^{\transp}\vb{c}=\nu_N$ minus twice the converted charge
weight $\vb{c}^{\transp}\Pi_{\mathcal{I}}\vb{c}$, where
$\nu_n=n/(1+2pn)$ as in \cref{sec:generic-phase-structure}.

Let $\vb{1}_n$ denote the $n$-component vector of ones. Since
\begin{align}
 V_{\mathcal{I}}^{\transp}\vb{c}&=\vb{1}_n,\\
 V_{\mathcal{I}}^{\transp}V_{\mathcal{I}}&=\mathds{1}_n+2p\vb{1}_n\vb{1}_n^{\transp},\\
 (V_{\mathcal{I}}^{\transp}V_{\mathcal{I}})\vb{1}_n&=(1+2pn)\vb{1}_n,
\end{align}
the projection has weight
\begin{align}
 \vb{c}^{\transp}\Pi_{\mathcal{I}}\vb{c}
 &=\frac{\vb{1}_n^{\transp}\vb{1}_n}{1+2pn}
 =\nu_n,\nonumber\\
 \frac{G}{G_{\rm Q}}&=\nu_N-2\nu_n.
 \label{appeq:jain-subset-charge-weight}
\end{align}
For $n=0$ the projector is zero and the response is wholly normal.
The full CAR value $G/G_{\rm Q}=-\nu_N$ requires that the converted
subspace contain the entire source direction $\vb{c}$. Equivalently,
\begin{align}
 \Km_{\rm J}^{-1}\vb{q}_{\rm f}
 &\in\operatorname{span}_{\mathbb{R}}(E_{\mathcal{I}}).
 \label{eq:charged-span-condition}
\end{align}
For a Jain block,
$\Km_{\rm J}^{-1}\vb{q}_{\rm f}=\vb{q}_{\rm f}/(1+2pN)$ has a
non-zero component at every index, so this requires $n=N$.

\subsection{Neutral clusters in the presence of a pinned charged operator}
\label{app:neutral-cluster-response}

additional neutral gapped modes is
\begin{align}
 P&=\sum_C(\abs{C}-1),
\end{align}
where the sum is over the disjoint components on unmatched indices;
$\abs{C}$ is their size on either one edge.

For a CAR phase, a $T$-type component $C$ requires a surviving CAR
operator with indices $(j,k)$ to satisfy
\begin{align}
 (\vb{e}_j+\vb{e}_k)^{\transp}\vb{r}&=0.
 \label{appeq:neutral-cluster-compatibility}
\end{align}
This must hold for every difference vector $\vb{r}$ within $C$.
Equivalently, the vector $\vb{e}_j+\vb{e}_k$, which records the
indices involved in the CAR process, must be constant on $C$.
If $\abs{C}=2$, crossed CAR using its two distinct indices survives.
If $\abs{C}\geq3$, no elementary CAR involving that component survives.
Let $B$ count the indices on one edge in these $T$-type components
of size at least $3$, and let $\mathcal{I}$ be the complementary
set of $n=N-B$ indices after relabeling.

When all surviving CAR processes are saturated, the converted charge
weight is $\nu_n$. 
Thus
\begin{align}
 \frac{G_A}{G_{\rm Q}}&=\nu_N-2\nu_{N-B},\\
 \frac{G_T}{G_{\rm Q}}&=\nu_N.
 \label{appeq:neutral-cluster-conductance}
\end{align}
The normal-phase result follows because all remaining transport
conserves charge and there is no dissipation. For example, for $p=1$, $N=4$, a
pinned operator $A_{11}$ and two pinned neutral operators
$V_{\mathrm{N},T;23,23}$, $V_{\mathrm{N},T;34,34}$ give $B=3$ and $G_A/G_{\rm Q}=-2/9$;
without such a component, the same saturated limit gives $-4/9$.

\subsection{Several competing conductance coefficients}
\label{app:jain-hierarchy-method}
For the downstream Jain block $\Km_{\rm J}$, consider the gapless
theory ($M=P=0$) with mirror symmetry between the two edges. Two processes obtained by interchanging its top and bottom
indices have equal conductance coefficients. We group such processes
into a family, labeled $f$; a process unchanged by this interchange,
up to an overall sign of its vertex vector, is counted only once.

We first specify the limit used below. Retain a finite set of families
with non-zero coefficients. For example, distinct temperature powers
can be written as
\begin{align}
 \gamma_f(T)&=\bar{\gamma}_f
 \left(\frac{T}{T_0}\right)^{\eta_f}[1+o(1)],\\
 \eta_1&<\eta_2<\cdots<\eta_{n_{\rm fam}},
 \label{appeq:strict-temperature-hierarchy}
\end{align}
where $T_0$ is a fixed reference temperature and every
$\bar{\gamma}_f$ is finite and positive. Thus
$\gamma_{f+1}(T)/\gamma_f(T)\to0$ as $T\to0$. We call this condition a strict hierarchy, and we denote it by
$\gamma_1\gg\gamma_2\gg\cdots$. 

At each non-zero temperature, first take the long-finger limit so
that every non-zero relaxation mode needed for the boundary response
has decayed. Only then take $T\to0$. Scattering within a coherent
segment remains weak, as in \cref{sec:plateau_in_general}.

To calculate the response at finite coefficient ratios, write
$\vb{u}_r=\Km_{\rm J}^{-1/2}\vb{m}_{{\rm t},r}$ and
$\vb{v}_r=\Km_{\rm J}^{-1/2}\vb{m}_{{\rm b},r}$, and define
\begin{align}
 F&=\frac{1}{4}\sum_r\gamma_r
 (\vb{u}_r+\vb{v}_r)(\vb{u}_r+\vb{v}_r)^{\transp},\nonumber\\
 H&=\frac{1}{4}\sum_r\gamma_r
 (\vb{u}_r-\vb{v}_r)(\vb{u}_r-\vb{v}_r)^{\transp}.
 \label{appeq:jain-rate-pair}
\end{align}
These positive semidefinite matrices collect the contributions of
the symmetric and antisymmetric process vectors. Define
\begin{align}
 \vb{x}&=\Km_{\rm J}^{-1/2}(\vb*{\mu}_{\rm t}+\vb*{\mu}_{\rm b}),\\
 \vb{y}&=\Km_{\rm J}^{-1/2}(\vb*{\mu}_{\rm t}-\vb*{\mu}_{\rm b}).
\end{align}
The continuum version of
\cref{appeq:voltage-recurrence} gives
\begin{align}
 \partial_x\vb{x}
 &=-\frac{2}{G_{\rm Q}}F\vb{y}
 -\frac{1}{2G_{\rm Q}}\sum_r\gamma_r
 (\vb{u}_r+\vb{v}_r)(\vb{u}_r-\vb{v}_r)^{\transp}\vb{x},\\
 \partial_x\vb{y}
 &=-\frac{2}{G_{\rm Q}}H\vb{x}
 -\frac{1}{2G_{\rm Q}}\sum_r\gamma_r
 (\vb{u}_r-\vb{v}_r)(\vb{u}_r+\vb{v}_r)^{\transp}\vb{y}.
 \label{appeq:jain-flow-before-mirror-symmetry}
\end{align}
The last term in each equation contains a mixed outer product of
the symmetric vector $\vb{u}_r+\vb{v}_r$ and the antisymmetric
vector $\vb{u}_r-\vb{v}_r$, and cancels out due to mirror symmetry. Thus
\cref{appeq:jain-flow-before-mirror-symmetry} reduces to
\begin{align}
 \partial_x\vb{x}&=-\frac{2}{G_{\rm Q}}F\vb{y},\\
 \partial_x\vb{y}&=-\frac{2}{G_{\rm Q}}H\vb{x}.
 \label{appeq:jain-even-odd-flow}
\end{align}
For CAR and tunneling with equal top and bottom indices, define
$D_X=\operatorname{diag}(\gamma_{X,11},\ldots,\gamma_{X,NN})$
for $X=A,T$, and let
$D_\alpha=\operatorname{diag}(\gamma_{\alpha,1},\ldots,\gamma_{\alpha,N})$
contain the dissipation coefficients on either edge. In this case,
\begin{align}
 F&=\Km_{\rm J}^{1/2}(D_T+D_\alpha/2)\Km_{\rm J}^{1/2},\\
 H&=\Km_{\rm J}^{1/2}(D_A+D_\alpha/2)\Km_{\rm J}^{1/2}.
\end{align}

For positive-definite $F,H$, the solutions that decay toward the tip
form a subspace of the form $\vb{y}=\mathsf{S}\vb{x}$.
Substitution in \cref{appeq:jain-even-odd-flow} gives
$\mathsf{S}F\mathsf{S}=H$. The positive-definite solution selects
the decaying modes:
\begin{align}
 \mathsf{S}&=F^{-1/2}(F^{1/2}HF^{1/2})^{1/2}F^{-1/2}.
 \label{appeq:jain-decaying-graph}
\end{align}
Since the normalized top and bottom potentials are
$(\vb{x}+\vb{y})/2$ and $(\vb{x}-\vb{y})/2$, respectively, the
boundary response defined in \cref{appeq:normalized-boundary-response}
is
\begin{align}
 \mathsf{R}&=(\mathds{1}-\mathsf{S})(\mathds{1}+\mathsf{S})^{-1}.
 \label{appeq:jain-impedance-response}
\end{align}
Its conductance follows from \cref{appeq:normalized-charge-response}.
If the leading contributions make $F$ or $H$ singular, keep the
physical subleading coefficients until after forming the boundary
response, and then take the limit defined in \cref{appeq:strict-temperature-hierarchy}.
A direction in the common kernel of the full $F,H$ is unaffected by
all retained processes; the tip condition gives $\mathsf{R}=\mathds{1}$
on that ballistic subspace.

For example, retain only the four CAR and tunneling coefficients
with equal top and bottom indices at $N=2$, $p=1$. Consider first
\begin{align}
 \gamma_{A,11}&\gg\gamma_{T,22}\gg\gamma_{A,22}\gg\gamma_{T,11}.
 \label{appeq:jain-first-ordering}
\end{align}
For two-dimensional positive matrices, the square-root expression
in \cref{appeq:jain-decaying-graph} also gives
\begin{align}
 \mathsf{S}^{-1}
 &=\frac{F+\sqrt{\det(FH)}H^{-1}}{\sqrt{\operatorname{tr}(FH)+2\sqrt{\det(FH)}}}.
\end{align}
For \cref{appeq:jain-first-ordering}, it follows that
\begin{align}
 \norm{\mathsf{S}^{-1}}
 &=\mathcal{O}\left(
 \sqrt{\frac{\gamma_{T,22}}{\gamma_{A,11}}}
 +\sqrt{\frac{\gamma_{T,11}}{\gamma_{A,22}}}\right)
 \longrightarrow0.
\end{align}
Both eigenvalues of $\mathsf{S}$ therefore diverge, so
\cref{appeq:jain-impedance-response} gives
$\mathsf{R}\to-\mathds{1}$ and $G/G_{\rm Q}=-2/5$.
Now interchange the two slowest coefficients:
\begin{align}
 \gamma_{A,11}&\gg\gamma_{T,22}\gg\gamma_{T,11}\gg\gamma_{A,22}.
 \label{appeq:jain-ordered-example}
\end{align}
In this case the eigenvalues behave as
\begin{align}
 \lambda_1(\mathsf{S})&\sim\frac{3}{2}
 \sqrt{\frac{\gamma_{A,11}}{\gamma_{T,22}}}\longrightarrow\infty,\\
 \lambda_2(\mathsf{S})&\sim\frac{2}{3}
 \sqrt{\frac{\gamma_{A,22}}{\gamma_{T,11}}}\longrightarrow0.
 \label{appeq:jain-example-eigenvalues}
\end{align}
The first therefore gives response $-1$ and the second response $+1$.
The negative-response direction approaches
$\vb{u}=\Km_{\rm J}^{1/2}\vb{e}_1$, whose electrical weight is
\begin{align}
 \frac{(\vb{c}^{\transp}\vb{u})^2}{\vb{u}^{\transp}\vb{u}}
 &=\frac{1}{3}.
\end{align}
The orthogonal direction carries the remaining weight
$\vb{c}^{\transp}\vb{c}-1/3=1/15$, so
\begin{align}
 \frac{G}{G_{\rm Q}}&=-\frac{1}{3}+\frac{1}{15}=-\frac{4}{15}.
\end{align}
Thus the same two leading processes can give different conductances
because the slower processes determine the response in the remaining
direction.

\subsection{Rational response in a strict hierarchy}
\label{app:strict-hierarchy-proof}

We now prove the rationality statement of \cref{sec:plateau_in_general}
for a non-singular rational symmetric single-edge matrix
$\Km_{\rm edge}$. 

The proof proceeds in four steps. First, we separate the process
vectors into their even and odd parts under exchange of the edges
and apply block Gram--Schmidt in order of decreasing conductance
coefficients. This constructs a simpler transport problem with
independent two-dimensional blocks. Second, we show that this
simplification preserves the limiting boundary response, even when
different relaxation rates vanish at different powers of temperature.
Third, we solve each block and show that its decaying direction
approaches a rational vector independent of finite bare prefactors.
Finally, imposing the rational contact conditions requires only
rational linear algebra, giving a rational $G/G_{\rm Q}$.

We retain the notation
$\Km=\operatorname{diag}(\Km_{\rm edge},-\Km_{\rm edge})$ for the
doubled system. 
Let $J$ interchange the two edges:
\begin{align}
 J&=\begin{pmatrix}0&\mathds{1}\\\mathds{1}&0\end{pmatrix}.
\end{align}
A family consists of a rational process vector $\vb{m}_f$ and its
mirror $J\vb{m}_f$, with the same coefficient $\gamma_f(T)$.
If $J\vb{m}_f=\pm\vb{m}_f$, the two outer products describe the
same process, which we count only once. Define
\begin{align}
 C_f&=\begin{cases}
 \vb{m}_f\vb{m}_f^{\transp}
 +J\vb{m}_f\vb{m}_f^{\transp}J,&J\vb{m}_f\ne\pm\vb{m}_f,\\
 \vb{m}_f\vb{m}_f^{\transp},&J\vb{m}_f=\pm\vb{m}_f,
 \end{cases}\\
 C(T)&=\sum_f\gamma_f(T)C_f.
 \label{appeq:hierarchy-covariance}
\end{align}
The continuum transport equation is now
\begin{align}
 \partial_x\vb*{\mu}
 &=-\frac{1}{G_{\rm Q}}C(T)\Km^{-1}\vb*{\mu}.
 \label{appeq:hierarchy-transport}
\end{align}
Thus $C(T)\Km^{-1}=-G_{\rm Q}\mathcal{A}(T)$ in the notation of
\cref{eq:recurrence}. We assume the strict ordering
and order of limits defined in \cref{appeq:strict-temperature-hierarchy}.

Suppose that the perturbation vectors span the full space, so that
$C(T)$ is positive definite at every $T>0$. In the long-finger limit,
the entrance potentials lie in the stable subspace of
$-C(T)\Km^{-1}$, meaning the span of solutions that decay toward the
tip. Assume rational contact conditions and electrical projections,
and that these contact conditions select a unique solution in the
limiting stable subspace. We show that this subspace has a
rational basis determined only by the order in the hierarchy. Solving the
contact equations then requires only rational matrix inversion.
This is why the limiting $G/G_{\rm Q}$ is rational and independent of the finite prefactors. 

\paragraph{Mirror parity and block Gram--Schmidt orthogonalization.}
A vector has even or odd mirror parity if $J\vb{v}=\vb{v}$ or
$J\vb{v}=-\vb{v}$, respectively. The identity
$J\Km^{-1}J=-\Km^{-1}$ implies
$\vb{v}^{\transp}\Km^{-1}\vb{w}=0$ whenever $\vb{v}$ and
$\vb{w}$ have the same parity. For a family with two distinct
mirror partners, decompose $C_f$ into
\begin{align}
 \vb{v}_+&=\vb{m}_f+J\vb{m}_f,\\
 \vb{v}_-&=\vb{m}_f-J\vb{m}_f,\\
 C_f&=\frac{1}{2}
 (\vb{v}_+\vb{v}_+^{\transp}+\vb{v}_-\vb{v}_-^{\transp}).
 \label{appeq:mirror-parity-covariance}
\end{align}
If $J\vb{m}_f=\pm\vb{m}_f$, keep just $\vb{m}_f$, which already
has definite parity. Order the resulting vectors by the hierarchy
$\gamma_1\gg\gamma_2\gg\cdots$ in
\cref{appeq:strict-temperature-hierarchy}: first the vector or two
vectors from the family with coefficient $\gamma_1$, then those
with coefficient $\gamma_2$, and so on. Within a family contributing
two vectors, list the even vector first and the odd vector second.
We apply a block Gram--Schmidt procedure with respect to $\Km^{-1}$,
following this list from beginning to end.

Choose the first non-zero remaining vector $\vb{v}$ and the first
remaining vector $\vb{w}$ with
$b=\vb{v}^{\transp}\Km^{-1}\vb{w}\ne0$. Such a vector exists
because the bilinear form defined by $\Km^{-1}$ remains
nondegenerate on the remaining space. The selected vectors have
opposite parity. Each has zero product with itself under $\Km^{-1}$,
so the usual one-vector Gram--Schmidt step would divide by zero.
We instead project out their two-dimensional span together. The
matrix of their products under $\Km^{-1}$ is
$\left(\begin{smallmatrix}0&b\\b&0\end{smallmatrix}\right)$,
whose determinant is $-b^2\ne0$. The block Gram--Schmidt step for
every other vector is therefore
\begin{align}
 \vb{u}\longmapsto\vb{u}
 -\frac{\vb{u}^{\transp}\Km^{-1}\vb{w}}{b}\vb{v}
 -\frac{\vb{u}^{\transp}\Km^{-1}\vb{v}}{b}\vb{w}.
 \label{appeq:rational-hyperbolic-elimination}
\end{align}
The updated vector has the same parity and satisfies both
$\vb{u}^{\transp}\Km^{-1}\vb{v}=0$ and
$\vb{u}^{\transp}\Km^{-1}\vb{w}=0$. Repeat on the subspace
defined by these two conditions. We omit normalization, so every
operation remains rational. The procedure produces independent
two-dimensional planes. Vectors in different
planes have zero product under $\Km^{-1}$.

Every non-zero subtraction in
\cref{appeq:rational-hyperbolic-elimination} uses a vector from a
family with a larger coefficient in the strict hierarchy. Indeed,
if $\vb{u}$ precedes $\vb{w}$, then
$\vb{u}^{\transp}\Km^{-1}\vb{v}=0$ by the choice of $\vb{w}$.
The only other vector from the same family as $\vb{v}$ has opposite
parity: it is either chosen as $\vb{w}$ immediately, or its product
with $\vb{v}$ under $\Km^{-1}$ is zero and it is unchanged.
The unremoved vector from the same family as $\vb{w}$ has the
parity of $\vb{v}$, so it can receive only a subtraction proportional
to the earlier vector $\vb{v}$. Thus vectors from the same family
are never mixed in this elimination.

Let $V$ contain the original parity vectors as columns in the order
just specified. Define a diagonal matrix $D$ whose entry for each
column is its coefficient in $C(T)$: $\gamma_f/2$ for either of the
two vectors from family $f$, or $\gamma_f$ if that family contributes
only one vector. Then $C=VDV^{\transp}$.
Let $\widehat{V}$ contain the resulting columns after elimination,
retaining zero columns for linear dependencies. The preceding property gives
$V=\widehat{V}U$ with a rational triangular matrix $U$ whose diagonal
is one. Every non-zero off-diagonal entry $U_{ij}$ connects column
$i$ to column $j$ from a later family with an asymptotically smaller
coefficient. Define
\begin{align}
 E&=D^{-1/2}UD^{1/2}-\mathds{1},\\
 C_0&=\widehat{V}D\widehat{V}^{\transp}.
\end{align}
Every non-zero off-diagonal entry $E_{ij}$ contains
$\sqrt{D_{jj}/D_{ii}}$, a constant multiple of
$\sqrt{\gamma_g/\gamma_f}$ with $g>f$. This tends to zero by
\cref{appeq:strict-temperature-hierarchy}, so the operator norm
$\delta=\norm{E}$ tends to zero. Since
$C=\widehat{V}D^{1/2}(\mathds{1}+E)(\mathds{1}+E)^{\transp}
D^{1/2}\widehat{V}^{\transp}$, bounding the singular values of
$\mathds{1}+E$ gives
\begin{align}
 (1-\delta)^2C_0&\preceq C\preceq(1+\delta)^2C_0.
 \label{appeq:relative-hierarchy-comparison}
\end{align}
Here $A\preceq B$ means that $B-A$ is positive semidefinite. The matrix $C_0$ is still mirror
symmetric.

\paragraph{Preservation of the limiting response.}
We must check that the relative error in
\cref{appeq:relative-hierarchy-comparison} does not change the
limiting stable subspace. For a general $\Km_{\rm edge}$, use the rational change of basis
\begin{align}
 \mathsf{P}&=\begin{pmatrix}
 \mathds{1}&\Km_{\rm edge}/2\\
 \mathds{1}&-\Km_{\rm edge}/2
 \end{pmatrix},\\
 \mathsf{P}^{\transp}\Km^{-1}\mathsf{P}
 &=\begin{pmatrix}0&\mathds{1}\\\mathds{1}&0\end{pmatrix}.
 \label{appeq:rational-parity-basis}
\end{align}
The first and second sets of columns have even and odd mirror parity.
Consequently $\mathsf{P}^{-1}C\mathsf{P}^{-\transp}$ is block
diagonal. Denote its two positive blocks by $F,H$. These play the
same algebraic role as in \cref{appeq:jain-rate-pair}, but are now
expressed in the rational basis \cref{appeq:rational-parity-basis}.
Writing $\vb*{\mu}=\mathsf{P}(\vb{x},\vb{y})^{\transp}$, the
stable subspace again has $\vb{y}=\mathsf{S}\vb{x}$, with
$\mathsf{S}$ given by \cref{appeq:jain-decaying-graph}.
For this comparison only, define
\begin{align}
 \mathsf{R}_{\rm can}
 &=(\mathds{1}-\mathsf{S})(\mathds{1}+\mathsf{S})^{-1}.
 \label{appeq:canonical-hierarchy-response}
\end{align}
It maps $\vb{x}+\vb{y}$ to $\vb{x}-\vb{y}$ and is a bounded
representation of the stable subspace. These coordinates are generally
different from the physical top and bottom potentials, so
$\mathsf{R}_{\rm can}$ is distinct from the Jain response
$\mathsf{R}$ in \cref{appeq:normalized-boundary-response}.

The positive matrix square root preserves positive-semidefinite
order. The expression in \cref{appeq:jain-decaying-graph} therefore
increases with $H$. It decreases with $F$, as follows by applying
the same expression to $\mathsf{S}^{-1}$ with $F,H$ interchanged.
Let $\mathsf{S}_0$ and $\mathsf{R}_{{\rm can},0}$ be the corresponding
matrices for $C_0$. The bounds in
\cref{appeq:relative-hierarchy-comparison} imply
\begin{align}
 k&=\frac{1+\delta}{1-\delta}\longrightarrow1,\\
 k^{-1}\mathsf{S}_0&\preceq\mathsf{S}\preceq k\mathsf{S}_0.
\end{align}
The map $\mathsf{S}\mapsto2(\mathds{1}+\mathsf{S})^{-1}-\mathds{1}$
reverses this order. Its scalar version $r(s)=(1-s)/(1+s)$ obeys
\begin{align}
 \sup_{s\geq0}\abs{r(ks)-r(s)}
 &=\frac{2(\sqrt{k}-1)}{\sqrt{k}+1}.
\end{align}
Applying this bound to the eigenvalues of $\mathsf{S}_0$ gives
\begin{align}
 \norm{\mathsf{R}_{\rm can}-\mathsf{R}_{{\rm can},0}}
 &\leq\frac{2(\sqrt{k}-1)}{\sqrt{k}+1}\longrightarrow0.
 \label{appeq:uniform-cayley-comparison}
\end{align}
The two problems therefore have the same limiting
stable subspace, even when individual relaxation eigenvalues vanish
at different powers of temperature.

\paragraph{Rational limiting subspace and contact matching.}
On one of the independent planes of $C_0$, let
$\vb{v},\vb{w}$ be columns $i,j$ of $\widehat{V}$. With
$b=\vb{v}^{\transp}\Km^{-1}\vb{w}$, the transport operator
$C_0\Km^{-1}$ acts as
\begin{align}
 C_0\Km^{-1}\vb{v}&=bD_{jj}\vb{w},\\
 C_0\Km^{-1}\vb{w}&=bD_{ii}\vb{v}.
\end{align}
Its positive eigenvalue is $\abs{b}\sqrt{D_{ii}D_{jj}}$, so the
corresponding decaying solution of \cref{appeq:hierarchy-transport}
has direction
\begin{align}
 \sqrt{D_{ii}}\vb{v}
 +\operatorname{sgn}(b)\sqrt{D_{jj}}\vb{w}.
 \label{appeq:hyperbolic-stable-vector}
\end{align}
If $\vb{v}$ comes from family $f$ and $\vb{w}$ from a later
family $g>f$, divide by $\sqrt{D_{ii}}$. The coefficient of
$\vb{w}$ is then a constant multiple of
$\sqrt{\gamma_g/\gamma_f}\to0$, leaving the rational direction
$\vb{v}$. If the two vectors belong to the same family,
$D_{ii}=D_{jj}=\gamma_f/2$, and the direction
is the rational vector $\vb{v}+\operatorname{sgn}(b)\vb{w}$.
In either case the limiting direction is independent of finite
positive prefactors. Taking one such direction from each independent
plane gives a rational basis $Z$ of the limiting stable subspace.

This subspace is nonnegative under $\Km^{-1}$, as a limit of the
positive stable subspaces at $T>0$. Its projection onto the canonical
$\vb{x}+\vb{y}$ coordinates is invertible: a non-zero vector with
$\vb{y}=-\vb{x}$ would satisfy
$\vb*{\mu}^{\transp}\Km^{-1}\vb*{\mu}
=-2\vb{x}^{\transp}\vb{x}<0$. Thus the comparison of the bounded
matrices in \cref{appeq:uniform-cayley-comparison} indeed identifies
the same limiting subspace in the original problem.

Write the physical contact equations as
$C_{\rm in}\vb*{\mu}(0)=eV_{\rm in}\vb{b}_{\rm in}$, where
$C_{\rm in}$ and $\vb{b}_{\rm in}$ are rational. Under the assumed nonsingularity of $C_{\rm in}Z$,
\begin{align}
 \frac{\vb*{\mu}(0)}{eV_{\rm in}}
 &=Z(C_{\rm in}Z)^{-1}\vb{b}_{\rm in}.
 \label{appeq:rational-contact-matching}
\end{align}
All entries on the right are rational and depend only on the ordered
families and contacts. The charge vector and
$\Km_{\rm edge}^{-1}$ in \cref{appeq:outgoing-current} are also
rational, proving the claim for $G/G_{\rm Q}$. A finite family set
has only finitely many strict orders and therefore gives only
finitely many such conductances. The KFP contact conditions used
here are rational in the microscopic potentials:
\begin{align}
 3\mu_{\rm t,I}-\mu_{\rm t,f}&=2eV_{\rm in},\\
 \mu_{\rm b,f}-\mu_{\rm b,I}&=0.
\end{align}
The square roots introduced by normalized charge--neutral coordinates
do not change this property.

\subsection{Explicit strict-hierarchy lists for \texorpdfstring{$p=1$}{p=1}}
\label{app:jain-hierarchy-lists}
Let $\mathcal{C}_N$ denote the possible values of $G/G_{\rm Q}$ obtained
by the finite kinetic reduction above, for $\Km_{\rm J}=\mathds{1}_N+2\vb{q}_{\rm f}\vb{q}_{\rm f}^{\transp}$.
Here $M=P=0$: no charged or neutral cosine has been pinned.
Every retained coefficient is non-zero, all active transport lengths are
saturated, and distinct process families have strictly ordered coefficients; the two
partners within a mirror family retain the same coefficient.
The retained operators include CAR, tunneling, single-electron dissipation, and the neutral vertices
$(\vb{r}_{ij},\pm\vb{r}_{kl})$ and $(\vb{r}_{ij},0)$ with their
mirror partners, where $\vb{r}_{ij}=\vb{e}_i-\vb{e}_j$.

We add the operators that increase the rank of
$F$ or $H$ in \cref{appeq:jain-rate-pair} until both ranks
reach $N$. Thus at most $2N$ family additions are needed; later
rank-dependent contributions do not change the strict limit.
Exact rational overlap elimination
and source matching give the counts in
\cref{tab:jain-hierarchy-counts}, with one producing hierarchy stored
for every value. 

\begin{table}[tb]
 \centering
 \begin{tabular}{cccc}
  \toprule
  $N$ & $\nu$ & Process families & $\abs{\mathcal{C}_N}$ \\
  \midrule
  1 & $1/3$ & 3 & 3 \\
  2 & $2/5$ & 11 & 17 \\
  3 & $3/7$ & 30 & 323 \\
  4 & $4/9$ & 72 & 17\,021 \\
  \bottomrule
 \end{tabular}
 \caption{Strict-hierarchy kinetic candidates at $p=1$ for the
 specified unpinned, mirror-symmetric operator set. These counts do
 not impose realizability of every hierarchy by one interaction matrix.}
 \label{tab:jain-hierarchy-counts}
\end{table}

Writing $\mathcal{C}_N=\{0\}\cup\mathcal{C}_N^+\cup(-\mathcal{C}_N^+)$,
the first two positive sets are
\begin{align}
 \mathcal{C}_1^+&=\left\{\frac{1}{3}\right\},\nonumber\\
 \mathcal{C}_2^+&=\qty{
 \frac{1}{15},\frac{2}{15},\frac{4}{15},\frac{2}{7},
 \frac{3}{10},\frac{8}{25},\frac{1}{3},\frac{2}{5} }.
 \label{appeq:jain-p1-n1-n2-list}
\end{align}

\section{RG flow for Jain states}

\subsection{Basis Transformation and Coupling Matrix}

We use the matrices and fields of \cref{seq:cross-scattering}:
$\Km_{\rm edge}$ describes one edge, $\Km_{\rm tot}$ describes both
sides, a hat denotes the charge--neutral basis, and a tilde denotes
the simultaneous diagonal basis obtained with $W$.
The common reference velocity is $u$, the mode velocities are $u_i$,
and $\ell$ is RG time. 

Consider the positive Jain branch
$\nu=N/(2pN+1)$ in the notation of \cref{eq:K-block-decomposition}, with
\begin{align}
    \Km_{\rm J}&=\mathds{1}_N+2p\mathsf{J}_N.
\end{align}
The orthogonal charge--neutral rotation $O$ \cite{kane1995impurity} in
\cref{eq:diagonalizing_o} gives
\begin{align}
    O^\transp\Km_{\rm J}O
    &=\mathrm{diag}\qty(\kappa_1,\ldots,\kappa_{N-1},\kappa_{\rm C}),
    \\
    \kappa_i&=
    \begin{cases}
        1,&i=1,\ldots,N-1,\\
        \kappa_{\rm C}=1+2pN,&i={\rm C}.
    \end{cases}
\end{align}

For the two sides of the finger,
$\Km_{\rm tot}=\Km_{\rm J}\oplus(-\Km_{\rm J})$. After applying
$\mathcal{R}=R\oplus R$, with $R=\mathds{1}_{N_{\rm I}}\oplus O$
as in the main text, and grouping the top and bottom component of each
sector, the purely fractional case $N_{\rm I}=0$ has $R=O$ and
its $K$-matrix consists of $N$ blocks
$\widehat{\Km}_i=\kappa_i\sigma_z$.

Let $\Pi$ be the permutation for which $\Pi^\transp$ sends the rotated order
$({\rm t},1),\ldots,({\rm t},{\rm C}),({\rm b},1),\ldots,({\rm b},{\rm C})$
to $({\rm t},1),({\rm b},1),\ldots,({\rm t},{\rm C}),({\rm b},{\rm C})$.
When integer modes are included, the same permutation groups their top--bottom pairs first.
Take $\widehat{\Um}_{\rm pair}=\Pi^\transp\mathcal{R}^\transp\Um_{\rm tot}\mathcal{R}\Pi$ to be block diagonal. Its $i$-th block is parameterized by $u_i$ and $g_i$ as
\begin{align}
    \widehat{\Um}_i
    &=\frac{\kappa_i u_i}{2}\qty[
        \qty(g_i+g_i^{-1})\mathds{1}_2
        -\qty(g_i-g_i^{-1})\sigma_x
    ].
    \label{appeq:param}
\end{align}

The simultaneous diagonalizer in this paired basis is $W_{\rm pair}=\bigoplus_iW_i$, with
\begin{align}
    W_i
    &=\frac{1}{\sqrt{\kappa_i}}\exp(-\theta_i\sigma_x),
    \\
    g_i&=\exp(-2\theta_i).
\end{align}
It obeys
\begin{align}
    W_i^\transp\widehat{\Km}_iW_i&=\sigma_z,
    \\
    W_i^\transp\widehat{\Um}_iW_i&=u_i\mathds{1}_2,
\end{align}
in agreement with \cref{eq:Wi-boost}. The full transform is
$W=\mathcal{R}\Pi W_{\rm pair}$, so
\begin{align}
    \tilde{\Km}&=W^\transp\Km_{\rm tot}W,\\
    \tilde{\Um}&=W^\transp\Um_{\rm tot}W
    =W_{\rm pair}^\transp\widehat{\Um}_{\rm pair}W_{\rm pair}.
\end{align}
For an original-basis vertex $\vb{m}$, its paired rotated vector is
$\widehat{\vb{m}}_{\rm pair}=\Pi^\transp\mathcal{R}^\transp\vb{m}$,
and its normalized vector is
$\vb*{\lambda}=W^\transp\vb{m}=W_{\rm pair}^\transp\widehat{\vb{m}}_{\rm pair}$.

\subsection{Perturbation Vectors and Matrix RG Flow}

Let $j=1,\ldots,N$ label a physical electron within the fractional block; its full-edge index is $N_{\rm I}+j$. In the purely fractional theory, its electron vector is
$\vb{m}_j=\Km_{\rm J}\vb{e}_j$. The two inter-side vectors are
\begin{align}
    \vb{m}_{A,j}
    &=\begin{pmatrix}\vb{m}_j\\-\vb{m}_j\end{pmatrix},
    \\
    \vb{m}_{T,j}
    &=\begin{pmatrix}\vb{m}_j\\\vb{m}_j\end{pmatrix}.
\end{align}
The embedded vectors for electron dissipation from a single side are
$\vb{m}^{\rm tot}_{{\rm t},j}=(\vb{m}_j,0)^\transp$ and
$\vb{m}^{\rm tot}_{{\rm b},j}=(0,-\vb{m}_j)^\transp$.

In the charge--neutral basis the
$i$-th blocks are
$\widehat{\vb{m}}_{A,j,i}=\kappa_iO_{ji}(1,-1)^\transp$ and
$\widehat{\vb{m}}_{T,j,i}=\kappa_iO_{ji}(1,1)^\transp$. Their components
in the diagonal basis are
\begin{align}
    \vb*{\lambda}_{A,j,i}
    &=\sqrt{\kappa_i}O_{ji}\exp(\theta_i)
      \begin{pmatrix}1\\-1\end{pmatrix},
    \\
    \vb*{\lambda}_{T,j,i}
    &=\sqrt{\kappa_i}O_{ji}\exp(-\theta_i)
      \begin{pmatrix}1\\1\end{pmatrix}.
\end{align}
For the neutral vertices with $(l,m)=(j,k)$ in
\cref{eq:neutral-inter-side-operators}, i.e., acting on the same pair of electrons on both sides, define
$b_{jk,i}=O_{ji}-O_{ki}$. Since $b_{jk,\mathrm C}=0$ and
$\kappa_i=1$ in the neutral sector, their normalized vectors are
\begin{align}
    \vb*{\lambda}_{\mathrm N,A;jk,i}
    &=b_{jk,i}g_i^{-1/2}\begin{pmatrix}1\\-1\end{pmatrix},\\
    \vb*{\lambda}_{\mathrm N,T;jk,i}
    &=b_{jk,i}g_i^{1/2}\begin{pmatrix}1\\1\end{pmatrix}.
    \label{appeq:neutral-normalized-vectors}
\end{align}
For the dissipative terms, the top-side and bottom-side vectors become
\begin{align}
    \vb*{\lambda}_{{\rm t},j,i}
    &=\sqrt{\kappa_i}O_{ji}
      \begin{pmatrix}\cosh(\theta_i)\\-\sinh(\theta_i)\end{pmatrix},
    \\
    \vb*{\lambda}_{{\rm b},j,i}
    &=-\sqrt{\kappa_i}O_{ji}
      \begin{pmatrix}-\sinh(\theta_i)\\\cosh(\theta_i)\end{pmatrix}.
\end{align}

\subsubsection{Quadratic action, cosine perturbations, and the RG step}
We first work at a common normal-mode velocity $u$, with
$\tilde{\Km}=\mathrm{diag}(\sigma_{\beta})$ and
$\tilde{\Um}=u\mathds{1}$. Here $\beta,\gamma$ label normal modes. The matrix
$W=\mathcal{R}\Pi W_{\rm pair}$ acts on the original fields. With
$\vb*{\lambda}_r=W^\transp\vb{m}_r$, we define
$\Theta_r=\vb*{\lambda}_r^\transp\tilde{\vb*{\phi}}$, and its
scaling dimension and conformal spin are
\begin{align}
    d_r&=\frac{1}{2}\sum_{\beta}\lambda_{r\beta}^2,\\
    s_r&=\frac{1}{2}\sum_{\beta}\sigma_{\beta}\lambda_{r\beta}^2.
\end{align}
Using Euclidean time $\tau$ and the area measure
$\dd^2\mathsf X=u\dd{\tau}\dd{x}$, separate the action into
its Gaussian and coherent perturbing parts:
\begin{align}
 S_0&=\frac{1}{4\pi}\int\dd{\tau}\dd{x}
 \sum_\beta\qty[i\sigma_\beta
 \partial_x\tilde\phi_\beta\partial_\tau\tilde\phi_\beta
 +u(\partial_x\tilde\phi_\beta)^2],
 \label{appeq:rg-gaussian-action}\\
 S_{\rm coh}&=\sum_r\frac{1}{2\pi a^2}\int\dd^2\mathsf X\,
 \qty[y_r V_{\vb m_r}+y_r^*V_{-\vb m_r}]
 \nonumber\\
 &=\sum_r\frac{\abs{y_r}}{\pi a^2}\int\dd^2\mathsf X\,
 \cos(\Theta_r+\arg y_r).
 \label{appeq:rg-cosine-action}
\end{align}
The Euclidean sign convention in \cref{appeq:rg-gaussian-action}
corresponds to chiral coordinates $u\tau-i\sigma_\beta x$. 
The vertices are normal ordered at cutoff $a$, with the normalization
of \cref{appeq:generic-perturbation}. 
To obtain the RG correction, split the fields into slow and fast
parts, $\tilde{\vb*{\phi}}=\tilde{\vb*{\phi}}_<+
\tilde{\vb*{\phi}}_>$, and integrate out the fast Gaussian fluctuations
in an infinitesimal cutoff shell. The effective action for the slow
fields follows from the connected cumulant expansion
\cite{cardy1996scaling}:
\begin{align}
 S_{\rm eff}&=S_0[\tilde{\vb*{\phi}}_<]
 -\ln\expval{e^{-S_{\rm coh}}}_>
 \nonumber\\
 &=S_0[\tilde{\vb*{\phi}}_<]+\expval{S_{\rm coh}}_>
 -\frac{1}{2}\qty[\expval{S_{\rm coh}^2}_>
 -\expval{S_{\rm coh}}_>^2]+\cdots .
 \label{appeq:rg-cumulant}
\end{align}
The average is over fast fields only; the remaining operators are
functions of the slow fields. At first order, the vertex acquires
$e^{-d_r\dd\ell}$, while rescaling the area contributes
$e^{2\dd\ell}$. Consequently,
\begin{align}
    \dot{y}_r&=(2-d_r)y_r,
    \label{appeq:rg-tree-general-spin}
\end{align}
where a dot denotes $\mathrm{d}/\mathrm{d}\ell$.

\subsubsection{Cosine OPE and the quadratic correction}
For two Euclidean points, define their center
$\mathsf X=(\mathsf X_1+\mathsf X_2)/2$ and separation
$\delta\mathsf X=\mathsf X_1-\mathsf X_2=(\tau,x)$, and write
$\mathsf X_\pm=\mathsf X\pm\delta\mathsf X/2$.
The complex separation is $z=u\tau-ix=\varrho e^{i\vartheta}$,
with $z_+=z$ and $z_-=\bar z$. The opposite-vector product has
zero total vertex vector and the OPE
\cite{cardy1996scaling,katzir2020superconducting}
\begin{align}
 V_{\vb m_r}(\mathsf X_+)V_{-\vb m_r}(\mathsf X_-)
 &=\qty(\frac{a}{\varrho})^{2d_r}e^{-2is_r\vartheta}
 \qty[1+i B_r-\frac{1}{2}B_r^2+\cdots],
 \label{appeq:rg-vertex-ope}\\
 B_r&=\sum_\beta\lambda_{r\beta}z_{\sigma_\beta}
 \partial_{z_{\sigma_\beta}}\tilde\phi_\beta(\mathsf X).
\end{align}
Here $B_r$ is the leading short-distance phase difference
$\Theta_r(\mathsf X_+)-\Theta_r(\mathsf X_-)$; omitted terms
contain higher derivatives. Reversing both vertex vectors changes
$B_r$ to $-B_r$.

The full cosine product also contains the two same-sign vertex
products. For compactness, let
$C_r(\mathsf X)=\cos[\Theta_r(\mathsf X)+\arg y_r]$. Combining
all four products gives
\begin{align}
 C_r(\mathsf X_+)C_r(\mathsf X_-)
 &=\frac{1}{2}\qty(\frac{a}{\varrho})^{2d_r}
 e^{-2is_r\vartheta}\qty[1-\frac{1}{2}B_r^2+\cdots]
 \nonumber\\
 &\quad+\frac{1}{2}\qty(\frac{\varrho}{a})^{2d_r}
 e^{2is_r\vartheta}
 \cos[2\Theta_r(\mathsf X)+2\arg y_r]+\cdots .
 \label{appeq:rg-cosine-ope}
\end{align}
The first line is the neutral contribution; its
$B_r^2$ term renormalizes the quadratic action. The second line is a
new cosine with vector $2\vb m_r$ and dimension $4d_r$, accompanied
by derivative terms not displayed. Products of distinct cosines
similarly contain vectors $\vb m_r\pm\vb m_{r'}$. Their shell
coefficients must also obey angular selection rules
\cite{katzir2020superconducting}.

We now retain the quadratic term in the first line of
\cref{appeq:rg-cosine-ope}, which gives
\begin{align}
 \delta S_0&=\sum_r\frac{\abs{y_r}^2}{8\pi^2a^4}
 \int\dd^2\mathsf X\int_{a<\varrho<ae^{\dd\ell}}
 \dd^2\delta\mathsf X\,
 \qty(\frac{a}{\varrho})^{2d_r}
 \Re\qty[e^{-2is_r\vartheta} B_r^2].
 \label{appeq:rg-action-shell}
\end{align}
Varying \cref{appeq:rg-gaussian-action} gives
$\partial_x(\partial_\tau-i\sigma_\beta u\partial_x) \tilde\phi_\beta=0$. After fixing the spatially uniform zero mode,
this is the chiral equation
$(\partial_\tau-i\sigma_\beta u\partial_x)\tilde\phi_\beta=0$.
Thus $\tilde\phi_\beta$ depends only on
$z_{\sigma_\beta}=u\tau-i\sigma_\beta x$, and not on
$z_{-\sigma_\beta}$ \cite{ginsparg1988applied}.
The chain rule therefore gives
\begin{align}
 \partial_x\tilde\phi_\beta &=-i\sigma_\beta\partial_{z_{\sigma_\beta}}\tilde\phi_\beta\\
\partial_{z_{\sigma_\beta}}\tilde\phi_\beta &=i\sigma_\beta\partial_x\tilde\phi_\beta.
 \label{appeq:rg-chiral-gradient}
\end{align}
The sign follows from our choice $z_{\sigma_\beta}=u\tau-i\sigma_\beta x$.
We use this free-field identity in the Gaussian OPE at separated
points. Substitution gives
\begin{align}
 B_r^2&=-\varrho^2\sum_{\beta\gamma}
 \sigma_\beta\sigma_\gamma\lambda_{r\beta}\lambda_{r\gamma}
 e^{i(\sigma_\beta+\sigma_\gamma)\vartheta}
 \partial_x\tilde\phi_\beta\partial_x\tilde\phi_\gamma.
 \label{appeq:rg-gradient-descendant}
\end{align}
The radial integral, including these two powers of $\varrho$, is
\begin{align}
 \int_a^{ae^{\dd\ell}}\dd{\varrho}\,
 \varrho^3\qty(\frac{a}{\varrho})^{2d_r}
 &=a^4\dd\ell+\mathcal O((\dd\ell)^2).
\end{align}
The angular integral projects onto the spin of the two-gradient
operator:
\begin{align}
 \int_0^{2\pi}\frac{\dd{\vartheta}}{2\pi}
 e^{i(\sigma_\beta+\sigma_\gamma-2s_r)\vartheta}
 &=\delta_{2s_r,\sigma_\beta+\sigma_\gamma}.
 \label{appeq:rg-spin-selection}
\end{align}
Finally, match the coefficient of each gradient product to
\begin{align}
 \delta S_0&=\frac{\dd\ell}{4\pi}\int\dd{\tau}\dd{x}
 \sum_{\beta\gamma}
 (W^\transp\dot{\Um}W)^{\rm coh}_{\beta\gamma}
 \partial_x\tilde\phi_\beta\partial_x\tilde\phi_\gamma.
 \label{appeq:rg-action-correction}
\end{align}
This yields
\begin{align}
 \qty(W^\transp\dot{\Um}W)^{\rm coh}_{\beta\gamma}
 &=-u\sum_r\abs{y_r}^2\sigma_\beta\sigma_\gamma
 \lambda_{r\beta}\lambda_{r\gamma}
 \delta_{2s_r,\sigma_\beta+\sigma_\gamma}.
 \label{appeq:rg-matrix-general-spin}
\end{align}
The factor $u$ comes from $\dd^2\mathsf X=u\dd\tau\dd x$.
The instantaneous $W$ is
also held fixed while projecting this infinitesimal correction;
$W^\transp\dot{\Um}W$ is not
$\mathrm{d}(W^\transp\Um W)/\mathrm{d}\ell$.

For $s_r=0$, only opposite-chirality quadratic entries survive, recovering the matrix equation used by \citet{katzir2020superconducting}. For $s_r=\pm1$, only entries coupling modes of chirality $\pm1$ survive. For $\abs{s_r}>1$, the
neutral self-contraction generates no quadratic, two-derivative term in an isotropic shell at equal velocities.

For unequal normal-mode velocities $u_{\beta}>0$, choose a reference
velocity $u$ to define the circular shell
$x=-\varrho\sin\vartheta$ and $u\tau=\varrho\cos\vartheta$.
The chiral separation of mode $\beta$ is then
\begin{align}
    z_{\beta}&=u_{\beta}\tau-i\sigma_{\beta} x
    =\varrho\qty(\frac{u_{\beta}}{u}\cos\vartheta
    +i\sigma_{\beta}\sin\vartheta).
\end{align}
Keeping these separations in both the correlator and the quadratic
descendant gives the explicit unequal-velocity version of
\cref{appeq:rg-matrix-general-spin}:
\begin{align}
    \qty(W^\transp\dot{\Um} W)_{\beta\gamma}^{\rm coh}
    &=-u\sum_r\abs{y_r}^2\sigma_{\beta}\sigma_{\gamma}\lambda_{r\beta}\lambda_{r\gamma}
    \Re\int_0^{2\pi}\frac{\dd{\vartheta}}{2\pi}
    \frac{
        \qty(\frac{u_{\beta}}{u}\cos\vartheta+i\sigma_{\beta}\sin\vartheta)
        \qty(\frac{u_{\gamma}}{u}\cos\vartheta+i\sigma_{\gamma}\sin\vartheta)
    }{
        \displaystyle\prod_{j}
        \qty(\frac{u_{j}}{u}\cos\vartheta+i\sigma_{j}\sin\vartheta)^{\lambda_{rj}^2}
    }.
    \label{appeq:rg-angular-kernel}
\end{align}
There is no sum over $\beta,\gamma$. The complex powers are continued
continuously from $\vartheta=0$ with the Euclidean time-ordering
prescription. 
At $u_{\beta}=u$ for every mode, the angular integral is exactly
$\delta_{2s_r,\sigma_{\beta}+\sigma_{\gamma}}$.

Unequal velocities break the rotational symmetry responsible for
that selection rule. Consequently, a non-zero quadratic
correction for $\abs{s_r}>1$ is possible already at
$\mathcal{O}(\abs{y_r}^2)$. 

\subsubsection{Evaluating the unequal-velocity integral}
The angular integral in \cref{appeq:rg-angular-kernel} can be
evaluated as an absolutely convergent series for arbitrary positive
velocities. Denote that integral by $F_{r,\beta\gamma}$, and, for fixed
$r,\beta,\gamma$, introduce
\begin{align*}
    \xi_{j}&=\frac{u-u_{j}}{u+u_{j}},\\
    b_{j}&=\frac{u+u_{j}}{2u},\\
    P_{j}&=\lambda_{rj}^2-\delta_{j\beta}-\delta_{j\gamma},\\
    k&=\frac{\sigma_{\beta}+\sigma_{\gamma}}{2}-s_r.
\end{align*}
The integer $k$ is the angular harmonic required for this matrix
entry. Factoring each velocity-dependent separation gives
\begin{align*}
    f_{j}(\vartheta)
    &=b_{j} e^{i\sigma_{j}\vartheta}
      (1-\xi_{j} e^{-2i\sigma_{j}\vartheta}).
\end{align*}
The integrand is $\prod_{j} f_{j}^{-P_{j}}$. Expand each factor using
$(1-z)^{-P}=\sum_{n\geq0}(P)_n z^n/n!$, where
$(P)_n=P(P+1)\cdots(P+n-1)$ and $(P)_0=1$. The angular
integration then selects a single Fourier harmonic:
\begin{align}
    F_{r,\beta\gamma}
    &=\qty(\prod_{j} b_{j}^{-P_{j}})
      \sum_{\substack{n_{j}\geq0\\ \sum_{j}\sigma_{j} n_{j}=k}}
      \prod_{j}\frac{(P_{j})_{n_{j}}}{n_{j}!}\xi_{j}^{n_{j}}.
    \label{appeq:rg-angular-series}
\end{align}
Since $\abs{\xi_{j}}<1$ for every finite positive velocity, the
binomial series converge uniformly and absolutely on the
integration circle. Thus, \cref{appeq:rg-angular-series} is an exact
evaluation, including cases with noninteger $P_{j}$. It can be
computed by convolving the positive- and negative-chirality
coefficient sequences and retaining the required harmonic.

For small velocity differences, set
$\xi_{\max}=\max_{j}\abs{\xi_{j}}$ and keep terms through second
order, leaving the known prefactor unexpanded:
\begin{align}
    F_{r,\beta\gamma}
    &=\qty(\prod_{j} b_{j}^{-P_{j}})\Biggl[
        \delta_{k,0}
        +\sum_{j} P_{j}\xi_{j}\delta_{k,\sigma_{j}}
        +\frac{1}{2}\sum_{j,j'}P_{j}(P_{j'}+\delta_{jj'})
          \xi_{j}\xi_{j'}\delta_{k,\sigma_{j}+\sigma_{j'}}
        +\mathcal{O}(\xi_{\max}^3)
      \Biggr].
    \label{appeq:rg-angular-expansion}
\end{align}
This expansion is at fixed normalized vertex components. The
constraint in \cref{appeq:rg-angular-series} requires at least
$\abs{k}$ powers of the velocity differences. In particular, an
opposite-chirality quadratic entry from a spin-$s_r$ vertex
starts no earlier than order $\xi_{\max}^{\abs{s_r}}$.

In a useful special case, all modes of a given chirality share
a velocity. Let $P_\pm=\sum_{\sigma_{j}=\pm1}P_{j}$, and use
$b_\pm,\xi_\pm$ for those two velocities. For $k\geq0$, the
sum reduces to the Gauss hypergeometric function:
\begin{align}
    F_{r,\beta\gamma}
    &=b_+^{-P_+}b_-^{-P_-}
      \frac{(P_+)_k}{k!}\xi_+^k\,
      {}_2F_1(P_++k,P_-;k+1;\xi_+\xi_-).
    \label{appeq:rg-angular-hypergeometric}
\end{align}
For $k<0$, exchange $+$ and $-$ and replace $k$ by $-k$.
If the velocity contrast makes its convergence
slow, periodic angular quadrature of \cref{appeq:rg-angular-kernel},
with the continuous logarithms specified there, provides an
independent numerical evaluation.

\subsubsection{The scalar inter-side terms and the local bath}
The same-mode CAR and tunneling vectors above satisfy
$s_{A,j}=s_{T,j}=0$, because their top and bottom contributions cancel.
In contrast, an edge electron entering the bath has
$s_{e,s,j}=\epsilon_s(2p+1)/2$. The bath calculation does not use
\cref{appeq:rg-spin-selection}: its two vertices have the same
position and different times. The measure
$\dd{x}\dd{\tau}_1\dd{\tau}_2$, the kernel
$1/(\tau_1-\tau_2)$, and the two edge vertices give
\begin{align}
    \dot{\alpha}_{s,j}&=(2-2d_{e,s,j})\alpha_{s,j}.
    \label{appeq:rg-bath-tree}
\end{align}
Here the scaling exponent is $3-1-2d_{e,s,j}$: three integration
coordinates, one inverse-time bath kernel, and two edge vertices.
For the quadratic correction, the neutral vertex product in
\cref{eq:diss_action} contains the phase difference
\begin{align}
 \Theta_{s,j}(x,\tau_1)-\Theta_{s,j}(x,\tau_2)
 &=(\tau_1-\tau_2)\vb*{\lambda}_{s,j}^{\transp}
 \partial_\tau\tilde{\vb*{\phi}}(x,\tau)+\cdots,
\end{align}
where $\tau=(\tau_1+\tau_2)/2$ and
$\Theta_{s,j}=\vb*{\lambda}_{s,j}^{\transp}\tilde{\vb*{\phi}}$.
The second-order expansion of its exponential gives
$-(\tau_1-\tau_2)^2
(\vb*{\lambda}_{s,j}^{\transp}\partial_\tau
\tilde{\vb*{\phi}})^2/2$. This is again a quadratic derivative
operator, now obtained at first order in the already bath-averaged
action, hence at order $\alpha_{s,j}\propto\abs{w_{s,j}}^2$.
Multiplying by the short-time edge correlator and bath kernel and integrating
$a<u\abs{\tau_1-\tau_2}<ae^{\mathrm{d}\ell}$, gives
\begin{align}
    \qty(W^\transp\dot{\Um} W)^{\rm bath}
    &=-2u\sum_{s,j}\alpha_{s,j}
    \tilde{\Km}\vb*{\lambda}_{s,j}\vb*{\lambda}_{s,j}^\transp\tilde{\Km},
    \label{appeq:rg-bath-matrix}
\end{align}
Here $\tilde{\Km}=\mathrm{diag}(\sigma_{\beta})$ is the normalized
$K$-matrix already defined above. We use the dimensionless bath
strength in the RG equations, absorbing the reference velocity into
$\alpha$. The two factors of $\tilde{\Km}$ follow from
$\partial_\tau\tilde{\vb*{\phi}}=iu\tilde{\Km}\partial_x\tilde{\vb*{\phi}}$.
This temporal shell has no angular selection rule, so it remains
non-zero for a spinful edge electron.

For the symmetric amplitudes $y_A,y_T,\alpha_{\rm f}$,
orthogonality of $O$ gives
\begin{align}
    \sum_jO_{ji}O_{jk}&=\delta_{ik}.
    \label{appeq:rg-orthogonality}
\end{align}
For the neutral couplings we likewise take
$y_{\mathrm N,X;jk}=y_{\mathrm N,X}$ for every $j<k$ and
$X\in\{A,T\}$. Equal magnitudes suffice for their quadratic
self-contractions. The identity
\begin{align}
    \sum_{j<k}b_{jk,i}b_{jk,l}
    &=N\qty(\delta_{il}-\delta_{i\mathrm C}\delta_{l\mathrm C})
    \label{appeq:neutral-root-projection}
\end{align}
follows from
$\sum_{j<k}(\vb e_j-\vb e_k)(\vb e_j-\vb e_k)^\transp
=N\mathds{1}_N-\mathsf J_N$ and
$O_{j\mathrm C}=1/\sqrt N$.
Consequently, these terms generate no off-block-diagonal correction in this symmetric approximation. 
In block $i$ the
electron products are
\begin{align}
    \sum_j\lambda_{A,j,i,+}\lambda_{A,j,i,-}
    &=-\kappa_i g_i^{-1},\\
    \sum_j\lambda_{T,j,i,+}\lambda_{T,j,i,-}
    &=\kappa_i g_i,\\
    \sum_{s,j}\sigma_z\vb*{\lambda}_{s,j,i}
       \vb*{\lambda}_{s,j,i}^\transp\sigma_z
    &=\frac{\kappa_i}{2}\qty[
        (g_i+g_i^{-1})\mathds{1}_2
        -(g_i-g_i^{-1})\sigma_x].
\end{align}
Substituting these expressions into
\cref{appeq:rg-matrix-general-spin,appeq:rg-bath-matrix} and
transforming back with $W_i^{-\transp}$ and $W_i^{-1}$ yields
\begin{align}
    \dot{\widehat{\Um}}_i
    &=\frac{u_i\kappa_i^2}{2}\qty{
        \begin{aligned}
            &\abs{y_A}^2\qty[
                (1-g_i^{-2})\mathds{1}_2
                -(1+g_i^{-2})\sigma_x]\cr
            &+\abs{y_T}^2\qty[
                (1-g_i^2)\mathds{1}_2
                +(1+g_i^2)\sigma_x]\cr
            &-2\alpha_{\rm f}\qty[
                (g_i^2+g_i^{-2})\mathds{1}_2
                -(g_i^2-g_i^{-2})\sigma_x]
        \end{aligned}}
    +\left.\dot{\widehat{\Um}}_i\right|_{\mathrm N}.
    \label{appeq:matrix_flow}
\end{align}
The neutral contribution is
\begin{align}
    \left.\dot{\widehat{\Um}}_i\right|_{\mathrm N}
    &=\frac{Nu_i}{2}(1-\delta_{i\mathrm C})\qty{
        \begin{aligned}
            &\abs{y_{\mathrm N,A}}^2\qty[
                (1-g_i^{-2})\mathds{1}_2
                -(1+g_i^{-2})\sigma_x]\cr
            &+\abs{y_{\mathrm N,T}}^2\qty[
                (1-g_i^2)\mathds{1}_2
                +(1+g_i^2)\sigma_x]
        \end{aligned}}.
    \label{appeq:neutral-matrix-flow}
\end{align}
The replacement of the common velocity by $u_i$ in each block is
understood within the shell-coefficient approximation just specified.
Same-side disorder is treated separately below; its variance cannot
in general be added to $\alpha_{\rm f}$ with unit coefficient.

\subsection{Extracting the symmetric RG flow}
Differentiating \cref{appeq:param} and projecting with the instantaneous
$W_i$ gives the simple identity
\begin{align}
    W_i^\transp\dot{\widehat{\Um}}_iW_i
    &=\dot{u}_i\mathds{1}_2
      -\frac{u_i\dot{g}_i}{g_i}\sigma_x.
    \label{appeq:rg-projected-derivative}
\end{align}
The trace and off-diagonal coefficient therefore give, respectively,
\begin{align}
    \dot{u}_i
    &=\frac{1}{2}\tr(W_i^\transp\dot{\widehat{\Um}}_iW_i),
    \label{appeq:trace}\\
    \dot{g}_i
    &=-\frac{g_i}{2u_i}
    \tr(\sigma_xW_i^\transp\dot{\widehat{\Um}}_iW_i).
    \label{appeq:traceless}
\end{align}
Insertion of \cref{appeq:matrix_flow} gives
\begin{align}
    \dot{u}_i&=-\kappa_i\alpha_{\rm f}u_i(g_i+g_i^{-1}),
    \label{appeq:flow_u}\\
    \dot{g}_i&=\kappa_i\qty[
    \abs{y_A}^2-\abs{y_T}^2g_i^2+\alpha_{\rm f}(1-g_i^2)]  +N(1-\delta_{i\mathrm C})\qty[
        \abs{y_{\mathrm N,A}}^2-g_i^2\abs{y_{\mathrm N,T}}^2].
    \label{appeq:flow_g}
\end{align}
The spin-zero coherent contributions have zero trace in the instantaneous normalized basis and therefore give no direct velocity correction in this equal-velocity approximation.
The neutral terms change $g_i$ only for $i\ne\mathrm C$.
These are \cref{eq:flow-ui-clean,eq:flow-gi-clean}.

\subsection{Single-mode limit and scaling dimensions}

For a Laughlin state $\nu=1/m$, set $N=1$ and
$p=(m-1)/2$. Then $\kappa_{\rm C}=m$, $g_{\rm C}=g$, and there is no
neutral disorder contribution. Equations
\eqref{appeq:flow_u} and \eqref{appeq:flow_g} become
\begin{align}
    \dv{u_{\rm C}}{\ell}
    &=-m\alpha_{\rm f}u_{\rm C}\qty(g+g^{-1}),
    \\
    \dv{g}{\ell}
    &=m\qty[
        \abs{y_A}^2-\abs{y_T}^2g^2+\alpha_{\rm f}\qty(1-g^2)
    ],
\end{align}
which is the single-mode flow of \citet{schiller2022interplay}.

Using
$\sum_{i=1}^{N-1}O_{ji}^2=(N-1)/N$ and
$O_{j{\rm C}}^2=1/N$, the dimensions are
\begin{align}
    d_A
    &=\frac{N-1}{N}g_{\rm N}^{-1}
      +\frac{1+2pN}{N}g_{\rm C}^{-1},
    \label{appeq:dim_A}\\
    d_T
    &=\frac{N-1}{N}g_{\rm N}
      +\frac{1+2pN}{N}g_{\rm C}.
    \label{appeq:dim_T}
\end{align}
These are \cref{eq:dim_A,eq:dim_T}. For $N=1$, they
reduce to $d_A=m/g$ and $d_T=mg$.

The fractional edge-electron scaling dimension is one quarter of their sum,
\begin{align}
    d_{e,\mathrm{f}}
    &=\frac{d_A+d_T}{4}
    \nonumber\\
    &=\frac{N-1}{4N}\qty(g_{\rm N}+g_{\rm N}^{-1})
      +\frac{1+2pN}{4N}\qty(g_{\rm C}+g_{\rm C}^{-1}).
\end{align}
Consequently,
$\min\qty(d_A,d_T)\leq2d_{e,\mathrm{f}}
\leq\max\qty(d_A,d_T)$, so dissipation of an electron in the fractional modes cannot
be strictly more relevant than both inter-side processes.

\subsection{Integer--fractional coupling for a Jain block}
\label{app:IF-spin-rg}
Let $s$ label the side of the integer electron and $\bar{s}$ the
opposite side. We take the chiral Jain block
$\Km_{\rm J}=\mathds{1}_N+2p\mathsf{J}_N$, with $N\geq1$ and
$p\geq1$, and set $y_{A,{\rm IF}}=0$ in this RG specialization.
The same-side and opposite-side normal IF transfers have spins
\begin{align}
    s_{E,{\rm IF},s}
    &=\frac{\epsilon_s}{2}\qty[1+\chi_{\rm f}(2p+1)],
    \label{appeq:IF-spin}\\
    s_{T,{\rm IF},s}
    &=\frac{\epsilon_s}{2}\qty[1-\chi_{\rm f}(2p+1)].
    \label{appeq:IF-opposite-spin}
\end{align}
Counter-propagating IF transfer therefore has $\abs{s_r}=p$:
it is same-side for $\chi_{\rm f}=-1$ and opposite-side for
$\chi_{\rm f}=+1$. Co-propagating transfer has $\abs{s_r}=p+1$.
In particular, the process involving counter-propagating modes at $\nu=4/3$ is
opposite-side tunneling.

On the block-diagonal Gaussian manifold the two transfers
involving fractional electron $j$ have the same dimension:
\begin{align}
    d_{{\rm IF},j}
    &=\frac{g_{\mathrm{I}}+g_{\mathrm{I}}^{-1}}{4}
      +\frac{1}{4}\sum_i\kappa_i O_{ji}^2(g_i+g_i^{-1}),\\
    \dot{y}_{X,{\rm IF},j}
    &=(2-d_{{\rm IF},j})y_{X,{\rm IF},j}.
\end{align}
Here $X\in\{E,T\}$. In the symmetric neutral sector this reduces
to \cref{eq:IF-dimension-general,eq:IF-flow}.

We now evaluate the neutral self-contractions at $g_{\mathrm{I}}=1$, with no
Gaussian integer--fractional mixing and equal IF amplitudes for all
physical fractional electrons. The normalized electron weight in
block $i$ is $\sqrt{\kappa_i}O_{ji}$. Two
identities fix, respectively, the block coefficient and conformal spin:
\begin{align}
    \sqrt{\kappa_i\kappa_k}\sum_jO_{ji}O_{jk}
    &=\kappa_i\delta_{ik},
    \label{appeq:IF-Jain-weights}\\
    \sum_i\kappa_iO_{ji}^2
    &=(\Km_{\rm J})_{jj}=2p+1.
\end{align}

First consider a shell at which all fractional normal modes have
a common velocity $u_{\rm f}$, while the integer velocity $u_{\mathrm{I}}$
can differ arbitrarily. Take the reference velocity $u=u_{\rm f}$,
and define
\begin{align}
    v&=\frac{u_{\mathrm{I}}}{u_{\rm f}},\\
    \xi&=\frac{1-v}{1+v}.
\end{align}
With equal amplitudes for the two side-exchanged orientations, let
\begin{align}
    Y_{\rm IF}
    &=\begin{cases}
        \abs{y_{E,{\rm IF}}}^2,&\chi_{\rm f}=-1,\\
        \abs{y_{T,{\rm IF}}}^2,&\chi_{\rm f}=+1.
    \end{cases}
    \label{appeq:IF-counter-strength}
\end{align}
Each amplitude is that of a single physical electron and orientation,
not a sum over species or sides.

To perform the angular integral, set $f_{\beta}=z_{\beta}/\varrho$. For the
orientation with a positive-chirality integer electron and a
negative-chirality fractional electron, the second identity above
gives the correlator factor
\begin{align}
    \prod_{j} f_{j}^{-\lambda_{rj}^2}
    &=\frac{2}{1+v}\,
      \frac{e^{2ip\vartheta}}{1-\xi e^{-2i\vartheta}}
    =\frac{2}{1+v}\sum_{n=0}^{\infty}
      \xi^n e^{2i(p-n)\vartheta}.
    \label{appeq:IF-angular-expansion}
\end{align}
The expansion converges for every $v>0$. The fractional numerator
$f_{\beta} f_{\gamma}$ selects $n=p+(\sigma_{\beta}+\sigma_{\gamma})/2$, namely
$p-1,p,p+1$. Numerators involving
an occupied integer component cancel its propagator factor and
have no constant harmonic for $p\geq1$. Thus these self-contractions
generate neither an integer-block nor an integer--fractional
quadratic correction under the stated assumptions.
Adding the side-exchanged orientation and summing over $j$ using
\cref{appeq:IF-Jain-weights} gives
\begin{align}
    \left.W_i^\transp\dot{\widehat{\Um}}_iW_i
    \right|_{\rm IF}
    &=-\frac{2\kappa_i u_{\rm f}Y_{\rm IF}}{1+v}\xi^{p-1}
      \qty[
        \qty(\cosh^2\theta_i+\xi^2\sinh^2\theta_i)\mathds{1}_2
        +2\xi\cosh\theta_i\sinh\theta_i\,\sigma_x
      ],
    \label{appeq:IF-projected-flow}
\end{align}
with no off-block-diagonal fractional correction.
Using \cref{appeq:rg-projected-derivative}, the two scalar
contributions are
\begin{align}
    \left.\dot{u}_i\right|_{\rm IF}
    &=-\kappa_i u_{\rm f}Y_{\rm IF}\xi^{p-1}
      \frac{(1+v^2)(g_i+g_i^{-1})+4v}{(1+v)^3},
    \label{appeq:IF-unequal-u-flow}\\
    \left.\dot{g}_i\right|_{\rm IF}
    &=\kappa_iY_{\rm IF}\frac{(1-v)^p}{(1+v)^{p+1}}(1-g_i^2).
    \label{appeq:IF-unequal-g-flow}
\end{align}
Both are direct $\mathcal{O}(\abs{y_{X,{\rm IF}}}^2)$ corrections.
At $v=1$, $\dot{g}_i|_{\rm IF}=0$ for every $p\geq1$.
The velocity correction survives only for $p=1$, where it is
$-\kappa_i u_{\rm f}Y_{\rm IF}\cosh^2\theta_i$.
For unequal velocities the interaction correction is generically
non-zero also for $p>1$. Near $v=1$,
\begin{align}
    \left.\dot{g}_i\right|_{\rm IF}
    &=\frac{\kappa_iY_{\rm IF}}{2^{p+1}}
      (1-v)^p(1-g_i^2)
      +\mathcal{O}\qty(Y_{\rm IF}\abs{1-v}^{p+1}).
    \label{appeq:IF-small-mismatch}
\end{align}
Thus only $p=1$ contributes at first order in the velocity
mismatch; $p=2$ starts at second order, and so on. These are all
second-order terms in the IF amplitude.
For $0<v<1$ they attract $g_i$ toward $1$ for every $p$.
For $v>1$ the sign reverses when $p$ is odd.
In the limit $v\to0^+$ the contributions for every $p\geq1$ reduce to
\begin{align}
    \left.\dot{u}_i\right|_{\rm IF}
    &\longrightarrow-\kappa_i u_{\rm f}Y_{\rm IF}(g_i+g_i^{-1}),\\
    \left.\dot{g}_i\right|_{\rm IF}
    &\longrightarrow\kappa_iY_{\rm IF}(1-g_i^2).
\end{align}
Only in this velocity limit do they have the scalar form obtained
by adding $Y_{\rm IF}$ to $\alpha_{\rm f}$.

For copropagating IF transfer, the corresponding correlator has
harmonics $e^{-2i(n+p+1)\vartheta}$. No quadratic descendant can
produce a constant harmonic, so their neutral self-contractions
give zero quadratic correction at this order for any $v>0$ on
this $g_{\mathrm{I}}=1$, common-fractional-velocity manifold. These operators
still enter the transport matrix.

The equal fractional velocities used in the closed formulas need
not remain equal under the flow: the different $\dot{u}_i$ above
can split them. Likewise, unequal $d_{{\rm IF},j}$ split
initially equal amplitudes. With distinct fractional velocities, nonuniform
IF amplitudes, $g_{\mathrm{I}}\ne1$, or Gaussian integer--fractional mixing,
use \cref{appeq:rg-angular-kernel,appeq:rg-angular-series} with the
instantaneous normalized vertex vectors. In general the resulting
matrix need not preserve the block-diagonal form. Alternatively,
using the common fractional velocity in subsequent shell
coefficients must be stated as an approximation.

Small fractional velocity splittings can also be expanded while
keeping $u_{\mathrm{I}}/u_{\rm f}$ arbitrary. At $g_{\mathrm{I}}=1$, the occupied integer
mode has $P_{\mathrm{I}}=1$ for a fractional matrix entry, so
$(P_{\mathrm{I}})_{n_{\mathrm{I}}}/n_{\mathrm{I}}!=1$. The Fourier constraint in
\cref{appeq:rg-angular-series} fixes $n_{\mathrm{I}}$ once the fractional
indices are chosen. Truncating only those fractional indices
therefore gives a systematic expansion in their velocity
splittings, without expanding the integer velocity ratio.
Mixed products that generate additional relevant vertices require
separate treatment.

The truncation throughout retains coherent quadratic terms of
order $\abs{y_X}^2$ and bath terms linear in
$\alpha_{s,j}\propto\abs{w_{s,j}}^2$. An additional integer-bath insertion
in an IF loop would be of order
$\alpha_{\mathrm{I}}\abs{y_{X,{\rm IF}}}^2$, fourth order in the microscopic
tunneling amplitudes, and is omitted. A calculation with the
integer bath fully resummed would require a separate expansion.
The velocity ratio in \cref{appeq:IF-unequal-u-flow,appeq:IF-unequal-g-flow}
is a parameter of the Gaussian action at the running scale; their
derivation uses no bath insertion.

\subsection{Random same-side processes}
For $N\geq2$, same-side transfer within a Jain block has neutral
vertex
$\vb{m}_{E,s;jk}=\vb{m}^{\rm tot}_{s,N_{\rm I}+j}-\vb{m}^{\rm tot}_{s,N_{\rm I}+k}$,
where $j,k=1,\ldots,N$ are fractional-local indices. Its only non-zero
block is $\epsilon_s\chi_{\rm f}(\vb{e}_j-\vb{e}_k)$ on the fractional
part of side $s$, with $\vb{e}_j\in\mathbb{R}^{N}$; all integer and
opposite-side entries vanish. Its spin is $\epsilon_s\chi_{\rm f}$, and it has no
charged-sector component after the $O$ rotation. With the
cutoff-normalized variance $D_E$ defined as in
\cref{eq:random-coupling-flow}, disorder averaging produces a
replicated term proportional to
\begin{align}
    D_E\sum_{\mathfrak{a},\mathfrak{b}}\int\dd{x}\dd{\tau}\dd{\tau'}\,
    V_{E,\mathfrak{a}}(x,\tau)V_{E,\mathfrak{b}}^\dagger(x,\tau').
\end{align}
Here $\mathfrak{a},\mathfrak{b}$ label replicas and $d_E=d_{\vb{m}_{E,s;jk}}$ is the scaling dimension for the specified side and mode pair; a common $d_E$ assumes equal dimensions for all included pairs. Power counting gives
\begin{align}
    \dot{D}_E&=(3-2d_E)D_E.
\end{align}
The quadratic correction is also obtained from a temporal shell at
fixed $x$, with the neutral electron-difference vector in place of
the electron vector. It therefore need not vanish for non-zero spin,
but its sector weights and normalization depend on the disorder
covariance. It must not be identified with an identical bath strength
in every block. No such term is included in the clean symmetric
matrix flow \eqref{appeq:matrix_flow}.

\end{widetext}

\end{document}